\documentclass[reprint,amsmath,amssymb,aps,]{revtex4-2}

\usepackage{xcolor}

\usepackage[utf8]{inputenc}
\usepackage{graphicx}

\usepackage{dcolumn}
\usepackage{bm}
\usepackage{soul}

\begin{document}

\title{Stimulated X-ray Raman scattering for selective preparation of \emph{dark} states bypassing optical selection rules}
\author{Francesco Montorsi}
\affiliation{Dipartimento di Chimica Industriale ``Toso Montanari'', Universit\`a di Bologna - Alma Mater Studiorum, Via Piero Gobetti 85, 40129 - Bologna, Italy}
\author{Shaul Mukamel}
\affiliation{Department of Chemistry and Department of Physics and Astronomy, University of California, Irvine, 92697, USA}
\author{Filippo Tamassia}
\author{Marco Garavelli}
\affiliation{Dipartimento di Chimica Industriale ``Toso Montanari'', Universit\`a di Bologna - Alma Mater Studiorum, Via Piero Gobetti 85, 40129 - Bologna, Italy}
\author{Francesco Segatta}
\email{francesco.segatta@unibo.it}
\affiliation{Dipartimento di Chimica Industriale ``Toso Montanari'', Universit\`a di Bologna - Alma Mater Studiorum, Via Piero Gobetti 85, 40129 - Bologna, Italy}
\author{Artur Nenov}
\email{artur.nenov@unibo.it}
\affiliation{Dipartimento di Chimica Industriale ``Toso Montanari'', Universit\`a di Bologna - Alma Mater Studiorum, Via Piero Gobetti 85, 40129 - Bologna, Italy}

\begin{abstract}
We present an X-ray based stimulated Raman approach to control the preparation of optically dark electronic states in generic molecular systems. Leveraging on the unique properties of core-level excited states, we demonstrate that  optically forbidden transitions between singlet states, or between singlet and triplet states are made accessible. Two molecular systems are studied as a test-bed of the proposed approach, and its experimental feasibility is eventually discussed.
\end{abstract}

\maketitle

\noindent Light-matter interaction is a fundamental process that has been extensively exploited to prepare and manipulate the evolution of desired electronic state(s) or superpositions thereof. A variety of photon energies can be employed, from the infrared to X-rays, triggering very diverse phenomena. The creation of the desired electronic state / superposition imposes the following requirements on the initial and final states: a) their energy difference needs to match the energy of the incident photon; b) they need to be dipole-coupled, i.e. the transition needs to be dipole-allowed. While the first requirement can be relatively easily satisfied by manipulating the laser source, the fulfillment of the second one depends on an intrinsic property of the system - the transition dipole moment - that can be altered only by manipulating the chemical system itself. Chemical systems often exhibit valence electronic excited states which are \emph{spectroscopically dark}, i.e. are characterized by vanishing small oscillator strength from the electronic ground state, leading to the inability of these states to undergo single-photon absorption/emission. 
Yet, dark states can mediate population transfer and are extremely relevant for a variety of processes such as ultra-fast internal conversion\cite{Nenov2018,Wolf2017}, energy and charge transfer,\cite{May2011} photoredox catalysis, photo-damage and thermally induced delayed fluorescence. One of the most prominent examples of the elusive involvement of dark states is the photoinduced internal conversion in DNA and RNA strands where the role of low-lying dark $n\pi^*$ has been the matter of debate for decades\cite{Improta2016}.

\noindent Preparing, manipulating and tracking the evolution of dark states via \emph{standard} spectroscopic techniques, is rather challenging. Transient absorption and photo-electron spectroscopy allows to track the fate of dark states as long as they give rise to absorption features in a background free spectral window. However, the inability to prepare selectively a dark state makes it difficult to identify its spectroscopic fingerprints next to those of the bright states. Therefore, experiments need to be accompanied by simulations parameterized with highly accurate electronic structure methods\cite{Conti2020}. A technique able to control the selective preparation of a molecule in a dark state would be of upmost importance for gaining unambiguous insight into their properties, deactivation mechanisms and reactivity. 

\noindent In this letter we show that stimulated X-ray Raman scattering (SXRS), one of the promising nonlinear X-ray techniques enabled by XFEL facilities,\cite{Chergui2023,Schwartz2023,Tanaka2002,Rohringer2019,Pfeifer2020} can be employed to selectively prepare molecular systems into a singlet or triplet state spectroscopically dark from the electronic ground state. To achieve this goal, we rely on dipole-coupling between valence and core-excited states, characterized by different selection rules with respect to \emph{optical} (UV/Vis) transitions. We also demonstrate how this scheme can be used to create electronic coherences between dark and bright states (e.g, singlet-singlet, singlet-triplet and triplet-triplet coherences) opening the possibility to study very uncommon and otherwise elusive states of matter. The technical implementation of the proposed SXRS experiment has been demonstrated recently by Cryan and co-workers,\cite{Cryan2020} who reported population of valence-excited states via impulsive SXRS at the NO oxygen K-edge, yet did not mention the potential application of the technique to bypass optical selection rules. 

\begin{figure}[t]
\centering
\includegraphics[width=0.46\textwidth,angle=0]{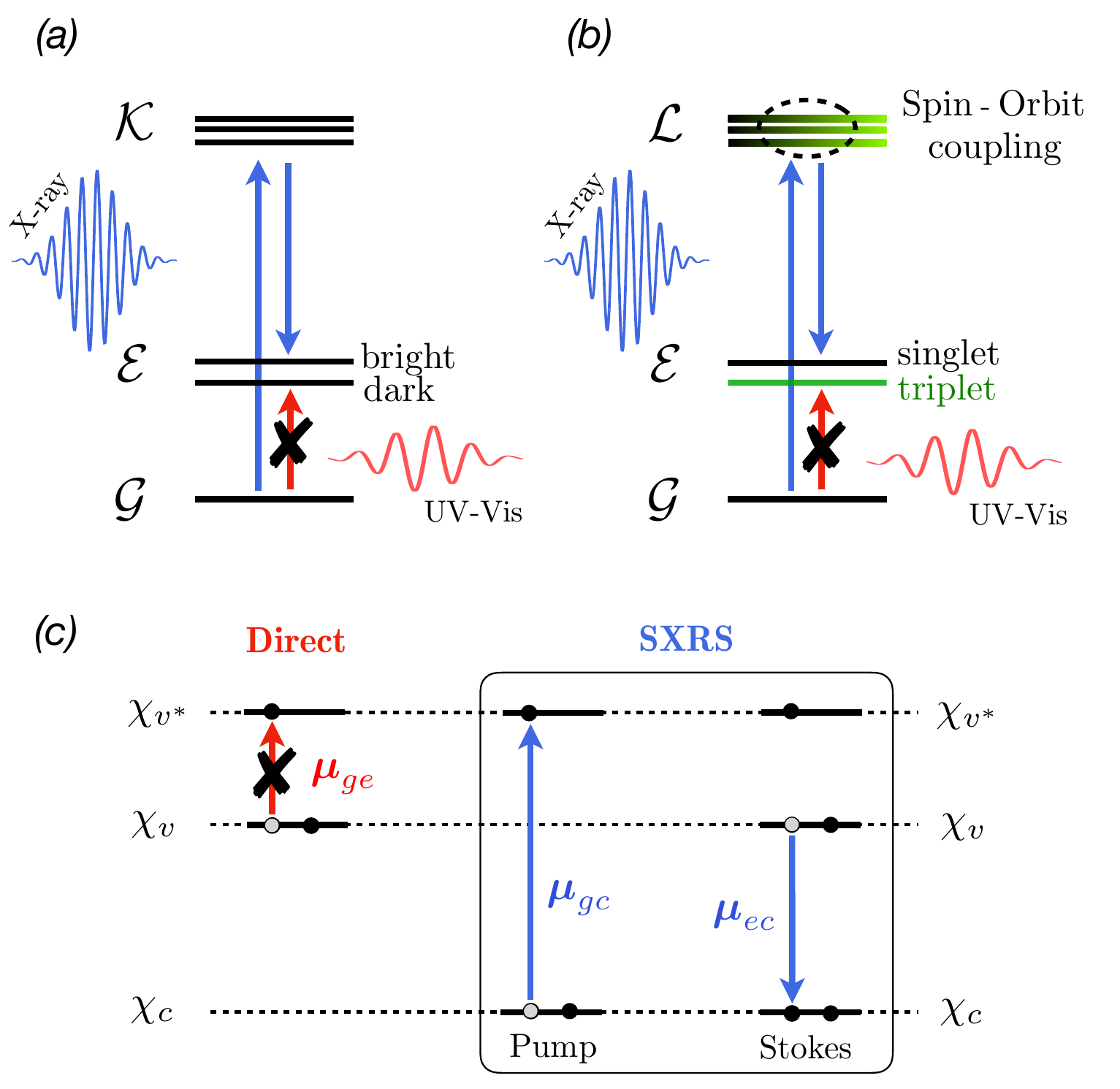}
\caption{Schematic of the proposed SXRS mechanism: (a) SXRS utilizing the K-edge manifold (denoted as $\mathcal{K}$) as a gateway for the population of dark singlet states in the valence manifold $\mathcal{E}$; (b) SXRS via the L-edge (denoted as $\mathcal{L}$) to selectively populate valence excited triplet states exploiting singlet-triplet mixing in the core-excited state manifold; (c) Schematic of the two step model that describes the SXRS process in the molecular orbital picture. Notably, the intensities of the transitions facilitated by the pump and Stokes depend on the spatial overlap between core and valence orbitals.}
\label{fig:scheme}
\end{figure}

\noindent A single attosecond X-ray pulse is here employed to promote the stimulated Raman process, resulting in the impulsive preparation of the desired state of matter. We explore the capabilities, requirements and limitations of this approach on two systems, namely: (a) \emph{trans}-azobenzene, for which we demonstrate that the dark $n\pi^*$ state can be populated with a near 100\% selectivity via nitrogen K-edge transitions; (b) thio-formaldehyde, for which the selective population of valence triplet states is demonstrated, capitalizing on the strong singlet-triplet mixing at the sulfur L-edge\cite{Kawerk2013}. A schematic description of these two mechanisms is depicted in Figure \ref{fig:scheme}-(a) and (b). 

\noindent High level ab-initio electronic structure calculations, based on multi-reference wave function theory (XMS-RASPT2)\cite{Malmqvist1990,Malmqvist2008} were performed to obtain the valence and core-excited singlet and triplet manifolds of the two systems\cite{Delcey2019}. Scalar relativistic effects are taken into account through a decoupling of the relativistic Dirac Hamiltonian via the exact-two-component (X2C) approach\cite{Peng2012}. Spin-orbit coupling is evaluated for the core excited states contributing to the L-edge\cite{Kasper2020}. All the electronic structure calculations are performed with the quantum chemistry software OpenMolcas.\cite{FdezGalvn2019,Aquilante2020,LiManni2023}  Details about the computational protocol are reported in the Supplemental Material.\cite{SupplementalMaterial}

\begin{figure}[b!]
\centering
\includegraphics[width=0.4\textwidth,angle=0]{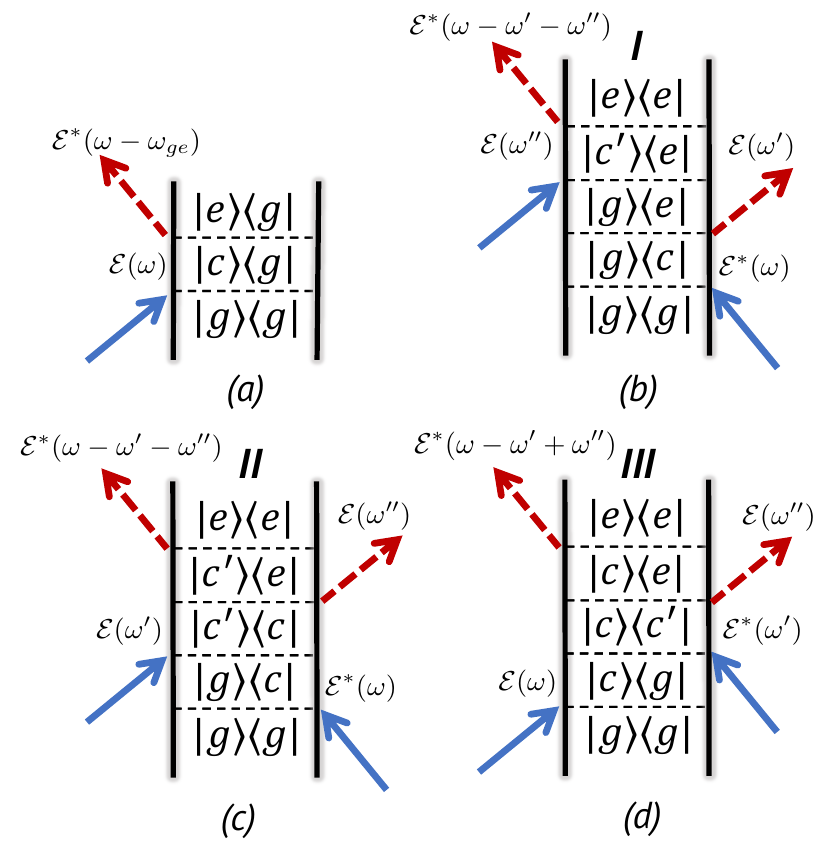}
\caption{Feynman diagrams for (a) the 2$^{nd}$ order and (b)-(c)-(d) the 4$^{th}$ order SXRS process, corresponding to preparation of coherences and populations, respectively. Electronic states are labeled as follows: $g$ is the ground state, $e$ is a valence excited state and $c$ is a core excited state. Only the components of the X-ray field $\boldsymbol{\mathcal{E}}(t)$ that survive after the rotating-wave approximation approximations are showed here by depicting their frequency domain envelope ($\mathcal{E}(\omega)$). Different colours for the arrows are used to depict excitation (\emph{pump}, blue arrow) and de-excitation (\emph{stokes}, red dashed arrows) interactions.}
\label{fig:FD}
\end{figure}

\noindent Excited state preparation via SXRS can be described by a two step process\cite{Mukamel2013}. A first interaction (named \emph{pump}) with an X-ray pulse promotes an electron from a \emph{core} orbital ($\chi_c$) into a \emph{virtual valence} orbital ($\chi_{v^*}$) populating a core excited state ($c$) denoted as $\chi_c^{[1]}\chi_{v^*}^{[1]}$ (molecular orbital occupation denoted in the square brackets). A second interaction (named \emph{stokes}) with the X-ray field drives an electron from an \emph{occupied valence} orbital ($\chi_v$) into the core vacancy, resulting in the population of a valence excited ($e$) state $\chi_{v}^{[1]}\chi_{v^*}^{[1]}$. A graphical description of this scheme is reported in Figure \ref{fig:scheme}-(c). The state of the system accessed at the end of such a process can be evaluated using a perturbation expansion of the molecular density matrix $\hat{\rho}(t)$ in powers of the X-ray field.\cite{Gelin2022} A diagrammatic approach can be used to visualize the different pathways that are initialized by the radiation-matter interaction. In particular, we will focus on two light-induced processes: a) generation of a coherence ($\rho_{ge}$) between the ground state ($g$) and a valence excited state ($e$), which is effectively described by second order perturbation theory (as depicted by the Feynman diagrams of Figure \ref{fig:FD}-(a)); b) generation of a valence state population ($\rho_{ee}$) where a perturbation expansion of the molecular density matrix up to the fourth order is required (Figure \ref{fig:FD}-(b),(c) and (d)).\footnote{Note that the process that brings the electron back to the ground state, captured by $\rho_{gg}$, is also possible. Nonetheless, since the focus here is on excited states, we will not explicitly show this pathway in what follows.} 

\noindent Let us denote with  $\omega_{ab}$ the transition energy between electronic states $a$ and $b$, where $a$ and $b$ are generic labels comprising the states from the three manifold of states here considered, namely, $\mathcal{G}$ (ground state), $\mathcal{E}$ (valence excited manifold), and $\mathcal{C}$ (core-excited manifold). We will indicate states of $\mathcal{G}$, $\mathcal{E}$ and $\mathcal{C}$ manifolds with the indexes $g$, $e$ and $c$, respectively. We also denote $\gamma_{ab}$ as the $a-b$ coherence dephasing time, and $\tilde\omega_{ab} = \displaystyle \omega_{ab}-i\gamma_{ab}/2$ as the complex $a-b$ frequency. The coherence prepared by the SXRS, in the dipole approximation\cite{MukamelBook}, reads:
\begin{equation}
    \rho^{(2)}_{eg}(t) =  -\frac{i}{\hbar^2} e^{-i\tilde{\omega}_{ge}t}\alpha^{(2)}_{eg}
 \label{coherence}
\end{equation}
where the term $\alpha^{(2)}_{eg}$ is the $e-g$ matrix element of the molecular polarizability operator $\hat{\alpha}$. Working in the impulsive limit (ISXRS), where all the required field matter interactions occurs under the time envelop of the same attosecond X-ray pulse, this expression reads:
\begin{equation}
    \begin{aligned}
    \alpha^{(2)}_{eg} = \sum_{c} (\boldsymbol{\hat{\epsilon}}^*\cdot\boldsymbol{\mu}_{ec})(\boldsymbol{\hat{\epsilon}}\cdot\boldsymbol{\mu}^{\dagger}_{gc}) \int_{-\infty}^{+\infty}\frac{d\omega}{2\pi} 
    \frac{\mathcal{E}(\omega) \mathcal{E}^*(\omega-\omega_{ge})}{\omega +  \omega_0 - \tilde{\omega}_{gc}}
    \end{aligned}
    \label{VPeg}
\end{equation}
where $\hat{\boldsymbol{\epsilon}}$ is the field polarization vector and $\boldsymbol{\hat{\mu}}_{ab}$ is the transition dipole moment of the $a\rightarrow b$ transition. $\mathcal{E}(\omega)$ and $\omega_0$ define the X-ray field frequency domain envelop and central frequency, respectively.
The  term  that describes the population of a valence state via the Raman process, can be  similarly obtained: here, three pathways need to be considered (see the Feynman diagrams reported in Figure \ref{fig:FD}-(b),(c) and (d)), so that the corresponding density matrix element, $\rho_{ee}^{(4)}(t)$, is the sum of three terms, i.e.:
\begin{equation}
    \rho_{ee}^{(4)}(t) = \frac{1}{\hbar^4}e^{-\gamma_{ee}t}2\Re\left[\alpha_{ee,I}^{(4)}+\alpha_{ee,II}^{(4)}+\alpha_{ee,III}^{(4)}\right]
    \label{population}
\end{equation}

\noindent and $\gamma_{ee}$ is the lifetime of the prepared valence excited state. In the impulsive limit $\alpha_{ee}^{(4)}$ reads:
\begin{widetext}
\begin{align}
    \label{alpha_eeA}
    \alpha_{ee,I}^{(4)} = & \sum_{c,c^\prime}\int_{-\infty}^{+\infty} \frac{d\omega}{2\pi} \int_{-\infty}^{+\infty} \frac{d\omega^\prime}{2\pi}\int_{-\infty}^{+\infty} \frac{d\omega^{\prime\prime}}{2\pi}\frac{(\boldsymbol{\hat{\epsilon}}^*\cdot\boldsymbol{\mu}_{gc})(\boldsymbol{\hat{\epsilon}}\cdot\boldsymbol{\mu}^\dagger_{ec})(\boldsymbol{\hat{\epsilon}}\cdot\boldsymbol{\mu}^\dagger_{gc^\prime})(\boldsymbol{\hat{\epsilon}}^*\cdot\boldsymbol{\mu}_{ec^\prime})\mathcal{E}^*(\omega)\mathcal{E}(\omega^\prime)\mathcal{E}(\omega^{\prime\prime})\mathcal{E}^*(\omega - \omega^{\prime}-\omega^{\prime\prime})}{i(\omega^{\prime\prime}+\omega^{\prime}-\omega+\omega_0-\tilde{\omega}_{ec^\prime})(\omega^\prime-\omega+\tilde{\omega}_{ge})(\omega+\omega_0-\tilde{\omega}_{gc})}\\
    \label{alpha_eeB}
    \alpha_{ee,II}^{(4)} = & \sum_{c,c^\prime} \int_{-\infty}^{+\infty} \frac{d\omega}{2\pi} \int_{-\infty}^{+\infty} \frac{d\omega^\prime}{2\pi} \int_{-\infty}^{+\infty} \frac{d\omega^{\prime\prime}}{2\pi} \frac{(\boldsymbol{\hat{\epsilon}}^*\cdot\boldsymbol{\mu}_{gc})(\boldsymbol{\hat{\epsilon}}\cdot\boldsymbol{\mu}^\dagger_{gc^\prime})(\boldsymbol{\hat{\epsilon}}\cdot\boldsymbol{\mu}^\dagger_{ec})(\boldsymbol{\hat{\epsilon}}^*\cdot\boldsymbol{\mu}_{ec^\prime})\mathcal{E}^*(\omega)\mathcal{E}(\omega^\prime)\mathcal{E}(\omega^{\prime\prime})\mathcal{E}^*(\omega - \omega^{\prime}-\omega^{\prime\prime})}{i(\omega^{\prime\prime}+\omega^{\prime}-\omega+\omega_0-\tilde{\omega}_{ec^\prime})(\omega^\prime-\omega-\tilde{\omega}_{cc^\prime})(\omega+\omega_0-\tilde{\omega}_{gc})}\\
    \label{alpha_eeC}
    \alpha_{ee,III}^{(4)} = & \sum_{c,c^\prime}\int_{-\infty}^{+\infty} \frac{d\omega}{2\pi} \int_{-\infty}^{+\infty} \frac{d\omega^\prime}{2\pi}\int_{-\infty}^{+\infty} \frac{d\omega^{\prime\prime}}{2\pi}\frac{(\boldsymbol{\hat{\epsilon}}\cdot\boldsymbol{\mu}^\dagger_{gc})(\boldsymbol{\hat{\epsilon}}^*\cdot\boldsymbol{\mu}_{gc^\prime})(\boldsymbol{\hat{\epsilon}}\cdot\boldsymbol{\mu}^\dagger_{ec^\prime})(\boldsymbol{\hat{\epsilon}}^*\cdot\boldsymbol{\mu}_{ec})\mathcal{E}(\omega)\mathcal{E}^*(\omega^\prime)\mathcal{E}(\omega^{\prime\prime})\mathcal{E}^*(\omega - \omega^{\prime}+\omega^{\prime\prime})}{i(\omega^{\prime\prime}-\omega^{\prime}+\omega+\omega_0-\tilde{\omega}_{ec})(\omega-\omega^\prime+\tilde{\omega}_{cc^\prime})(\omega+\omega_0-\tilde{\omega}_{gc})}
\end{align}
\end{widetext}

\noindent A complete derivation of eqs. \ref{VPeg}, \ref{alpha_eeA}, \ref{alpha_eeB} and \ref{alpha_eeC} is reported in the Supplemental Material.\cite{SupplementalMaterial} Notably, the reported expressions for the molecular polarizability are general and can be applied in both \emph{resonant} and \emph{off-resonant} regimes\cite{Mukamel2013,MukamelBook}.

\noindent The $\alpha_{ab}$ terms are sum-over states expressions which depend both on field properties (i.e., $\mathcal{E}(\omega)$,  $\omega_0$, $\boldsymbol{\hat{\epsilon}}$) and on molecular properties (i.e., $\boldsymbol{\mu}_{ab}$, $\tilde{\omega}_{ab}$). While control of the former can be exploited to enhance the ability of preparing the desired state of matter, in this letter we will focus on the latter and in particular on the role of transition dipole moments, since they are the most crucial factor determining the ability to access the dark valence state(s). Nevertheless, it is worth noticing that a minimal requirement to promote the desired ISXRS process is that the pulse spectral bandwidth must be grater than the energy gap $\omega_{ge}$ between the ground and desired valence $e$ state (see eq. \ref{VPeg}). 

 \noindent The direct access to valence states from the ground state relies on the $\boldsymbol{\mu}_{ge}$ magnitude, i.e. on the brightness of the $g\rightarrow e$ transition. Such transitions are forbidden in direct one-photon absorption due to: (i) spin change (e.g., singlet to triplet transitions); (ii) vanishing small \emph{spatial overlap}\footnote{Note that the term \emph{overlap} is here used to highlight a situation in which the two orbitals occupy the same region of the 3D space. It does not refer to the overlap integral which, by construction, is zero.} between $\chi_v$ and $\chi_{v^*}$ (e.g., charge transfer transitions); (iii) symmetry restrictions (e.g., $n\rightarrow\pi^*$ transitions in many conjugate and aromatic organic molecules). ISXRS, at variance, involves intermediate core excited states via the product of transition dipoles $\boldsymbol{\mu}_{gc}$ and $\boldsymbol{\mu}_{ce}$ (eqs. \ref{VPeg} to \ref{alpha_eeC}) whose magnitude is dictated by the extent of the \emph{spatial overlap} of the highly localized core and delocalized valence orbitals ($\chi_c/\chi_v$ and $\chi_c/\chi_{v^*}$) and by the \emph{local} symmetry around the atomic core-hole site\cite{Manne1970,Wolf2017}. For states belonging to the K-edge ($\mathcal{K}$), $\chi_c$ is a $1s$ orbital whose spherical shape guarantees symmetry allowed transitions toward any kind of orbital having a $p$ character. Valence \emph{frontier} orbitals in organic molecules ($v\in \{\sigma, \pi, n\}$; $v^* \in \{\sigma^*, \pi^*\}$) are mainly described by linear combination of $p$ atomic orbitals making all core transitions allowed. Therefore, the magnitude of both $\boldsymbol{\mu}_{gc}$ and $\boldsymbol{\mu}_{ce}$ is essentially dictated by the \emph{spatial overlap} between the valence molecular orbitals and the $1s$ core orbital: it is larger for valence orbitals with significant coefficients at the core-excited atoms (such as lone pairs, denoted $n$, localized on heteroatoms), and seller for valence orbitals delocalized over large portions of the molecular system (such as $\sigma/\sigma^*$ and $\pi/\pi^*$ orbitals).

\begin{widetext}

\begin{figure}[t]
\centering
\includegraphics[width=1.0\textwidth,angle=0]{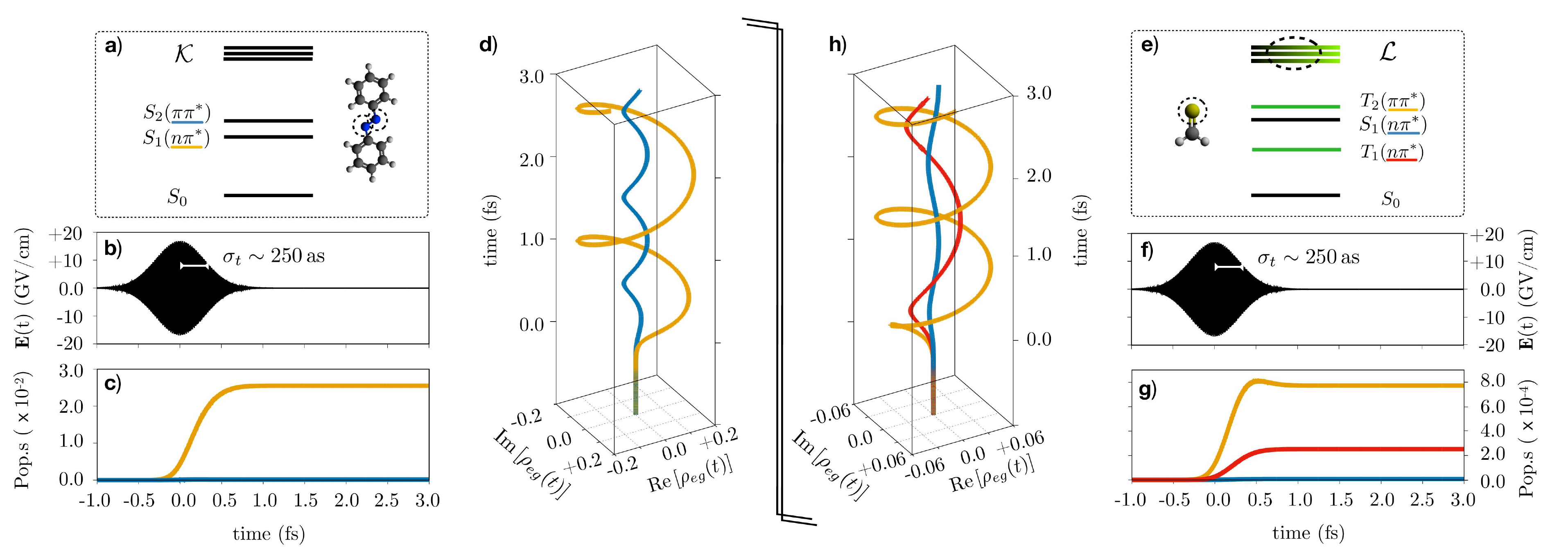}
\caption{Examples of dark singlet and triplet valence excited state preparation. (a/e) Scheme of the \emph{trans}-azobenzene and thio-formaldehyde electronic structure. (b/f) Gaussian shaped X-ray field used to induce the ISXRS process. (c/g) Evolution of the excited state populations $\rho_{ee}(t)$ during and after the interaction with the X-ray pulse. (d/h) Evolution of the real and imaginary parts of the coherences $\rho_{eg}(t)$. Throughout the figure, energy levels, populations and coherences involving the various excited states are consistently depicted with different colors.} 
\label{fig:SingletTriplet}
\end{figure}
\end{widetext}

\noindent The observation made suggests to use the K-edge of heteroatoms as the gateway to selectively access dark $n\pi^*$ states. This idea is here put into practice for the specific case of \emph{trans}-azobenzene, which exhibits a dark $n\pi^*$ state, placed around 2.7 eV, and a bright $\pi\pi^*$ state around 4.0 eV which dominates its linear absorption spectrum\cite{Nenov2018}. A schematic description of the electronic structure is reported in Figure \ref{fig:SingletTriplet}-(a). The system field-driven dynamics during the ISXRS process is here simulated solving the time-dependent Schr\"odinger equation for the electronic degrees of freedom through the QuTiP python package\cite{Johansson2012,Johansson2013}. To do so, a Gaussian shaped X-ray pulse characterized by a FWHM of 500\,as and an intensity of about $10^{+17}$\,Wcm$^{-2}$  (which mimics the pulses experimentally employed in ref. \citenum{Cryan2020}) has been introduced into the electronic Hamiltonian of the system and employed in the simulations (see Figure \ref{fig:SingletTriplet}-(b)). Population decay and coherence dephasing terms are assumed to be negligibly small in the few-fs time-scale explored (see the Supplemental Material\cite{SupplementalMaterial} for additional details).

\noindent The ISXRS process selectively prepares the dark $n\pi^*$ state in a two-step process $1s^{[2]}n^{[2]}\pi^{*[0]}\rightarrow 1s^{[1]}n^{[2]}\pi^{*[1]} \rightarrow 1s^{[2]}n^{[1]}\pi^{*[1]}$. This is documented in Figure \ref{fig:SingletTriplet}-(d) which shows the time evolution of coherence term $\rho_{eg}(t)$ for azobenzene, with $e$ being either the $\pi\pi^*$ or the $n\pi^*$ state. Notably, the magnitude of $\alpha_{eg}^{(2)}$ for $n\pi^*$ is about one order of magnitude larger than the one for the $\pi\pi^*$ state. A similar trend is observed when computing the excited state populations $\rho_{ee}(t)$, as reported in Figure \ref{fig:SingletTriplet}-(c). Here, the ISXRS process induced by the employed X-ray field produces an $n\pi^*$ population of about 3\% of the initial ground state population, while the $\pi\pi^*$ population is about two orders of magnitude lower.

\noindent 
In passing note that: a) the proposed Raman scheme can be generalized to target other manifolds as the gateway to prepare dark valence states, such as $\sigma\pi^*$ states via deep-UV pumping. Even if such transitions might be brighter than core-level ones, due to the high density of states it would be more challenging to achieve high selectivity in this energy window; b) the coupling between electronic and nuclear degrees of freedom may lower the molecular symmetry and symmetry forbidden transitions may acquire slight brightness. This is the case of \emph{trans}-azobenzene, for which a weak $n\pi^*$ band could be also directly excited from the ground-state, allowing to benchmark the proposed ISXRS process. 

\noindent Direct population of electronically excited states involving spin change is a much harder challenge since one-photon absorption is forbidden for these states by the so-called spin selection rule. We now demonstrate that a suitably designed ISXRS process can be used to selectively populate also the triplet manifold.

\noindent First, we note that triplet states in the $\mathcal{E}$ manifold might be in principle accessible in case of a large spin-orbit couplings (as in metal-organic complexes). This effect is manifested when the mass of the nuclei constituting the molecule are large enough to induce relativistic effects. However, the vast majority of organic molecules are characterized by atoms with relatively low mass, so that only very few of them display a significant spin-orbit coupling among the valence states. Third (or higher) row elements have access to the L-edge states ($\mathcal{L}$) which are characterized by a pronounced spin-orbit coupling (up to hundreds of meV) due to the direct involvement of (\emph{strongly coupled}) inner $2p$ orbitals. Utilizing $\mathcal{L}$ states as intermediates in the ISXRS process makes it possible to bypass the spin selection rule, by creating a superposition of core-excited states with different spin via the \emph{pump} interaction, which can than be selectively converted to pure triplet valence state via the \emph{stokes} interaction (see Figure \ref{fig:scheme}-(b)). The magnitude of both $\boldsymbol{\mu}_{gc}$ and $\boldsymbol{\mu}_{ce}$ depends on symmetry and \emph{spatial overlap} constrains (in analogy with the K-edge case). However, $\chi_c$ in this case corresponds to a triad of $2p$ orbitals (i.e. $2p_x$, $2p_y$ and $2p_z$) whose different orientation in the 3D space gives rise to a more complex set of system-dependent selection rules, that are discussed in the Supplemental Material.\cite{SupplementalMaterial}

\noindent This population of triplet states concept is here demonstrated on the example of thio-formaldehyde. Its lowest three electronic states are a singlet ($S_1$) state of  $n\pi^*$ character positioned in between two triplet states ($T_1$ and $T_2$) of $n\pi^*$ and $\pi\pi^*$ character, respectively (Figure-\ref{fig:SingletTriplet}-(e)). States in the $\mathcal{E}$ manifold display an almost negligible spin-orbit coupling, making the direct population of the aforementioned triplet states virtually impossible. Nevertheless, strong relativistic effects are manifested at the sulfur L-edge, leading to a strong singlet/triplet mixing among the $\mathcal{L}$ states.

\noindent Figure \ref{fig:SingletTriplet} reports the results obtained solving the time-dependent Schr\"odinger equation in the same condition discussed in the previous example, employing an X-ray pulse near resonant with the sulfur L-edge (as depicted in figure \ref{fig:SingletTriplet}-(f)). In particular, Figure \ref{fig:SingletTriplet}-(h) shows the $\rho_{eg}(t)$ prepared by the ISXRS process, which demonstrates the possibility to create a coherence between the  ground state and the valence excited triplet states. Notably, $g-T_1$ and $g-T_2$ coherences are prepared without any background from the $S_1$ state (see the Supplemental Material\cite{SupplementalMaterial} for additional details). A similar relative trend is observed for the triplet state population process, as evidenced by Figure \ref{fig:SingletTriplet}-(g). The slightly lower population achieved in this case ($\sim $1\% and $\sim $0.2\% for $T_1$ and $T_2$ states, respectively) is ascribed by the lower values of the thio-formaldehyde core-to-valence transition dipole moments (reported in the Supplemental Material\cite{SupplementalMaterial}).
 
\noindent The selectivity of the ISXRS can still be enhanced exploiting the dependence of eq. \ref{VPeg} to \ref{alpha_eeC} upon field parameters. Optimal control theory\cite{Somli1993} can indeed be used to increase the $\alpha_{ab}$ value of a desired (triplet or singlet) state even when the electronic structure parameters (transition energy and transition dipole moments) are not ideal per se for achieving selectivity.

\noindent To conclude, in this letter, we propose an impulsive stimulated Raman experiment for the selective population of \emph{optically} dark valence excited states. Notably, the proposed scheme can be used to create coherences and populations. In the same spirit, an approach to populate dark autoionizing singlet states has recently appeared in the literature\cite{Singh2023}. 
The approach presented here is more general as it allows to access a wider range of dark states below the ionization threshold. It also bypasses the spin selection rule relying on the unique properties (spin-orbit coupling) that characterize L-edge states. 

\noindent The selectivity of the ISXRS process towards different \emph{prototypical} valence excited states of organic cromophores is here investigated providing an intuitive \emph{rule of thumb} based on the: (i) \emph{local} symmetry of valence orbital around the excited core-hole; (ii) the \emph{spatial overlap} between valence and core molecular orbitals. This simple, but effective, picture is the manifestation of the atomic character of core excitation/de-excitation processes.

\noindent The ability of the proposed scheme to prepare a system in a superposition of singlet and triplet states directly at the Franck-Condon geometry allows to access  and shape coherences that may be exploited as  molecular electronic-spin qubits,\cite{Jirovec2021} paving the way for intriguing applications of the proposed technique in quantum computing. 

\noindent Interestingly, the yield of the proposed processes can be straightforwardly enhanced by leveraging over the linear (quadratic) dependence over the field intensity for coherences (populations). Strictly speaking this power law will be respected only in the perturbative regime (which is the case for a large range of intensities, as shown in the Supplemental Material\cite{SupplementalMaterial}), and when competing channels (ionization loss, Rabi flopping, Stark effects) are not dominant.\cite{Cryan2022bookchap}

Methods to detect such Raman prepared states, typically but not limited to using additional pulse(s), have been experimentally proven.\cite{Eichmann2020,Cryan2020} If such probe is an attosecond X-ray pulse,\cite{Duris2019} one would be able to study the electronic density evolution of completely unexplored states of matter.

\begin{acknowledgments}
F.S. acknowledges financial support from the ``Ecosystem for Sustainable Transition in Emilia-Romagna'' (ECOSISTER) project, code no. ECS00000033. 
Support from the U.S. Department of Energy, Office of Science, Office of Basic Energy Sciences, Chemical Sciences, Geosciences and Biosciences Division under award no. DE-SC0022225 is gratefully acknowledged.
\end{acknowledgments}

\bibliography{bib}

\begin{thebibliography}{35}%
\makeatletter
\providecommand \@ifxundefined [1]{%
 \@ifx{#1\undefined}
}%
\providecommand \@ifnum [1]{%
 \ifnum #1\expandafter \@firstoftwo
 \else \expandafter \@secondoftwo
 \fi
}%
\providecommand \@ifx [1]{%
 \ifx #1\expandafter \@firstoftwo
 \else \expandafter \@secondoftwo
 \fi
}%
\providecommand \natexlab [1]{#1}%
\providecommand \enquote  [1]{``#1''}%
\providecommand \bibnamefont  [1]{#1}%
\providecommand \bibfnamefont [1]{#1}%
\providecommand \citenamefont [1]{#1}%
\providecommand \href@noop [0]{\@secondoftwo}%
\providecommand \href [0]{\begingroup \@sanitize@url \@href}%
\providecommand \@href[1]{\@@startlink{#1}\@@href}%
\providecommand \@@href[1]{\endgroup#1\@@endlink}%
\providecommand \@sanitize@url [0]{\catcode `\\12\catcode `\$12\catcode
  `\&12\catcode `\#12\catcode `\^12\catcode `\_12\catcode `\%12\relax}%
\providecommand \@@startlink[1]{}%
\providecommand \@@endlink[0]{}%
\providecommand \url  [0]{\begingroup\@sanitize@url \@url }%
\providecommand \@url [1]{\endgroup\@href {#1}{\urlprefix }}%
\providecommand \urlprefix  [0]{URL }%
\providecommand \Eprint [0]{\href }%
\providecommand \doibase [0]{https://doi.org/}%
\providecommand \selectlanguage [0]{\@gobble}%
\providecommand \bibinfo  [0]{\@secondoftwo}%
\providecommand \bibfield  [0]{\@secondoftwo}%
\providecommand \translation [1]{[#1]}%
\providecommand \BibitemOpen [0]{}%
\providecommand \bibitemStop [0]{}%
\providecommand \bibitemNoStop [0]{.\EOS\space}%
\providecommand \EOS [0]{\spacefactor3000\relax}%
\providecommand \BibitemShut  [1]{\csname bibitem#1\endcsname}%
\let\auto@bib@innerbib\@empty
\bibitem [{\citenamefont {Nenov}\ \emph {et~al.}(2018)\citenamefont {Nenov},
  \citenamefont {Borrego-Varillas}, \citenamefont {Oriana}, \citenamefont
  {Ganzer}, \citenamefont {Segatta}, \citenamefont {Conti}, \citenamefont
  {Segarra-Marti}, \citenamefont {Omachi}, \citenamefont {Dapor}, \citenamefont
  {Taioli}, \citenamefont {Manzoni}, \citenamefont {Mukamel}, \citenamefont
  {Cerullo},\ and\ \citenamefont {Garavelli}}]{Nenov2018}%
  \BibitemOpen
  \bibfield  {author} {\bibinfo {author} {\bibfnamefont {A.}~\bibnamefont
  {Nenov}}, \bibinfo {author} {\bibfnamefont {R.}~\bibnamefont
  {Borrego-Varillas}}, \bibinfo {author} {\bibfnamefont {A.}~\bibnamefont
  {Oriana}}, \bibinfo {author} {\bibfnamefont {L.}~\bibnamefont {Ganzer}},
  \bibinfo {author} {\bibfnamefont {F.}~\bibnamefont {Segatta}}, \bibinfo
  {author} {\bibfnamefont {I.}~\bibnamefont {Conti}}, \bibinfo {author}
  {\bibfnamefont {J.}~\bibnamefont {Segarra-Marti}}, \bibinfo {author}
  {\bibfnamefont {J.}~\bibnamefont {Omachi}}, \bibinfo {author} {\bibfnamefont
  {M.}~\bibnamefont {Dapor}}, \bibinfo {author} {\bibfnamefont
  {S.}~\bibnamefont {Taioli}}, \bibinfo {author} {\bibfnamefont
  {C.}~\bibnamefont {Manzoni}}, \bibinfo {author} {\bibfnamefont
  {S.}~\bibnamefont {Mukamel}}, \bibinfo {author} {\bibfnamefont
  {G.}~\bibnamefont {Cerullo}},\ and\ \bibinfo {author} {\bibfnamefont
  {M.}~\bibnamefont {Garavelli}},\ }\bibfield  {title} {\bibinfo {title}
  {{UV}-light-induced vibrational coherences: The key to understand kasha rule
  violation in $trans$-azobenzene},\ }\href
  {https://doi.org/10.1021/acs.jpclett.8b00152} {\bibfield  {journal} {\bibinfo
   {journal} {The Journal of Physical Chemistry Letters}\ }\textbf {\bibinfo
  {volume} {9}},\ \bibinfo {pages} {1534} (\bibinfo {year} {2018})}\BibitemShut
  {NoStop}%
\bibitem [{\citenamefont {Wolf}\ \emph {et~al.}(2017)\citenamefont {Wolf},
  \citenamefont {Myhre}, \citenamefont {Cryan}, \citenamefont {Coriani},
  \citenamefont {Squibb}, \citenamefont {Battistoni}, \citenamefont {Berrah},
  \citenamefont {Bostedt}, \citenamefont {Bucksbaum}, \citenamefont
  {Coslovich}, \citenamefont {Feifel}, \citenamefont {Gaffney}, \citenamefont
  {Grilj}, \citenamefont {Martinez}, \citenamefont {Miyabe}, \citenamefont
  {Moeller}, \citenamefont {Mucke}, \citenamefont {Natan}, \citenamefont
  {Obaid}, \citenamefont {Osipov}, \citenamefont {Plekan}, \citenamefont
  {Wang}, \citenamefont {Koch},\ and\ \citenamefont {G\"{u}hr}}]{Wolf2017}%
  \BibitemOpen
  \bibfield  {author} {\bibinfo {author} {\bibfnamefont {T.~J.~A.}\
  \bibnamefont {Wolf}}, \bibinfo {author} {\bibfnamefont {R.~H.}\ \bibnamefont
  {Myhre}}, \bibinfo {author} {\bibfnamefont {J.~P.}\ \bibnamefont {Cryan}},
  \bibinfo {author} {\bibfnamefont {S.}~\bibnamefont {Coriani}}, \bibinfo
  {author} {\bibfnamefont {R.~J.}\ \bibnamefont {Squibb}}, \bibinfo {author}
  {\bibfnamefont {A.}~\bibnamefont {Battistoni}}, \bibinfo {author}
  {\bibfnamefont {N.}~\bibnamefont {Berrah}}, \bibinfo {author} {\bibfnamefont
  {C.}~\bibnamefont {Bostedt}}, \bibinfo {author} {\bibfnamefont
  {P.}~\bibnamefont {Bucksbaum}}, \bibinfo {author} {\bibfnamefont
  {G.}~\bibnamefont {Coslovich}}, \bibinfo {author} {\bibfnamefont
  {R.}~\bibnamefont {Feifel}}, \bibinfo {author} {\bibfnamefont {K.~J.}\
  \bibnamefont {Gaffney}}, \bibinfo {author} {\bibfnamefont {J.}~\bibnamefont
  {Grilj}}, \bibinfo {author} {\bibfnamefont {T.~J.}\ \bibnamefont {Martinez}},
  \bibinfo {author} {\bibfnamefont {S.}~\bibnamefont {Miyabe}}, \bibinfo
  {author} {\bibfnamefont {S.~P.}\ \bibnamefont {Moeller}}, \bibinfo {author}
  {\bibfnamefont {M.}~\bibnamefont {Mucke}}, \bibinfo {author} {\bibfnamefont
  {A.}~\bibnamefont {Natan}}, \bibinfo {author} {\bibfnamefont
  {R.}~\bibnamefont {Obaid}}, \bibinfo {author} {\bibfnamefont
  {T.}~\bibnamefont {Osipov}}, \bibinfo {author} {\bibfnamefont
  {O.}~\bibnamefont {Plekan}}, \bibinfo {author} {\bibfnamefont
  {S.}~\bibnamefont {Wang}}, \bibinfo {author} {\bibfnamefont {H.}~\bibnamefont
  {Koch}},\ and\ \bibinfo {author} {\bibfnamefont {M.}~\bibnamefont
  {G\"{u}hr}},\ }\bibfield  {title} {\bibinfo {title} {Probing ultrafast
  $\pi\pi^*$/$n\pi^*$ internal conversion in organic chromophores via k-edge
  resonant absorption},\ }\bibfield  {journal} {\bibinfo  {journal} {Nature
  Communications}\ }\textbf {\bibinfo {volume} {8}},\ \href
  {https://doi.org/10.1038/s41467-017-00069-7} {10.1038/s41467-017-00069-7}
  (\bibinfo {year} {2017})\BibitemShut {NoStop}%
\bibitem [{\citenamefont {May}\ and\ \citenamefont {K\"{u}hn}(2011)}]{May2011}%
  \BibitemOpen
  \bibfield  {author} {\bibinfo {author} {\bibfnamefont {V.}~\bibnamefont
  {May}}\ and\ \bibinfo {author} {\bibfnamefont {O.}~\bibnamefont {K\"{u}hn}},\
  }\href {https://doi.org/10.1002/9783527633791} {\emph {\bibinfo {title}
  {Charge and Energy Transfer Dynamics in Molecular Systems}}}\ (\bibinfo
  {publisher} {Wiley},\ \bibinfo {year} {2011})\BibitemShut {NoStop}%
\bibitem [{\citenamefont {Improta}\ \emph {et~al.}(2016)\citenamefont
  {Improta}, \citenamefont {Santoro},\ and\ \citenamefont
  {Blancafort}}]{Improta2016}%
  \BibitemOpen
  \bibfield  {author} {\bibinfo {author} {\bibfnamefont {R.}~\bibnamefont
  {Improta}}, \bibinfo {author} {\bibfnamefont {F.}~\bibnamefont {Santoro}},\
  and\ \bibinfo {author} {\bibfnamefont {L.}~\bibnamefont {Blancafort}},\
  }\bibfield  {title} {\bibinfo {title} {Quantum mechanical studies on the
  photophysics and the photochemistry of nucleic acids and nucleobases},\
  }\href {https://doi.org/10.1021/acs.chemrev.5b00444} {\bibfield  {journal}
  {\bibinfo  {journal} {Chemical Reviews}\ }\textbf {\bibinfo {volume} {116}},\
  \bibinfo {pages} {3540–3593} (\bibinfo {year} {2016})}\BibitemShut
  {NoStop}%
\bibitem [{\citenamefont {Conti}\ \emph {et~al.}(2020)\citenamefont {Conti},
  \citenamefont {Cerullo}, \citenamefont {Nenov},\ and\ \citenamefont
  {Garavelli}}]{Conti2020}%
  \BibitemOpen
  \bibfield  {author} {\bibinfo {author} {\bibfnamefont {I.}~\bibnamefont
  {Conti}}, \bibinfo {author} {\bibfnamefont {G.}~\bibnamefont {Cerullo}},
  \bibinfo {author} {\bibfnamefont {A.}~\bibnamefont {Nenov}},\ and\ \bibinfo
  {author} {\bibfnamefont {M.}~\bibnamefont {Garavelli}},\ }\bibfield  {title}
  {\bibinfo {title} {Ultrafast spectroscopy of photoactive molecular systems
  from first principles: Where we stand today and where we are going},\ }\href
  {https://doi.org/10.1021/jacs.0c04952} {\bibfield  {journal} {\bibinfo
  {journal} {Journal of the American Chemical Society}\ }\textbf {\bibinfo
  {volume} {142}},\ \bibinfo {pages} {16117} (\bibinfo {year}
  {2020})}\BibitemShut {NoStop}%
\bibitem [{\citenamefont {Chergui}\ \emph {et~al.}(2023)\citenamefont
  {Chergui}, \citenamefont {Beye}, \citenamefont {Mukamel}, \citenamefont
  {Svetina},\ and\ \citenamefont {Masciovecchio}}]{Chergui2023}%
  \BibitemOpen
  \bibfield  {author} {\bibinfo {author} {\bibfnamefont {M.}~\bibnamefont
  {Chergui}}, \bibinfo {author} {\bibfnamefont {M.}~\bibnamefont {Beye}},
  \bibinfo {author} {\bibfnamefont {S.}~\bibnamefont {Mukamel}}, \bibinfo
  {author} {\bibfnamefont {C.}~\bibnamefont {Svetina}},\ and\ \bibinfo {author}
  {\bibfnamefont {C.}~\bibnamefont {Masciovecchio}},\ }\bibfield  {title}
  {\bibinfo {title} {Progress and prospects in nonlinear extreme-ultraviolet
  and x-ray optics and spectroscopy},\ }\href
  {https://doi.org/10.1038/s42254-023-00643-7} {\bibfield  {journal} {\bibinfo
  {journal} {Nature Reviews Physics}\ }\textbf {\bibinfo {volume} {5}},\
  \bibinfo {pages} {578–596} (\bibinfo {year} {2023})}\BibitemShut {NoStop}%
\bibitem [{\citenamefont {Schwartz}\ and\ \citenamefont
  {Drisdell}(2023)}]{Schwartz2023}%
  \BibitemOpen
  \bibfield  {author} {\bibinfo {author} {\bibfnamefont {C.~P.}\ \bibnamefont
  {Schwartz}}\ and\ \bibinfo {author} {\bibfnamefont {W.~S.}\ \bibnamefont
  {Drisdell}},\ }\bibinfo {title} {Nonlinear soft x-ray spectroscopy},\ in\
  \href {https://doi.org/10.1007/978-981-99-6714-8_4} {\emph {\bibinfo
  {booktitle} {Nonlinear X-Ray Spectroscopy for Materials Science}}}\ (\bibinfo
   {publisher} {Springer Nature Singapore},\ \bibinfo {year} {2023})\ p.\
  \bibinfo {pages} {83–118}\BibitemShut {NoStop}%
\bibitem [{\citenamefont {Tanaka}\ and\ \citenamefont
  {Mukamel}(2002)}]{Tanaka2002}%
  \BibitemOpen
  \bibfield  {author} {\bibinfo {author} {\bibfnamefont {S.}~\bibnamefont
  {Tanaka}}\ and\ \bibinfo {author} {\bibfnamefont {S.}~\bibnamefont
  {Mukamel}},\ }\bibfield  {title} {\bibinfo {title} {Coherent x-ray raman
  spectroscopy: A nonlinear local probe for electronic excitations},\
  }\bibfield  {journal} {\bibinfo  {journal} {Physical Review Letters}\
  }\textbf {\bibinfo {volume} {89}},\ \href
  {https://doi.org/10.1103/physrevlett.89.043001}
  {10.1103/physrevlett.89.043001} (\bibinfo {year} {2002})\BibitemShut
  {NoStop}%
\bibitem [{\citenamefont {Rohringer}(2019)}]{Rohringer2019}%
  \BibitemOpen
  \bibfield  {author} {\bibinfo {author} {\bibfnamefont {N.}~\bibnamefont
  {Rohringer}},\ }\bibfield  {title} {\bibinfo {title} {X-ray raman scattering:
  a building block for nonlinear spectroscopy},\ }\href
  {https://doi.org/10.1098/rsta.2017.0471} {\bibfield  {journal} {\bibinfo
  {journal} {Philosophical Transactions of the Royal Society A: Mathematical,
  Physical and Engineering Sciences}\ }\textbf {\bibinfo {volume} {377}},\
  \bibinfo {pages} {20170471} (\bibinfo {year} {2019})}\BibitemShut {NoStop}%
\bibitem [{\citenamefont {Pfeifer}(2020)}]{Pfeifer2020}%
  \BibitemOpen
  \bibfield  {author} {\bibinfo {author} {\bibfnamefont {T.}~\bibnamefont
  {Pfeifer}},\ }\bibfield  {title} {\bibinfo {title} {Intense x-rays can be
  (slightly) exciting},\ }\href {https://doi.org/10.1126/science.abd6168}
  {\bibfield  {journal} {\bibinfo  {journal} {Science}\ }\textbf {\bibinfo
  {volume} {369}},\ \bibinfo {pages} {1568–1569} (\bibinfo {year}
  {2020})}\BibitemShut {NoStop}%
\bibitem [{\citenamefont {O'Neal}\ \emph {et~al.}(2020)\citenamefont {O'Neal},
  \citenamefont {Champenois}, \citenamefont {Oberli}, \citenamefont {Obaid},
  \citenamefont {Al-Haddad}, \citenamefont {Barnard}, \citenamefont {Berrah},
  \citenamefont {Coffee}, \citenamefont {Duris}, \citenamefont {Galinis},
  \citenamefont {Garratt}, \citenamefont {Glownia}, \citenamefont {Haxton},
  \citenamefont {Ho}, \citenamefont {Li}, \citenamefont {Li}, \citenamefont
  {MacArthur}, \citenamefont {Marangos}, \citenamefont {Natan}, \citenamefont
  {Shivaram}, \citenamefont {Slaughter}, \citenamefont {Walter}, \citenamefont
  {Wandel}, \citenamefont {Young}, \citenamefont {Bostedt}, \citenamefont
  {Bucksbaum}, \citenamefont {Pic{\'{o}}n}, \citenamefont {Marinelli},\ and\
  \citenamefont {Cryan}}]{Cryan2020}%
  \BibitemOpen
  \bibfield  {author} {\bibinfo {author} {\bibfnamefont {J.~T.}\ \bibnamefont
  {O'Neal}}, \bibinfo {author} {\bibfnamefont {E.~G.}\ \bibnamefont
  {Champenois}}, \bibinfo {author} {\bibfnamefont {S.}~\bibnamefont {Oberli}},
  \bibinfo {author} {\bibfnamefont {R.}~\bibnamefont {Obaid}}, \bibinfo
  {author} {\bibfnamefont {A.}~\bibnamefont {Al-Haddad}}, \bibinfo {author}
  {\bibfnamefont {J.}~\bibnamefont {Barnard}}, \bibinfo {author} {\bibfnamefont
  {N.}~\bibnamefont {Berrah}}, \bibinfo {author} {\bibfnamefont
  {R.}~\bibnamefont {Coffee}}, \bibinfo {author} {\bibfnamefont
  {J.}~\bibnamefont {Duris}}, \bibinfo {author} {\bibfnamefont
  {G.}~\bibnamefont {Galinis}}, \bibinfo {author} {\bibfnamefont
  {D.}~\bibnamefont {Garratt}}, \bibinfo {author} {\bibfnamefont {J.~M.}\
  \bibnamefont {Glownia}}, \bibinfo {author} {\bibfnamefont {D.}~\bibnamefont
  {Haxton}}, \bibinfo {author} {\bibfnamefont {P.}~\bibnamefont {Ho}}, \bibinfo
  {author} {\bibfnamefont {S.}~\bibnamefont {Li}}, \bibinfo {author}
  {\bibfnamefont {X.}~\bibnamefont {Li}}, \bibinfo {author} {\bibfnamefont
  {J.}~\bibnamefont {MacArthur}}, \bibinfo {author} {\bibfnamefont {J.~P.}\
  \bibnamefont {Marangos}}, \bibinfo {author} {\bibfnamefont {A.}~\bibnamefont
  {Natan}}, \bibinfo {author} {\bibfnamefont {N.}~\bibnamefont {Shivaram}},
  \bibinfo {author} {\bibfnamefont {D.~S.}\ \bibnamefont {Slaughter}}, \bibinfo
  {author} {\bibfnamefont {P.}~\bibnamefont {Walter}}, \bibinfo {author}
  {\bibfnamefont {S.}~\bibnamefont {Wandel}}, \bibinfo {author} {\bibfnamefont
  {L.}~\bibnamefont {Young}}, \bibinfo {author} {\bibfnamefont
  {C.}~\bibnamefont {Bostedt}}, \bibinfo {author} {\bibfnamefont {P.~H.}\
  \bibnamefont {Bucksbaum}}, \bibinfo {author} {\bibfnamefont {A.}~\bibnamefont
  {Pic{\'{o}}n}}, \bibinfo {author} {\bibfnamefont {A.}~\bibnamefont
  {Marinelli}},\ and\ \bibinfo {author} {\bibfnamefont {J.~P.}\ \bibnamefont
  {Cryan}},\ }\bibfield  {title} {\bibinfo {title} {Electronic population
  transfer via impulsive stimulated x-ray raman scattering with attosecond
  soft-x-ray pulses},\ }\bibfield  {journal} {\bibinfo  {journal} {Physical
  Review Letters}\ }\textbf {\bibinfo {volume} {125}},\ \href
  {https://doi.org/10.1103/physrevlett.125.073203}
  {10.1103/physrevlett.125.073203} (\bibinfo {year} {2020})\BibitemShut
  {NoStop}%
\bibitem [{\citenamefont {Kawerk}\ \emph {et~al.}(2013)\citenamefont {Kawerk},
  \citenamefont {Carniato}, \citenamefont {Iwayama}, \citenamefont {Shigemasa},
  \citenamefont {Piancastelli}, \citenamefont {Wassaf}, \citenamefont
  {Khoury},\ and\ \citenamefont {Simon}}]{Kawerk2013}%
  \BibitemOpen
  \bibfield  {author} {\bibinfo {author} {\bibfnamefont {E.}~\bibnamefont
  {Kawerk}}, \bibinfo {author} {\bibfnamefont {S.}~\bibnamefont {Carniato}},
  \bibinfo {author} {\bibfnamefont {H.}~\bibnamefont {Iwayama}}, \bibinfo
  {author} {\bibfnamefont {E.}~\bibnamefont {Shigemasa}}, \bibinfo {author}
  {\bibfnamefont {M.~N.}\ \bibnamefont {Piancastelli}}, \bibinfo {author}
  {\bibfnamefont {J.}~\bibnamefont {Wassaf}}, \bibinfo {author} {\bibfnamefont
  {A.}~\bibnamefont {Khoury}},\ and\ \bibinfo {author} {\bibfnamefont
  {M.}~\bibnamefont {Simon}},\ }\bibfield  {title} {\bibinfo {title}
  {Experimental and theoretical study of x-ray absorption around the chlorine l
  edge in vinyl chloride},\ }\href
  {https://doi.org/10.1016/j.elspec.2013.01.018} {\bibfield  {journal}
  {\bibinfo  {journal} {Journal of Electron Spectroscopy and Related
  Phenomena}\ }\textbf {\bibinfo {volume} {186}},\ \bibinfo {pages} {1}
  (\bibinfo {year} {2013})}\BibitemShut {NoStop}%
\bibitem [{\citenamefont {Malmqvist}\ \emph {et~al.}(1990)\citenamefont
  {Malmqvist}, \citenamefont {Rendell},\ and\ \citenamefont
  {Roos}}]{Malmqvist1990}%
  \BibitemOpen
  \bibfield  {author} {\bibinfo {author} {\bibfnamefont {P.~A.}\ \bibnamefont
  {Malmqvist}}, \bibinfo {author} {\bibfnamefont {A.}~\bibnamefont {Rendell}},\
  and\ \bibinfo {author} {\bibfnamefont {B.~O.}\ \bibnamefont {Roos}},\
  }\bibfield  {title} {\bibinfo {title} {The restricted active space
  self-consistent-field method, implemented with a split graph unitary group
  approach},\ }\href {https://doi.org/10.1021/j100377a011} {\bibfield
  {journal} {\bibinfo  {journal} {The Journal of Physical Chemistry}\ }\textbf
  {\bibinfo {volume} {94}},\ \bibinfo {pages} {5477} (\bibinfo {year}
  {1990})}\BibitemShut {NoStop}%
\bibitem [{\citenamefont {Malmqvist}\ \emph {et~al.}(2008)\citenamefont
  {Malmqvist}, \citenamefont {Pierloot}, \citenamefont {Shahi}, \citenamefont
  {Cramer},\ and\ \citenamefont {Gagliardi}}]{Malmqvist2008}%
  \BibitemOpen
  \bibfield  {author} {\bibinfo {author} {\bibfnamefont {P.~{\AA}.}\
  \bibnamefont {Malmqvist}}, \bibinfo {author} {\bibfnamefont {K.}~\bibnamefont
  {Pierloot}}, \bibinfo {author} {\bibfnamefont {A.~R.~M.}\ \bibnamefont
  {Shahi}}, \bibinfo {author} {\bibfnamefont {C.~J.}\ \bibnamefont {Cramer}},\
  and\ \bibinfo {author} {\bibfnamefont {L.}~\bibnamefont {Gagliardi}},\
  }\bibfield  {title} {\bibinfo {title} {The restricted active space followed
  by second-order perturbation theory method: Theory and application to the
  study of {CuO}2 and cu2o2 systems},\ }\href
  {https://doi.org/10.1063/1.2920188} {\bibfield  {journal} {\bibinfo
  {journal} {The Journal of Chemical Physics}\ }\textbf {\bibinfo {volume}
  {128}},\ \bibinfo {pages} {204109} (\bibinfo {year} {2008})}\BibitemShut
  {NoStop}%
\bibitem [{\citenamefont {Delcey}\ \emph {et~al.}(2019)\citenamefont {Delcey},
  \citenamefont {S{\o}rensen}, \citenamefont {Vacher}, \citenamefont {Couto},\
  and\ \citenamefont {Lundberg}}]{Delcey2019}%
  \BibitemOpen
  \bibfield  {author} {\bibinfo {author} {\bibfnamefont {M.~G.}\ \bibnamefont
  {Delcey}}, \bibinfo {author} {\bibfnamefont {L.~K.}\ \bibnamefont
  {S{\o}rensen}}, \bibinfo {author} {\bibfnamefont {M.}~\bibnamefont {Vacher}},
  \bibinfo {author} {\bibfnamefont {R.~C.}\ \bibnamefont {Couto}},\ and\
  \bibinfo {author} {\bibfnamefont {M.}~\bibnamefont {Lundberg}},\ }\bibfield
  {title} {\bibinfo {title} {Efficient calculations of a large number of highly
  excited states for multiconfigurational wavefunctions},\ }\href
  {https://doi.org/10.1002/jcc.25832} {\bibfield  {journal} {\bibinfo
  {journal} {Journal of Computational Chemistry}\ }\textbf {\bibinfo {volume}
  {40}},\ \bibinfo {pages} {1789} (\bibinfo {year} {2019})}\BibitemShut
  {NoStop}%
\bibitem [{\citenamefont {Peng}\ and\ \citenamefont {Reiher}(2012)}]{Peng2012}%
  \BibitemOpen
  \bibfield  {author} {\bibinfo {author} {\bibfnamefont {D.}~\bibnamefont
  {Peng}}\ and\ \bibinfo {author} {\bibfnamefont {M.}~\bibnamefont {Reiher}},\
  }\bibfield  {title} {\bibinfo {title} {Exact decoupling of the relativistic
  fock operator},\ }\bibfield  {journal} {\bibinfo  {journal} {Theoretical
  Chemistry Accounts}\ }\textbf {\bibinfo {volume} {131}},\ \href
  {https://doi.org/10.1007/s00214-011-1081-y} {10.1007/s00214-011-1081-y}
  (\bibinfo {year} {2012})\BibitemShut {NoStop}%
\bibitem [{\citenamefont {Kasper}\ \emph {et~al.}(2020)\citenamefont {Kasper},
  \citenamefont {Stetina}, \citenamefont {Jenkins},\ and\ \citenamefont
  {Li}}]{Kasper2020}%
  \BibitemOpen
  \bibfield  {author} {\bibinfo {author} {\bibfnamefont {J.~M.}\ \bibnamefont
  {Kasper}}, \bibinfo {author} {\bibfnamefont {T.~F.}\ \bibnamefont {Stetina}},
  \bibinfo {author} {\bibfnamefont {A.~J.}\ \bibnamefont {Jenkins}},\ and\
  \bibinfo {author} {\bibfnamefont {X.}~\bibnamefont {Li}},\ }\bibfield
  {title} {\bibinfo {title} {Ab initio methods for l-edge x-ray absorption
  spectroscopy},\ }\bibfield  {journal} {\bibinfo  {journal} {Chemical Physics
  Reviews}\ }\textbf {\bibinfo {volume} {1}},\ \href
  {https://doi.org/10.1063/5.0029725} {10.1063/5.0029725} (\bibinfo {year}
  {2020})\BibitemShut {NoStop}%
\bibitem [{\citenamefont {Galv{\'{a}}n}\ \emph {et~al.}(2019)\citenamefont
  {Galv{\'{a}}n}, \citenamefont {Vacher}, \citenamefont {Alavi}, \citenamefont
  {Angeli}, \citenamefont {Aquilante}, \citenamefont {Autschbach},
  \citenamefont {Bao}, \citenamefont {Bokarev}, \citenamefont {Bogdanov},
  \citenamefont {Carlson}, \citenamefont {Chibotaru}, \citenamefont
  {Creutzberg}, \citenamefont {Dattani}, \citenamefont {Delcey}, \citenamefont
  {Dong}, \citenamefont {Dreuw}, \citenamefont {Freitag}, \citenamefont
  {Frutos}, \citenamefont {Gagliardi}, \citenamefont {Gendron}, \citenamefont
  {Giussani}, \citenamefont {Gonz{\'{a}}lez}, \citenamefont {Grell},
  \citenamefont {Guo}, \citenamefont {Hoyer}, \citenamefont {Johansson},
  \citenamefont {Keller}, \citenamefont {Knecht}, \citenamefont
  {Kova{\v{c}}evi{\'{c}}}, \citenamefont {K\"{a}llman}, \citenamefont {Manni},
  \citenamefont {Lundberg}, \citenamefont {Ma}, \citenamefont {Mai},
  \citenamefont {Malhado}, \citenamefont {Malmqvist}, \citenamefont
  {Marquetand}, \citenamefont {Mewes}, \citenamefont {Norell}, \citenamefont
  {Olivucci}, \citenamefont {Oppel}, \citenamefont {Phung}, \citenamefont
  {Pierloot}, \citenamefont {Plasser}, \citenamefont {Reiher}, \citenamefont
  {Sand}, \citenamefont {Schapiro}, \citenamefont {Sharma}, \citenamefont
  {Stein}, \citenamefont {S{\o}rensen}, \citenamefont {Truhlar}, \citenamefont
  {Ugandi}, \citenamefont {Ungur}, \citenamefont {Valentini}, \citenamefont
  {Vancoillie}, \citenamefont {Veryazov}, \citenamefont {Weser}, \citenamefont
  {Weso{\l}owski}, \citenamefont {Widmark}, \citenamefont {Wouters},
  \citenamefont {Zech}, \citenamefont {Zobel},\ and\ \citenamefont
  {Lindh}}]{FdezGalvn2019}%
  \BibitemOpen
  \bibfield  {author} {\bibinfo {author} {\bibfnamefont {I.~F.}\ \bibnamefont
  {Galv{\'{a}}n}}, \bibinfo {author} {\bibfnamefont {M.}~\bibnamefont
  {Vacher}}, \bibinfo {author} {\bibfnamefont {A.}~\bibnamefont {Alavi}},
  \bibinfo {author} {\bibfnamefont {C.}~\bibnamefont {Angeli}}, \bibinfo
  {author} {\bibfnamefont {F.}~\bibnamefont {Aquilante}}, \bibinfo {author}
  {\bibfnamefont {J.}~\bibnamefont {Autschbach}}, \bibinfo {author}
  {\bibfnamefont {J.~J.}\ \bibnamefont {Bao}}, \bibinfo {author} {\bibfnamefont
  {S.~I.}\ \bibnamefont {Bokarev}}, \bibinfo {author} {\bibfnamefont {N.~A.}\
  \bibnamefont {Bogdanov}}, \bibinfo {author} {\bibfnamefont {R.~K.}\
  \bibnamefont {Carlson}}, \bibinfo {author} {\bibfnamefont {L.~F.}\
  \bibnamefont {Chibotaru}}, \bibinfo {author} {\bibfnamefont {J.}~\bibnamefont
  {Creutzberg}}, \bibinfo {author} {\bibfnamefont {N.}~\bibnamefont {Dattani}},
  \bibinfo {author} {\bibfnamefont {M.~G.}\ \bibnamefont {Delcey}}, \bibinfo
  {author} {\bibfnamefont {S.~S.}\ \bibnamefont {Dong}}, \bibinfo {author}
  {\bibfnamefont {A.}~\bibnamefont {Dreuw}}, \bibinfo {author} {\bibfnamefont
  {L.}~\bibnamefont {Freitag}}, \bibinfo {author} {\bibfnamefont {L.~M.}\
  \bibnamefont {Frutos}}, \bibinfo {author} {\bibfnamefont {L.}~\bibnamefont
  {Gagliardi}}, \bibinfo {author} {\bibfnamefont {F.}~\bibnamefont {Gendron}},
  \bibinfo {author} {\bibfnamefont {A.}~\bibnamefont {Giussani}}, \bibinfo
  {author} {\bibfnamefont {L.}~\bibnamefont {Gonz{\'{a}}lez}}, \bibinfo
  {author} {\bibfnamefont {G.}~\bibnamefont {Grell}}, \bibinfo {author}
  {\bibfnamefont {M.}~\bibnamefont {Guo}}, \bibinfo {author} {\bibfnamefont
  {C.~E.}\ \bibnamefont {Hoyer}}, \bibinfo {author} {\bibfnamefont
  {M.}~\bibnamefont {Johansson}}, \bibinfo {author} {\bibfnamefont
  {S.}~\bibnamefont {Keller}}, \bibinfo {author} {\bibfnamefont
  {S.}~\bibnamefont {Knecht}}, \bibinfo {author} {\bibfnamefont
  {G.}~\bibnamefont {Kova{\v{c}}evi{\'{c}}}}, \bibinfo {author} {\bibfnamefont
  {E.}~\bibnamefont {K\"{a}llman}}, \bibinfo {author} {\bibfnamefont {G.~L.}\
  \bibnamefont {Manni}}, \bibinfo {author} {\bibfnamefont {M.}~\bibnamefont
  {Lundberg}}, \bibinfo {author} {\bibfnamefont {Y.}~\bibnamefont {Ma}},
  \bibinfo {author} {\bibfnamefont {S.}~\bibnamefont {Mai}}, \bibinfo {author}
  {\bibfnamefont {J.~P.}\ \bibnamefont {Malhado}}, \bibinfo {author}
  {\bibfnamefont {P.~{\AA}.}\ \bibnamefont {Malmqvist}}, \bibinfo {author}
  {\bibfnamefont {P.}~\bibnamefont {Marquetand}}, \bibinfo {author}
  {\bibfnamefont {S.~A.}\ \bibnamefont {Mewes}}, \bibinfo {author}
  {\bibfnamefont {J.}~\bibnamefont {Norell}}, \bibinfo {author} {\bibfnamefont
  {M.}~\bibnamefont {Olivucci}}, \bibinfo {author} {\bibfnamefont
  {M.}~\bibnamefont {Oppel}}, \bibinfo {author} {\bibfnamefont {Q.~M.}\
  \bibnamefont {Phung}}, \bibinfo {author} {\bibfnamefont {K.}~\bibnamefont
  {Pierloot}}, \bibinfo {author} {\bibfnamefont {F.}~\bibnamefont {Plasser}},
  \bibinfo {author} {\bibfnamefont {M.}~\bibnamefont {Reiher}}, \bibinfo
  {author} {\bibfnamefont {A.~M.}\ \bibnamefont {Sand}}, \bibinfo {author}
  {\bibfnamefont {I.}~\bibnamefont {Schapiro}}, \bibinfo {author}
  {\bibfnamefont {P.}~\bibnamefont {Sharma}}, \bibinfo {author} {\bibfnamefont
  {C.~J.}\ \bibnamefont {Stein}}, \bibinfo {author} {\bibfnamefont {L.~K.}\
  \bibnamefont {S{\o}rensen}}, \bibinfo {author} {\bibfnamefont {D.~G.}\
  \bibnamefont {Truhlar}}, \bibinfo {author} {\bibfnamefont {M.}~\bibnamefont
  {Ugandi}}, \bibinfo {author} {\bibfnamefont {L.}~\bibnamefont {Ungur}},
  \bibinfo {author} {\bibfnamefont {A.}~\bibnamefont {Valentini}}, \bibinfo
  {author} {\bibfnamefont {S.}~\bibnamefont {Vancoillie}}, \bibinfo {author}
  {\bibfnamefont {V.}~\bibnamefont {Veryazov}}, \bibinfo {author}
  {\bibfnamefont {O.}~\bibnamefont {Weser}}, \bibinfo {author} {\bibfnamefont
  {T.~A.}\ \bibnamefont {Weso{\l}owski}}, \bibinfo {author} {\bibfnamefont
  {P.-O.}\ \bibnamefont {Widmark}}, \bibinfo {author} {\bibfnamefont
  {S.}~\bibnamefont {Wouters}}, \bibinfo {author} {\bibfnamefont
  {A.}~\bibnamefont {Zech}}, \bibinfo {author} {\bibfnamefont {J.~P.}\
  \bibnamefont {Zobel}},\ and\ \bibinfo {author} {\bibfnamefont
  {R.}~\bibnamefont {Lindh}},\ }\bibfield  {title} {\bibinfo {title}
  {{OpenMolcas}: From source code to insight},\ }\href
  {https://doi.org/10.1021/acs.jctc.9b00532} {\bibfield  {journal} {\bibinfo
  {journal} {Journal of Chemical Theory and Computation}\ }\textbf {\bibinfo
  {volume} {15}},\ \bibinfo {pages} {5925} (\bibinfo {year}
  {2019})}\BibitemShut {NoStop}%
\bibitem [{\citenamefont {Aquilante}\ \emph {et~al.}(2020)\citenamefont
  {Aquilante}, \citenamefont {Autschbach}, \citenamefont {Baiardi},
  \citenamefont {Battaglia}, \citenamefont {Borin}, \citenamefont {Chibotaru},
  \citenamefont {Conti}, \citenamefont {Vico}, \citenamefont {Delcey},
  \citenamefont {Galv{\'{a}}n}, \citenamefont {Ferr{\'{e}}}, \citenamefont
  {Freitag}, \citenamefont {Garavelli}, \citenamefont {Gong}, \citenamefont
  {Knecht}, \citenamefont {Larsson}, \citenamefont {Lindh}, \citenamefont
  {Lundberg}, \citenamefont {Malmqvist}, \citenamefont {Nenov}, \citenamefont
  {Norell}, \citenamefont {Odelius}, \citenamefont {Olivucci}, \citenamefont
  {Pedersen}, \citenamefont {Pedraza-Gonz{\'{a}}lez}, \citenamefont {Phung},
  \citenamefont {Pierloot}, \citenamefont {Reiher}, \citenamefont {Schapiro},
  \citenamefont {Segarra-Mart{\'{\i}}}, \citenamefont {Segatta}, \citenamefont
  {Seijo}, \citenamefont {Sen}, \citenamefont {Sergentu}, \citenamefont
  {Stein}, \citenamefont {Ungur}, \citenamefont {Vacher}, \citenamefont
  {Valentini},\ and\ \citenamefont {Veryazov}}]{Aquilante2020}%
  \BibitemOpen
  \bibfield  {author} {\bibinfo {author} {\bibfnamefont {F.}~\bibnamefont
  {Aquilante}}, \bibinfo {author} {\bibfnamefont {J.}~\bibnamefont
  {Autschbach}}, \bibinfo {author} {\bibfnamefont {A.}~\bibnamefont {Baiardi}},
  \bibinfo {author} {\bibfnamefont {S.}~\bibnamefont {Battaglia}}, \bibinfo
  {author} {\bibfnamefont {V.~A.}\ \bibnamefont {Borin}}, \bibinfo {author}
  {\bibfnamefont {L.~F.}\ \bibnamefont {Chibotaru}}, \bibinfo {author}
  {\bibfnamefont {I.}~\bibnamefont {Conti}}, \bibinfo {author} {\bibfnamefont
  {L.~D.}\ \bibnamefont {Vico}}, \bibinfo {author} {\bibfnamefont
  {M.}~\bibnamefont {Delcey}}, \bibinfo {author} {\bibfnamefont {I.~F.}\
  \bibnamefont {Galv{\'{a}}n}}, \bibinfo {author} {\bibfnamefont
  {N.}~\bibnamefont {Ferr{\'{e}}}}, \bibinfo {author} {\bibfnamefont
  {L.}~\bibnamefont {Freitag}}, \bibinfo {author} {\bibfnamefont
  {M.}~\bibnamefont {Garavelli}}, \bibinfo {author} {\bibfnamefont
  {X.}~\bibnamefont {Gong}}, \bibinfo {author} {\bibfnamefont {S.}~\bibnamefont
  {Knecht}}, \bibinfo {author} {\bibfnamefont {E.~D.}\ \bibnamefont {Larsson}},
  \bibinfo {author} {\bibfnamefont {R.}~\bibnamefont {Lindh}}, \bibinfo
  {author} {\bibfnamefont {M.}~\bibnamefont {Lundberg}}, \bibinfo {author}
  {\bibfnamefont {P.~{\AA}.}\ \bibnamefont {Malmqvist}}, \bibinfo {author}
  {\bibfnamefont {A.}~\bibnamefont {Nenov}}, \bibinfo {author} {\bibfnamefont
  {J.}~\bibnamefont {Norell}}, \bibinfo {author} {\bibfnamefont
  {M.}~\bibnamefont {Odelius}}, \bibinfo {author} {\bibfnamefont
  {M.}~\bibnamefont {Olivucci}}, \bibinfo {author} {\bibfnamefont {T.~B.}\
  \bibnamefont {Pedersen}}, \bibinfo {author} {\bibfnamefont {L.}~\bibnamefont
  {Pedraza-Gonz{\'{a}}lez}}, \bibinfo {author} {\bibfnamefont {Q.~M.}\
  \bibnamefont {Phung}}, \bibinfo {author} {\bibfnamefont {K.}~\bibnamefont
  {Pierloot}}, \bibinfo {author} {\bibfnamefont {M.}~\bibnamefont {Reiher}},
  \bibinfo {author} {\bibfnamefont {I.}~\bibnamefont {Schapiro}}, \bibinfo
  {author} {\bibfnamefont {J.}~\bibnamefont {Segarra-Mart{\'{\i}}}}, \bibinfo
  {author} {\bibfnamefont {F.}~\bibnamefont {Segatta}}, \bibinfo {author}
  {\bibfnamefont {L.}~\bibnamefont {Seijo}}, \bibinfo {author} {\bibfnamefont
  {S.}~\bibnamefont {Sen}}, \bibinfo {author} {\bibfnamefont {D.-C.}\
  \bibnamefont {Sergentu}}, \bibinfo {author} {\bibfnamefont {C.~J.}\
  \bibnamefont {Stein}}, \bibinfo {author} {\bibfnamefont {L.}~\bibnamefont
  {Ungur}}, \bibinfo {author} {\bibfnamefont {M.}~\bibnamefont {Vacher}},
  \bibinfo {author} {\bibfnamefont {A.}~\bibnamefont {Valentini}},\ and\
  \bibinfo {author} {\bibfnamefont {V.}~\bibnamefont {Veryazov}},\ }\bibfield
  {title} {\bibinfo {title} {Modern quantum chemistry with [open]molcas},\
  }\href {https://doi.org/10.1063/5.0004835} {\bibfield  {journal} {\bibinfo
  {journal} {The Journal of Chemical Physics}\ }\textbf {\bibinfo {volume}
  {152}},\ \bibinfo {pages} {214117} (\bibinfo {year} {2020})}\BibitemShut
  {NoStop}%
\bibitem [{\citenamefont {Li~Manni}\ \emph {et~al.}(2023)\citenamefont
  {Li~Manni}, \citenamefont {Fdez.~Galván}, \citenamefont {Alavi},
  \citenamefont {Aleotti}, \citenamefont {Aquilante}, \citenamefont
  {Autschbach}, \citenamefont {Avagliano}, \citenamefont {Baiardi},
  \citenamefont {Bao}, \citenamefont {Battaglia}, \citenamefont {Birnoschi},
  \citenamefont {Blanco-González}, \citenamefont {Bokarev}, \citenamefont
  {Broer}, \citenamefont {Cacciari}, \citenamefont {Calio}, \citenamefont
  {Carlson}, \citenamefont {Carvalho~Couto}, \citenamefont {Cerdán},
  \citenamefont {Chibotaru}, \citenamefont {Chilton}, \citenamefont {Church},
  \citenamefont {Conti}, \citenamefont {Coriani}, \citenamefont
  {Cuéllar-Zuquin}, \citenamefont {Daoud}, \citenamefont {Dattani},
  \citenamefont {Decleva}, \citenamefont {de~Graaf}, \citenamefont {Delcey},
  \citenamefont {De~Vico}, \citenamefont {Dobrautz}, \citenamefont {Dong},
  \citenamefont {Feng}, \citenamefont {Ferré}, \citenamefont {Filatov(Gulak)},
  \citenamefont {Gagliardi}, \citenamefont {Garavelli}, \citenamefont
  {González}, \citenamefont {Guan}, \citenamefont {Guo}, \citenamefont
  {Hennefarth}, \citenamefont {Hermes}, \citenamefont {Hoyer}, \citenamefont
  {Huix-Rotllant}, \citenamefont {Jaiswal}, \citenamefont {Kaiser},
  \citenamefont {Kaliakin}, \citenamefont {Khamesian}, \citenamefont {King},
  \citenamefont {Kochetov}, \citenamefont {Krośnicki}, \citenamefont {Kumaar},
  \citenamefont {Larsson}, \citenamefont {Lehtola}, \citenamefont {Lepetit},
  \citenamefont {Lischka}, \citenamefont {López~Ríos}, \citenamefont
  {Lundberg}, \citenamefont {Ma}, \citenamefont {Mai}, \citenamefont
  {Marquetand}, \citenamefont {Merritt}, \citenamefont {Montorsi},
  \citenamefont {M\"{o}rchen}, \citenamefont {Nenov}, \citenamefont {Nguyen},
  \citenamefont {Nishimoto}, \citenamefont {Oakley}, \citenamefont {Olivucci},
  \citenamefont {Oppel}, \citenamefont {Padula}, \citenamefont {Pandharkar},
  \citenamefont {Phung}, \citenamefont {Plasser}, \citenamefont {Raggi},
  \citenamefont {Rebolini}, \citenamefont {Reiher}, \citenamefont {Rivalta},
  \citenamefont {Roca-Sanjuán}, \citenamefont {Romig}, \citenamefont {Safari},
  \citenamefont {Sánchez-Mansilla}, \citenamefont {Sand}, \citenamefont
  {Schapiro}, \citenamefont {Scott}, \citenamefont {Segarra-Martí},
  \citenamefont {Segatta}, \citenamefont {Sergentu}, \citenamefont {Sharma},
  \citenamefont {Shepard}, \citenamefont {Shu}, \citenamefont {Staab},
  \citenamefont {Straatsma}, \citenamefont {Sørensen}, \citenamefont
  {Tenorio}, \citenamefont {Truhlar}, \citenamefont {Ungur}, \citenamefont
  {Vacher}, \citenamefont {Veryazov}, \citenamefont {Voß}, \citenamefont
  {Weser}, \citenamefont {Wu}, \citenamefont {Yang}, \citenamefont {Yarkony},
  \citenamefont {Zhou}, \citenamefont {Zobel},\ and\ \citenamefont
  {Lindh}}]{LiManni2023}%
  \BibitemOpen
  \bibfield  {author} {\bibinfo {author} {\bibfnamefont {G.}~\bibnamefont
  {Li~Manni}}, \bibinfo {author} {\bibfnamefont {I.}~\bibnamefont
  {Fdez.~Galván}}, \bibinfo {author} {\bibfnamefont {A.}~\bibnamefont
  {Alavi}}, \bibinfo {author} {\bibfnamefont {F.}~\bibnamefont {Aleotti}},
  \bibinfo {author} {\bibfnamefont {F.}~\bibnamefont {Aquilante}}, \bibinfo
  {author} {\bibfnamefont {J.}~\bibnamefont {Autschbach}}, \bibinfo {author}
  {\bibfnamefont {D.}~\bibnamefont {Avagliano}}, \bibinfo {author}
  {\bibfnamefont {A.}~\bibnamefont {Baiardi}}, \bibinfo {author} {\bibfnamefont
  {J.~J.}\ \bibnamefont {Bao}}, \bibinfo {author} {\bibfnamefont
  {S.}~\bibnamefont {Battaglia}}, \bibinfo {author} {\bibfnamefont
  {L.}~\bibnamefont {Birnoschi}}, \bibinfo {author} {\bibfnamefont
  {A.}~\bibnamefont {Blanco-González}}, \bibinfo {author} {\bibfnamefont
  {S.~I.}\ \bibnamefont {Bokarev}}, \bibinfo {author} {\bibfnamefont
  {R.}~\bibnamefont {Broer}}, \bibinfo {author} {\bibfnamefont
  {R.}~\bibnamefont {Cacciari}}, \bibinfo {author} {\bibfnamefont {P.~B.}\
  \bibnamefont {Calio}}, \bibinfo {author} {\bibfnamefont {R.~K.}\ \bibnamefont
  {Carlson}}, \bibinfo {author} {\bibfnamefont {R.}~\bibnamefont
  {Carvalho~Couto}}, \bibinfo {author} {\bibfnamefont {L.}~\bibnamefont
  {Cerdán}}, \bibinfo {author} {\bibfnamefont {L.~F.}\ \bibnamefont
  {Chibotaru}}, \bibinfo {author} {\bibfnamefont {N.~F.}\ \bibnamefont
  {Chilton}}, \bibinfo {author} {\bibfnamefont {J.~R.}\ \bibnamefont {Church}},
  \bibinfo {author} {\bibfnamefont {I.}~\bibnamefont {Conti}}, \bibinfo
  {author} {\bibfnamefont {S.}~\bibnamefont {Coriani}}, \bibinfo {author}
  {\bibfnamefont {J.}~\bibnamefont {Cuéllar-Zuquin}}, \bibinfo {author}
  {\bibfnamefont {R.~E.}\ \bibnamefont {Daoud}}, \bibinfo {author}
  {\bibfnamefont {N.}~\bibnamefont {Dattani}}, \bibinfo {author} {\bibfnamefont
  {P.}~\bibnamefont {Decleva}}, \bibinfo {author} {\bibfnamefont
  {C.}~\bibnamefont {de~Graaf}}, \bibinfo {author} {\bibfnamefont {M.~G.}\
  \bibnamefont {Delcey}}, \bibinfo {author} {\bibfnamefont {L.}~\bibnamefont
  {De~Vico}}, \bibinfo {author} {\bibfnamefont {W.}~\bibnamefont {Dobrautz}},
  \bibinfo {author} {\bibfnamefont {S.~S.}\ \bibnamefont {Dong}}, \bibinfo
  {author} {\bibfnamefont {R.}~\bibnamefont {Feng}}, \bibinfo {author}
  {\bibfnamefont {N.}~\bibnamefont {Ferré}}, \bibinfo {author} {\bibfnamefont
  {M.}~\bibnamefont {Filatov(Gulak)}}, \bibinfo {author} {\bibfnamefont
  {L.}~\bibnamefont {Gagliardi}}, \bibinfo {author} {\bibfnamefont
  {M.}~\bibnamefont {Garavelli}}, \bibinfo {author} {\bibfnamefont
  {L.}~\bibnamefont {González}}, \bibinfo {author} {\bibfnamefont
  {Y.}~\bibnamefont {Guan}}, \bibinfo {author} {\bibfnamefont {M.}~\bibnamefont
  {Guo}}, \bibinfo {author} {\bibfnamefont {M.~R.}\ \bibnamefont {Hennefarth}},
  \bibinfo {author} {\bibfnamefont {M.~R.}\ \bibnamefont {Hermes}}, \bibinfo
  {author} {\bibfnamefont {C.~E.}\ \bibnamefont {Hoyer}}, \bibinfo {author}
  {\bibfnamefont {M.}~\bibnamefont {Huix-Rotllant}}, \bibinfo {author}
  {\bibfnamefont {V.~K.}\ \bibnamefont {Jaiswal}}, \bibinfo {author}
  {\bibfnamefont {A.}~\bibnamefont {Kaiser}}, \bibinfo {author} {\bibfnamefont
  {D.~S.}\ \bibnamefont {Kaliakin}}, \bibinfo {author} {\bibfnamefont
  {M.}~\bibnamefont {Khamesian}}, \bibinfo {author} {\bibfnamefont {D.~S.}\
  \bibnamefont {King}}, \bibinfo {author} {\bibfnamefont {V.}~\bibnamefont
  {Kochetov}}, \bibinfo {author} {\bibfnamefont {M.}~\bibnamefont
  {Krośnicki}}, \bibinfo {author} {\bibfnamefont {A.~A.}\ \bibnamefont
  {Kumaar}}, \bibinfo {author} {\bibfnamefont {E.~D.}\ \bibnamefont {Larsson}},
  \bibinfo {author} {\bibfnamefont {S.}~\bibnamefont {Lehtola}}, \bibinfo
  {author} {\bibfnamefont {M.-B.}\ \bibnamefont {Lepetit}}, \bibinfo {author}
  {\bibfnamefont {H.}~\bibnamefont {Lischka}}, \bibinfo {author} {\bibfnamefont
  {P.}~\bibnamefont {López~Ríos}}, \bibinfo {author} {\bibfnamefont
  {M.}~\bibnamefont {Lundberg}}, \bibinfo {author} {\bibfnamefont
  {D.}~\bibnamefont {Ma}}, \bibinfo {author} {\bibfnamefont {S.}~\bibnamefont
  {Mai}}, \bibinfo {author} {\bibfnamefont {P.}~\bibnamefont {Marquetand}},
  \bibinfo {author} {\bibfnamefont {I.~C.~D.}\ \bibnamefont {Merritt}},
  \bibinfo {author} {\bibfnamefont {F.}~\bibnamefont {Montorsi}}, \bibinfo
  {author} {\bibfnamefont {M.}~\bibnamefont {M\"{o}rchen}}, \bibinfo {author}
  {\bibfnamefont {A.}~\bibnamefont {Nenov}}, \bibinfo {author} {\bibfnamefont
  {V.~H.~A.}\ \bibnamefont {Nguyen}}, \bibinfo {author} {\bibfnamefont
  {Y.}~\bibnamefont {Nishimoto}}, \bibinfo {author} {\bibfnamefont {M.~S.}\
  \bibnamefont {Oakley}}, \bibinfo {author} {\bibfnamefont {M.}~\bibnamefont
  {Olivucci}}, \bibinfo {author} {\bibfnamefont {M.}~\bibnamefont {Oppel}},
  \bibinfo {author} {\bibfnamefont {D.}~\bibnamefont {Padula}}, \bibinfo
  {author} {\bibfnamefont {R.}~\bibnamefont {Pandharkar}}, \bibinfo {author}
  {\bibfnamefont {Q.~M.}\ \bibnamefont {Phung}}, \bibinfo {author}
  {\bibfnamefont {F.}~\bibnamefont {Plasser}}, \bibinfo {author} {\bibfnamefont
  {G.}~\bibnamefont {Raggi}}, \bibinfo {author} {\bibfnamefont
  {E.}~\bibnamefont {Rebolini}}, \bibinfo {author} {\bibfnamefont
  {M.}~\bibnamefont {Reiher}}, \bibinfo {author} {\bibfnamefont
  {I.}~\bibnamefont {Rivalta}}, \bibinfo {author} {\bibfnamefont
  {D.}~\bibnamefont {Roca-Sanjuán}}, \bibinfo {author} {\bibfnamefont
  {T.}~\bibnamefont {Romig}}, \bibinfo {author} {\bibfnamefont {A.~A.}\
  \bibnamefont {Safari}}, \bibinfo {author} {\bibfnamefont {A.}~\bibnamefont
  {Sánchez-Mansilla}}, \bibinfo {author} {\bibfnamefont {A.~M.}\ \bibnamefont
  {Sand}}, \bibinfo {author} {\bibfnamefont {I.}~\bibnamefont {Schapiro}},
  \bibinfo {author} {\bibfnamefont {T.~R.}\ \bibnamefont {Scott}}, \bibinfo
  {author} {\bibfnamefont {J.}~\bibnamefont {Segarra-Martí}}, \bibinfo
  {author} {\bibfnamefont {F.}~\bibnamefont {Segatta}}, \bibinfo {author}
  {\bibfnamefont {D.-C.}\ \bibnamefont {Sergentu}}, \bibinfo {author}
  {\bibfnamefont {P.}~\bibnamefont {Sharma}}, \bibinfo {author} {\bibfnamefont
  {R.}~\bibnamefont {Shepard}}, \bibinfo {author} {\bibfnamefont
  {Y.}~\bibnamefont {Shu}}, \bibinfo {author} {\bibfnamefont {J.~K.}\
  \bibnamefont {Staab}}, \bibinfo {author} {\bibfnamefont {T.~P.}\ \bibnamefont
  {Straatsma}}, \bibinfo {author} {\bibfnamefont {L.~K.}\ \bibnamefont
  {Sørensen}}, \bibinfo {author} {\bibfnamefont {B.~N.~C.}\ \bibnamefont
  {Tenorio}}, \bibinfo {author} {\bibfnamefont {D.~G.}\ \bibnamefont
  {Truhlar}}, \bibinfo {author} {\bibfnamefont {L.}~\bibnamefont {Ungur}},
  \bibinfo {author} {\bibfnamefont {M.}~\bibnamefont {Vacher}}, \bibinfo
  {author} {\bibfnamefont {V.}~\bibnamefont {Veryazov}}, \bibinfo {author}
  {\bibfnamefont {T.~A.}\ \bibnamefont {Voß}}, \bibinfo {author}
  {\bibfnamefont {O.}~\bibnamefont {Weser}}, \bibinfo {author} {\bibfnamefont
  {D.}~\bibnamefont {Wu}}, \bibinfo {author} {\bibfnamefont {X.}~\bibnamefont
  {Yang}}, \bibinfo {author} {\bibfnamefont {D.}~\bibnamefont {Yarkony}},
  \bibinfo {author} {\bibfnamefont {C.}~\bibnamefont {Zhou}}, \bibinfo {author}
  {\bibfnamefont {J.~P.}\ \bibnamefont {Zobel}},\ and\ \bibinfo {author}
  {\bibfnamefont {R.}~\bibnamefont {Lindh}},\ }\bibfield  {title} {\bibinfo
  {title} {The openmolcas web: A community-driven approach to advancing
  computational chemistry},\ }\href {https://doi.org/10.1021/acs.jctc.3c00182}
  {\bibfield  {journal} {\bibinfo  {journal} {Journal of Chemical Theory and
  Computation}\ }\textbf {\bibinfo {volume} {19}},\ \bibinfo {pages}
  {6933–6991} (\bibinfo {year} {2023})}\BibitemShut {NoStop}%
\bibitem [{Sup()}]{SupplementalMaterial}%
  \BibitemOpen
  \href@noop {} {\bibinfo {title} {See supplemental material at [\dots url
  \dots] for model hamiltonian definition, derivation of the equations for both
  the coherence and population mechanisms, additional details and data about
  the electronic structure calculations, comparison of perturbative and
  non-perturbative field-matter interaction regimes in a range of field
  intensities.}}\BibitemShut {Stop}%
\bibitem [{\citenamefont {Mukamel}\ \emph {et~al.}(2013)\citenamefont
  {Mukamel}, \citenamefont {Healion}, \citenamefont {Zhang},\ and\
  \citenamefont {Biggs}}]{Mukamel2013}%
  \BibitemOpen
  \bibfield  {author} {\bibinfo {author} {\bibfnamefont {S.}~\bibnamefont
  {Mukamel}}, \bibinfo {author} {\bibfnamefont {D.}~\bibnamefont {Healion}},
  \bibinfo {author} {\bibfnamefont {Y.}~\bibnamefont {Zhang}},\ and\ \bibinfo
  {author} {\bibfnamefont {J.~D.}\ \bibnamefont {Biggs}},\ }\bibfield  {title}
  {\bibinfo {title} {Multidimensional attosecond resonant x-ray spectroscopy of
  molecules: Lessons from the optical regime},\ }\href
  {https://doi.org/10.1146/annurev-physchem-040412-110021} {\bibfield
  {journal} {\bibinfo  {journal} {Annual Review of Physical Chemistry}\
  }\textbf {\bibinfo {volume} {64}},\ \bibinfo {pages} {101} (\bibinfo {year}
  {2013})}\BibitemShut {NoStop}%
\bibitem [{\citenamefont {Gelin}\ \emph {et~al.}(2022)\citenamefont {Gelin},
  \citenamefont {Chen},\ and\ \citenamefont {Domcke}}]{Gelin2022}%
  \BibitemOpen
  \bibfield  {author} {\bibinfo {author} {\bibfnamefont {M.~F.}\ \bibnamefont
  {Gelin}}, \bibinfo {author} {\bibfnamefont {L.}~\bibnamefont {Chen}},\ and\
  \bibinfo {author} {\bibfnamefont {W.}~\bibnamefont {Domcke}},\ }\bibfield
  {title} {\bibinfo {title} {Equation-of-motion methods for the calculation of
  femtosecond time-resolved 4-wave-mixing and $n$-wave-mixing signals},\ }\href
  {https://doi.org/10.1021/acs.chemrev.2c00329} {\bibfield  {journal} {\bibinfo
   {journal} {Chemical Reviews}\ }\textbf {\bibinfo {volume} {122}},\ \bibinfo
  {pages} {17339} (\bibinfo {year} {2022})}\BibitemShut {NoStop}%
\bibitem [{Note1()}]{Note1}%
  \BibitemOpen
  \bibinfo {note} {Note that the process that brings the electron back to the
  ground state, captured by $\rho _{gg}$, is also possible. Nonetheless, since
  the focus here is on excited states, we will not explicitly show this pathway
  in what follows.}\BibitemShut {Stop}%
\bibitem [{\citenamefont {Mukamel}(1995)}]{MukamelBook}%
  \BibitemOpen
  \bibfield  {author} {\bibinfo {author} {\bibfnamefont {S.}~\bibnamefont
  {Mukamel}},\ }\href@noop {} {\emph {\bibinfo {title} {{Principles of
  Nonlinear Optical Spectroscopy}}}}\ (\bibinfo  {publisher} {Oxford University
  Press},\ \bibinfo {address} {New York},\ \bibinfo {year} {1995})\BibitemShut
  {NoStop}%
\bibitem [{Note2()}]{Note2}%
  \BibitemOpen
  \bibinfo {note} {Note that the term \protect \emph {overlap} is here used to
  highlight a situation in which the two orbitals occupy the same region of the
  3D space. It does not refer to the overlap integral which, by construction,
  is zero.}\BibitemShut {Stop}%
\bibitem [{\citenamefont {Manne}(1970)}]{Manne1970}%
  \BibitemOpen
  \bibfield  {author} {\bibinfo {author} {\bibfnamefont {R.}~\bibnamefont
  {Manne}},\ }\bibfield  {title} {\bibinfo {title} {Molecular orbital
  interpretation of x-ray emission spectra: Simple hydrocarbons and carbon
  oxides},\ }\href {https://doi.org/10.1063/1.1672852} {\bibfield  {journal}
  {\bibinfo  {journal} {The Journal of Chemical Physics}\ }\textbf {\bibinfo
  {volume} {52}},\ \bibinfo {pages} {5733} (\bibinfo {year}
  {1970})}\BibitemShut {NoStop}%
\bibitem [{\citenamefont {Johansson}\ \emph {et~al.}(2012)\citenamefont
  {Johansson}, \citenamefont {Nation},\ and\ \citenamefont
  {Nori}}]{Johansson2012}%
  \BibitemOpen
  \bibfield  {author} {\bibinfo {author} {\bibfnamefont {J.}~\bibnamefont
  {Johansson}}, \bibinfo {author} {\bibfnamefont {P.}~\bibnamefont {Nation}},\
  and\ \bibinfo {author} {\bibfnamefont {F.}~\bibnamefont {Nori}},\ }\bibfield
  {title} {\bibinfo {title} {{QuTiP}: An open-source python framework for the
  dynamics of open quantum systems},\ }\href
  {https://doi.org/10.1016/j.cpc.2012.02.021} {\bibfield  {journal} {\bibinfo
  {journal} {Computer Physics Communications}\ }\textbf {\bibinfo {volume}
  {183}},\ \bibinfo {pages} {1760} (\bibinfo {year} {2012})}\BibitemShut
  {NoStop}%
\bibitem [{\citenamefont {Johansson}\ \emph {et~al.}(2013)\citenamefont
  {Johansson}, \citenamefont {Nation},\ and\ \citenamefont
  {Nori}}]{Johansson2013}%
  \BibitemOpen
  \bibfield  {author} {\bibinfo {author} {\bibfnamefont {J.}~\bibnamefont
  {Johansson}}, \bibinfo {author} {\bibfnamefont {P.}~\bibnamefont {Nation}},\
  and\ \bibinfo {author} {\bibfnamefont {F.}~\bibnamefont {Nori}},\ }\bibfield
  {title} {\bibinfo {title} {{QuTiP} 2: A python framework for the dynamics of
  open quantum systems},\ }\href {https://doi.org/10.1016/j.cpc.2012.11.019}
  {\bibfield  {journal} {\bibinfo  {journal} {Computer Physics Communications}\
  }\textbf {\bibinfo {volume} {184}},\ \bibinfo {pages} {1234} (\bibinfo {year}
  {2013})}\BibitemShut {NoStop}%
\bibitem [{\citenamefont {Soml{\'{o}}i}\ \emph {et~al.}(1993)\citenamefont
  {Soml{\'{o}}i}, \citenamefont {Kazakov},\ and\ \citenamefont
  {Tannor}}]{Somli1993}%
  \BibitemOpen
  \bibfield  {author} {\bibinfo {author} {\bibfnamefont {J.}~\bibnamefont
  {Soml{\'{o}}i}}, \bibinfo {author} {\bibfnamefont {V.~A.}\ \bibnamefont
  {Kazakov}},\ and\ \bibinfo {author} {\bibfnamefont {D.~J.}\ \bibnamefont
  {Tannor}},\ }\bibfield  {title} {\bibinfo {title} {Controlled dissociation of
  i2 via optical transitions between the x and b electronic states},\ }\href
  {https://doi.org/10.1016/0301-0104(93)80108-l} {\bibfield  {journal}
  {\bibinfo  {journal} {Chemical Physics}\ }\textbf {\bibinfo {volume} {172}},\
  \bibinfo {pages} {85} (\bibinfo {year} {1993})}\BibitemShut {NoStop}%
\bibitem [{\citenamefont {Singh}\ \emph {et~al.}(2023)\citenamefont {Singh},
  \citenamefont {Fareed}, \citenamefont {Birulia}, \citenamefont {Magunov},
  \citenamefont {Grum-Grzhimailo}, \citenamefont {Lassonde}, \citenamefont
  {Laram{\'{e}}e}, \citenamefont {Marcelino}, \citenamefont {Shirinabadi},
  \citenamefont {L{\'{e}}gar{\'{e}}}, \citenamefont {Ozaki},\ and\
  \citenamefont {Strelkov}}]{Singh2023}%
  \BibitemOpen
  \bibfield  {author} {\bibinfo {author} {\bibfnamefont {M.}~\bibnamefont
  {Singh}}, \bibinfo {author} {\bibfnamefont {M.~A.}\ \bibnamefont {Fareed}},
  \bibinfo {author} {\bibfnamefont {V.}~\bibnamefont {Birulia}}, \bibinfo
  {author} {\bibfnamefont {A.}~\bibnamefont {Magunov}}, \bibinfo {author}
  {\bibfnamefont {A.~N.}\ \bibnamefont {Grum-Grzhimailo}}, \bibinfo {author}
  {\bibfnamefont {P.}~\bibnamefont {Lassonde}}, \bibinfo {author}
  {\bibfnamefont {A.}~\bibnamefont {Laram{\'{e}}e}}, \bibinfo {author}
  {\bibfnamefont {R.}~\bibnamefont {Marcelino}}, \bibinfo {author}
  {\bibfnamefont {R.~G.}\ \bibnamefont {Shirinabadi}}, \bibinfo {author}
  {\bibfnamefont {F.}~\bibnamefont {L{\'{e}}gar{\'{e}}}}, \bibinfo {author}
  {\bibfnamefont {T.}~\bibnamefont {Ozaki}},\ and\ \bibinfo {author}
  {\bibfnamefont {V.}~\bibnamefont {Strelkov}},\ }\bibfield  {title} {\bibinfo
  {title} {Ultrafast resonant state formation by the coupling of rydberg and
  dark autoionizing states},\ }\bibfield  {journal} {\bibinfo  {journal}
  {Physical Review Letters}\ }\textbf {\bibinfo {volume} {130}},\ \href
  {https://doi.org/10.1103/physrevlett.130.073201}
  {10.1103/physrevlett.130.073201} (\bibinfo {year} {2023})\BibitemShut
  {NoStop}%
\bibitem [{\citenamefont {Jirovec}\ \emph {et~al.}(2021)\citenamefont
  {Jirovec}, \citenamefont {Hofmann}, \citenamefont {Ballabio}, \citenamefont
  {Mutter}, \citenamefont {Tavani}, \citenamefont {Botifoll}, \citenamefont
  {Crippa}, \citenamefont {Kukucka}, \citenamefont {Sagi}, \citenamefont
  {Martins}, \citenamefont {Saez-Mollejo}, \citenamefont {Prieto},
  \citenamefont {Borovkov}, \citenamefont {Arbiol}, \citenamefont {Chrastina},
  \citenamefont {Isella},\ and\ \citenamefont {Katsaros}}]{Jirovec2021}%
  \BibitemOpen
  \bibfield  {author} {\bibinfo {author} {\bibfnamefont {D.}~\bibnamefont
  {Jirovec}}, \bibinfo {author} {\bibfnamefont {A.}~\bibnamefont {Hofmann}},
  \bibinfo {author} {\bibfnamefont {A.}~\bibnamefont {Ballabio}}, \bibinfo
  {author} {\bibfnamefont {P.~M.}\ \bibnamefont {Mutter}}, \bibinfo {author}
  {\bibfnamefont {G.}~\bibnamefont {Tavani}}, \bibinfo {author} {\bibfnamefont
  {M.}~\bibnamefont {Botifoll}}, \bibinfo {author} {\bibfnamefont
  {A.}~\bibnamefont {Crippa}}, \bibinfo {author} {\bibfnamefont
  {J.}~\bibnamefont {Kukucka}}, \bibinfo {author} {\bibfnamefont
  {O.}~\bibnamefont {Sagi}}, \bibinfo {author} {\bibfnamefont {F.}~\bibnamefont
  {Martins}}, \bibinfo {author} {\bibfnamefont {J.}~\bibnamefont
  {Saez-Mollejo}}, \bibinfo {author} {\bibfnamefont {I.}~\bibnamefont
  {Prieto}}, \bibinfo {author} {\bibfnamefont {M.}~\bibnamefont {Borovkov}},
  \bibinfo {author} {\bibfnamefont {J.}~\bibnamefont {Arbiol}}, \bibinfo
  {author} {\bibfnamefont {D.}~\bibnamefont {Chrastina}}, \bibinfo {author}
  {\bibfnamefont {G.}~\bibnamefont {Isella}},\ and\ \bibinfo {author}
  {\bibfnamefont {G.}~\bibnamefont {Katsaros}},\ }\bibfield  {title} {\bibinfo
  {title} {A singlet-triplet hole spin qubit in planar ge},\ }\href
  {https://doi.org/10.1038/s41563-021-01022-2} {\bibfield  {journal} {\bibinfo
  {journal} {Nature Materials}\ }\textbf {\bibinfo {volume} {20}},\ \bibinfo
  {pages} {1106} (\bibinfo {year} {2021})}\BibitemShut {NoStop}%
\bibitem [{\citenamefont {Cryan}\ \emph {et~al.}(2022)\citenamefont {Cryan},
  \citenamefont {Driver}, \citenamefont {Duris}, \citenamefont {Guo},
  \citenamefont {Li}, \citenamefont {O’Neal},\ and\ \citenamefont
  {Marinelli}}]{Cryan2022bookchap}%
  \BibitemOpen
  \bibfield  {author} {\bibinfo {author} {\bibfnamefont {J.~P.}\ \bibnamefont
  {Cryan}}, \bibinfo {author} {\bibfnamefont {T.}~\bibnamefont {Driver}},
  \bibinfo {author} {\bibfnamefont {J.}~\bibnamefont {Duris}}, \bibinfo
  {author} {\bibfnamefont {Z.}~\bibnamefont {Guo}}, \bibinfo {author}
  {\bibfnamefont {S.}~\bibnamefont {Li}}, \bibinfo {author} {\bibfnamefont
  {J.~T.}\ \bibnamefont {O’Neal}},\ and\ \bibinfo {author} {\bibfnamefont
  {A.}~\bibnamefont {Marinelli}},\ }\bibinfo {title} {The development of
  attosecond xfels for understanding ultrafast electron motion},\ in\ \href
  {https://doi.org/10.1016/bs.aamop.2022.05.001} {\emph {\bibinfo {booktitle}
  {Advances in Atomic, Molecular, and Optical Physics}}}\ (\bibinfo
  {publisher} {Elsevier},\ \bibinfo {year} {2022})\ p.\ \bibinfo {pages}
  {1–64}\BibitemShut {NoStop}%
\bibitem [{\citenamefont {Eichmann}\ \emph {et~al.}(2020)\citenamefont
  {Eichmann}, \citenamefont {Rottke}, \citenamefont {Meise}, \citenamefont
  {Rubensson}, \citenamefont {S\"{o}derstr\"{o}m}, \citenamefont {Agåker},
  \citenamefont {Såthe}, \citenamefont {Meyer}, \citenamefont {Baumann},
  \citenamefont {Boll}, \citenamefont {De~Fanis}, \citenamefont {Grychtol},
  \citenamefont {Ilchen}, \citenamefont {Mazza}, \citenamefont {Montano},
  \citenamefont {Music}, \citenamefont {Ovcharenko}, \citenamefont {Rivas},
  \citenamefont {Serkez}, \citenamefont {Wagner},\ and\ \citenamefont
  {Eisebitt}}]{Eichmann2020}%
  \BibitemOpen
  \bibfield  {author} {\bibinfo {author} {\bibfnamefont {U.}~\bibnamefont
  {Eichmann}}, \bibinfo {author} {\bibfnamefont {H.}~\bibnamefont {Rottke}},
  \bibinfo {author} {\bibfnamefont {S.}~\bibnamefont {Meise}}, \bibinfo
  {author} {\bibfnamefont {J.-E.}\ \bibnamefont {Rubensson}}, \bibinfo {author}
  {\bibfnamefont {J.}~\bibnamefont {S\"{o}derstr\"{o}m}}, \bibinfo {author}
  {\bibfnamefont {M.}~\bibnamefont {Agåker}}, \bibinfo {author} {\bibfnamefont
  {C.}~\bibnamefont {Såthe}}, \bibinfo {author} {\bibfnamefont
  {M.}~\bibnamefont {Meyer}}, \bibinfo {author} {\bibfnamefont {T.~M.}\
  \bibnamefont {Baumann}}, \bibinfo {author} {\bibfnamefont {R.}~\bibnamefont
  {Boll}}, \bibinfo {author} {\bibfnamefont {A.}~\bibnamefont {De~Fanis}},
  \bibinfo {author} {\bibfnamefont {P.}~\bibnamefont {Grychtol}}, \bibinfo
  {author} {\bibfnamefont {M.}~\bibnamefont {Ilchen}}, \bibinfo {author}
  {\bibfnamefont {T.}~\bibnamefont {Mazza}}, \bibinfo {author} {\bibfnamefont
  {J.}~\bibnamefont {Montano}}, \bibinfo {author} {\bibfnamefont
  {V.}~\bibnamefont {Music}}, \bibinfo {author} {\bibfnamefont
  {Y.}~\bibnamefont {Ovcharenko}}, \bibinfo {author} {\bibfnamefont {D.~E.}\
  \bibnamefont {Rivas}}, \bibinfo {author} {\bibfnamefont {S.}~\bibnamefont
  {Serkez}}, \bibinfo {author} {\bibfnamefont {R.}~\bibnamefont {Wagner}},\
  and\ \bibinfo {author} {\bibfnamefont {S.}~\bibnamefont {Eisebitt}},\
  }\bibfield  {title} {\bibinfo {title} {Photon-recoil imaging: Expanding the
  view of nonlinear x-ray physics},\ }\href
  {https://doi.org/10.1126/science.abc2622} {\bibfield  {journal} {\bibinfo
  {journal} {Science}\ }\textbf {\bibinfo {volume} {369}},\ \bibinfo {pages}
  {1630–1633} (\bibinfo {year} {2020})}\BibitemShut {NoStop}%
\bibitem [{\citenamefont {Duris}\ \emph {et~al.}(2019)\citenamefont {Duris},
  \citenamefont {Li}, \citenamefont {Driver}, \citenamefont {Champenois},
  \citenamefont {MacArthur}, \citenamefont {Lutman}, \citenamefont {Zhang},
  \citenamefont {Rosenberger}, \citenamefont {Aldrich}, \citenamefont {Coffee},
  \citenamefont {Coslovich}, \citenamefont {Decker}, \citenamefont {Glownia},
  \citenamefont {Hartmann}, \citenamefont {Helml}, \citenamefont {Kamalov},
  \citenamefont {Knurr}, \citenamefont {Krzywinski}, \citenamefont {Lin},
  \citenamefont {Marangos}, \citenamefont {Nantel}, \citenamefont {Natan},
  \citenamefont {O'Neal}, \citenamefont {Shivaram}, \citenamefont {Walter},
  \citenamefont {Wang}, \citenamefont {Welch}, \citenamefont {Wolf},
  \citenamefont {Xu}, \citenamefont {Kling}, \citenamefont {Bucksbaum},
  \citenamefont {Zholents}, \citenamefont {Huang}, \citenamefont {Cryan},\ and\
  \citenamefont {Marinelli}}]{Duris2019}%
  \BibitemOpen
  \bibfield  {author} {\bibinfo {author} {\bibfnamefont {J.}~\bibnamefont
  {Duris}}, \bibinfo {author} {\bibfnamefont {S.}~\bibnamefont {Li}}, \bibinfo
  {author} {\bibfnamefont {T.}~\bibnamefont {Driver}}, \bibinfo {author}
  {\bibfnamefont {E.~G.}\ \bibnamefont {Champenois}}, \bibinfo {author}
  {\bibfnamefont {J.~P.}\ \bibnamefont {MacArthur}}, \bibinfo {author}
  {\bibfnamefont {A.~A.}\ \bibnamefont {Lutman}}, \bibinfo {author}
  {\bibfnamefont {Z.}~\bibnamefont {Zhang}}, \bibinfo {author} {\bibfnamefont
  {P.}~\bibnamefont {Rosenberger}}, \bibinfo {author} {\bibfnamefont {J.~W.}\
  \bibnamefont {Aldrich}}, \bibinfo {author} {\bibfnamefont {R.}~\bibnamefont
  {Coffee}}, \bibinfo {author} {\bibfnamefont {G.}~\bibnamefont {Coslovich}},
  \bibinfo {author} {\bibfnamefont {F.-J.}\ \bibnamefont {Decker}}, \bibinfo
  {author} {\bibfnamefont {J.~M.}\ \bibnamefont {Glownia}}, \bibinfo {author}
  {\bibfnamefont {G.}~\bibnamefont {Hartmann}}, \bibinfo {author}
  {\bibfnamefont {W.}~\bibnamefont {Helml}}, \bibinfo {author} {\bibfnamefont
  {A.}~\bibnamefont {Kamalov}}, \bibinfo {author} {\bibfnamefont
  {J.}~\bibnamefont {Knurr}}, \bibinfo {author} {\bibfnamefont
  {J.}~\bibnamefont {Krzywinski}}, \bibinfo {author} {\bibfnamefont {M.-F.}\
  \bibnamefont {Lin}}, \bibinfo {author} {\bibfnamefont {J.~P.}\ \bibnamefont
  {Marangos}}, \bibinfo {author} {\bibfnamefont {M.}~\bibnamefont {Nantel}},
  \bibinfo {author} {\bibfnamefont {A.}~\bibnamefont {Natan}}, \bibinfo
  {author} {\bibfnamefont {J.~T.}\ \bibnamefont {O'Neal}}, \bibinfo {author}
  {\bibfnamefont {N.}~\bibnamefont {Shivaram}}, \bibinfo {author}
  {\bibfnamefont {P.}~\bibnamefont {Walter}}, \bibinfo {author} {\bibfnamefont
  {A.~L.}\ \bibnamefont {Wang}}, \bibinfo {author} {\bibfnamefont {J.~J.}\
  \bibnamefont {Welch}}, \bibinfo {author} {\bibfnamefont {T.~J.~A.}\
  \bibnamefont {Wolf}}, \bibinfo {author} {\bibfnamefont {J.~Z.}\ \bibnamefont
  {Xu}}, \bibinfo {author} {\bibfnamefont {M.~F.}\ \bibnamefont {Kling}},
  \bibinfo {author} {\bibfnamefont {P.~H.}\ \bibnamefont {Bucksbaum}}, \bibinfo
  {author} {\bibfnamefont {A.}~\bibnamefont {Zholents}}, \bibinfo {author}
  {\bibfnamefont {Z.}~\bibnamefont {Huang}}, \bibinfo {author} {\bibfnamefont
  {J.~P.}\ \bibnamefont {Cryan}},\ and\ \bibinfo {author} {\bibfnamefont
  {A.}~\bibnamefont {Marinelli}},\ }\bibfield  {title} {\bibinfo {title}
  {Tunable isolated attosecond x-ray pulses with gigawatt peak power from a
  free-electron laser},\ }\href {https://doi.org/10.1038/s41566-019-0549-5}
  {\bibfield  {journal} {\bibinfo  {journal} {Nature Photonics}\ }\textbf
  {\bibinfo {volume} {14}},\ \bibinfo {pages} {30} (\bibinfo {year}
  {2019})}\BibitemShut {NoStop}%
\end{thebibliography}%


\begin{thebibliography}{16}%
\makeatletter
\providecommand \@ifxundefined [1]{%
 \@ifx{#1\undefined}
}%
\providecommand \@ifnum [1]{%
 \ifnum #1\expandafter \@firstoftwo
 \else \expandafter \@secondoftwo
 \fi
}%
\providecommand \@ifx [1]{%
 \ifx #1\expandafter \@firstoftwo
 \else \expandafter \@secondoftwo
 \fi
}%
\providecommand \natexlab [1]{#1}%
\providecommand \enquote  [1]{``#1''}%
\providecommand \bibnamefont  [1]{#1}%
\providecommand \bibfnamefont [1]{#1}%
\providecommand \citenamefont [1]{#1}%
\providecommand \href@noop [0]{\@secondoftwo}%
\providecommand \href [0]{\begingroup \@sanitize@url \@href}%
\providecommand \@href[1]{\@@startlink{#1}\@@href}%
\providecommand \@@href[1]{\endgroup#1\@@endlink}%
\providecommand \@sanitize@url [0]{\catcode `\\12\catcode `\$12\catcode
  `\&12\catcode `\#12\catcode `\^12\catcode `\_12\catcode `\%12\relax}%
\providecommand \@@startlink[1]{}%
\providecommand \@@endlink[0]{}%
\providecommand \url  [0]{\begingroup\@sanitize@url \@url }%
\providecommand \@url [1]{\endgroup\@href {#1}{\urlprefix }}%
\providecommand \urlprefix  [0]{URL }%
\providecommand \Eprint [0]{\href }%
\providecommand \doibase [0]{https://doi.org/}%
\providecommand \selectlanguage [0]{\@gobble}%
\providecommand \bibinfo  [0]{\@secondoftwo}%
\providecommand \bibfield  [0]{\@secondoftwo}%
\providecommand \translation [1]{[#1]}%
\providecommand \BibitemOpen [0]{}%
\providecommand \bibitemStop [0]{}%
\providecommand \bibitemNoStop [0]{.\EOS\space}%
\providecommand \EOS [0]{\spacefactor3000\relax}%
\providecommand \BibitemShut  [1]{\csname bibitem#1\endcsname}%
\let\auto@bib@innerbib\@empty
\bibitem [{\citenamefont {Mukamel}(1995)}]{MukamelBook}%
  \BibitemOpen
  \bibfield  {author} {\bibinfo {author} {\bibfnamefont {S.}~\bibnamefont
  {Mukamel}},\ }\href@noop {} {\emph {\bibinfo {title} {{Principles of
  Nonlinear Optical Spectroscopy}}}}\ (\bibinfo  {publisher} {Oxford University
  Press},\ \bibinfo {address} {New York},\ \bibinfo {year} {1995})\BibitemShut
  {NoStop}%
\bibitem [{\citenamefont {Johansson}\ \emph {et~al.}(2012)\citenamefont
  {Johansson}, \citenamefont {Nation},\ and\ \citenamefont
  {Nori}}]{Johansson2012}%
  \BibitemOpen
  \bibfield  {author} {\bibinfo {author} {\bibfnamefont {J.}~\bibnamefont
  {Johansson}}, \bibinfo {author} {\bibfnamefont {P.}~\bibnamefont {Nation}},\
  and\ \bibinfo {author} {\bibfnamefont {F.}~\bibnamefont {Nori}},\ }\bibfield
  {title} {\bibinfo {title} {{QuTiP}: An open-source python framework for the
  dynamics of open quantum systems},\ }\href
  {https://doi.org/10.1016/j.cpc.2012.02.021} {\bibfield  {journal} {\bibinfo
  {journal} {Computer Physics Communications}\ }\textbf {\bibinfo {volume}
  {183}},\ \bibinfo {pages} {1760} (\bibinfo {year} {2012})}\BibitemShut
  {NoStop}%
\bibitem [{\citenamefont {Johansson}\ \emph {et~al.}(2013)\citenamefont
  {Johansson}, \citenamefont {Nation},\ and\ \citenamefont
  {Nori}}]{Johansson2013}%
  \BibitemOpen
  \bibfield  {author} {\bibinfo {author} {\bibfnamefont {J.}~\bibnamefont
  {Johansson}}, \bibinfo {author} {\bibfnamefont {P.}~\bibnamefont {Nation}},\
  and\ \bibinfo {author} {\bibfnamefont {F.}~\bibnamefont {Nori}},\ }\bibfield
  {title} {\bibinfo {title} {{QuTiP} 2: A python framework for the dynamics of
  open quantum systems},\ }\href {https://doi.org/10.1016/j.cpc.2012.11.019}
  {\bibfield  {journal} {\bibinfo  {journal} {Computer Physics Communications}\
  }\textbf {\bibinfo {volume} {184}},\ \bibinfo {pages} {1234} (\bibinfo {year}
  {2013})}\BibitemShut {NoStop}%
\bibitem [{\citenamefont {O'Neal}\ \emph {et~al.}(2020)\citenamefont {O'Neal},
  \citenamefont {Champenois}, \citenamefont {Oberli}, \citenamefont {Obaid},
  \citenamefont {Al-Haddad}, \citenamefont {Barnard}, \citenamefont {Berrah},
  \citenamefont {Coffee}, \citenamefont {Duris}, \citenamefont {Galinis},
  \citenamefont {Garratt}, \citenamefont {Glownia}, \citenamefont {Haxton},
  \citenamefont {Ho}, \citenamefont {Li}, \citenamefont {Li}, \citenamefont
  {MacArthur}, \citenamefont {Marangos}, \citenamefont {Natan}, \citenamefont
  {Shivaram}, \citenamefont {Slaughter}, \citenamefont {Walter}, \citenamefont
  {Wandel}, \citenamefont {Young}, \citenamefont {Bostedt}, \citenamefont
  {Bucksbaum}, \citenamefont {Pic{\'{o}}n}, \citenamefont {Marinelli},\ and\
  \citenamefont {Cryan}}]{Cryan2020}%
  \BibitemOpen
  \bibfield  {author} {\bibinfo {author} {\bibfnamefont {J.~T.}\ \bibnamefont
  {O'Neal}}, \bibinfo {author} {\bibfnamefont {E.~G.}\ \bibnamefont
  {Champenois}}, \bibinfo {author} {\bibfnamefont {S.}~\bibnamefont {Oberli}},
  \bibinfo {author} {\bibfnamefont {R.}~\bibnamefont {Obaid}}, \bibinfo
  {author} {\bibfnamefont {A.}~\bibnamefont {Al-Haddad}}, \bibinfo {author}
  {\bibfnamefont {J.}~\bibnamefont {Barnard}}, \bibinfo {author} {\bibfnamefont
  {N.}~\bibnamefont {Berrah}}, \bibinfo {author} {\bibfnamefont
  {R.}~\bibnamefont {Coffee}}, \bibinfo {author} {\bibfnamefont
  {J.}~\bibnamefont {Duris}}, \bibinfo {author} {\bibfnamefont
  {G.}~\bibnamefont {Galinis}}, \bibinfo {author} {\bibfnamefont
  {D.}~\bibnamefont {Garratt}}, \bibinfo {author} {\bibfnamefont {J.~M.}\
  \bibnamefont {Glownia}}, \bibinfo {author} {\bibfnamefont {D.}~\bibnamefont
  {Haxton}}, \bibinfo {author} {\bibfnamefont {P.}~\bibnamefont {Ho}}, \bibinfo
  {author} {\bibfnamefont {S.}~\bibnamefont {Li}}, \bibinfo {author}
  {\bibfnamefont {X.}~\bibnamefont {Li}}, \bibinfo {author} {\bibfnamefont
  {J.}~\bibnamefont {MacArthur}}, \bibinfo {author} {\bibfnamefont {J.~P.}\
  \bibnamefont {Marangos}}, \bibinfo {author} {\bibfnamefont {A.}~\bibnamefont
  {Natan}}, \bibinfo {author} {\bibfnamefont {N.}~\bibnamefont {Shivaram}},
  \bibinfo {author} {\bibfnamefont {D.~S.}\ \bibnamefont {Slaughter}}, \bibinfo
  {author} {\bibfnamefont {P.}~\bibnamefont {Walter}}, \bibinfo {author}
  {\bibfnamefont {S.}~\bibnamefont {Wandel}}, \bibinfo {author} {\bibfnamefont
  {L.}~\bibnamefont {Young}}, \bibinfo {author} {\bibfnamefont
  {C.}~\bibnamefont {Bostedt}}, \bibinfo {author} {\bibfnamefont {P.~H.}\
  \bibnamefont {Bucksbaum}}, \bibinfo {author} {\bibfnamefont {A.}~\bibnamefont
  {Pic{\'{o}}n}}, \bibinfo {author} {\bibfnamefont {A.}~\bibnamefont
  {Marinelli}},\ and\ \bibinfo {author} {\bibfnamefont {J.~P.}\ \bibnamefont
  {Cryan}},\ }\bibfield  {title} {\bibinfo {title} {Electronic population
  transfer via impulsive stimulated x-ray raman scattering with attosecond
  soft-x-ray pulses},\ }\bibfield  {journal} {\bibinfo  {journal} {Physical
  Review Letters}\ }\textbf {\bibinfo {volume} {125}},\ \href
  {https://doi.org/10.1103/physrevlett.125.073203}
  {10.1103/physrevlett.125.073203} (\bibinfo {year} {2020})\BibitemShut
  {NoStop}%
\bibitem [{Note1()}]{Note1}%
  \BibitemOpen
  \bibinfo {note} {Note that a similar physical motivation is used to introduce
  the so-called Doorway-Window approximation in nonlinear spectroscopy.\cite
  {MukamelBook}}\BibitemShut {NoStop}%
\bibitem [{\citenamefont {Mukamel}\ \emph {et~al.}(2013)\citenamefont
  {Mukamel}, \citenamefont {Healion}, \citenamefont {Zhang},\ and\
  \citenamefont {Biggs}}]{Mukamel2013}%
  \BibitemOpen
  \bibfield  {author} {\bibinfo {author} {\bibfnamefont {S.}~\bibnamefont
  {Mukamel}}, \bibinfo {author} {\bibfnamefont {D.}~\bibnamefont {Healion}},
  \bibinfo {author} {\bibfnamefont {Y.}~\bibnamefont {Zhang}},\ and\ \bibinfo
  {author} {\bibfnamefont {J.~D.}\ \bibnamefont {Biggs}},\ }\bibfield  {title}
  {\bibinfo {title} {Multidimensional attosecond resonant x-ray spectroscopy of
  molecules: Lessons from the optical regime},\ }\href
  {https://doi.org/10.1146/annurev-physchem-040412-110021} {\bibfield
  {journal} {\bibinfo  {journal} {Annual Review of Physical Chemistry}\
  }\textbf {\bibinfo {volume} {64}},\ \bibinfo {pages} {101} (\bibinfo {year}
  {2013})}\BibitemShut {NoStop}%
\bibitem [{\citenamefont {Cavaletto}\ \emph {et~al.}(2023)\citenamefont
  {Cavaletto}, \citenamefont {Nam}, \citenamefont {Rouxel}, \citenamefont
  {Keefer}, \citenamefont {Yong},\ and\ \citenamefont
  {Mukamel}}]{Cavaletto2023}%
  \BibitemOpen
  \bibfield  {author} {\bibinfo {author} {\bibfnamefont {S.~M.}\ \bibnamefont
  {Cavaletto}}, \bibinfo {author} {\bibfnamefont {Y.}~\bibnamefont {Nam}},
  \bibinfo {author} {\bibfnamefont {J.~R.}\ \bibnamefont {Rouxel}}, \bibinfo
  {author} {\bibfnamefont {D.}~\bibnamefont {Keefer}}, \bibinfo {author}
  {\bibfnamefont {H.}~\bibnamefont {Yong}},\ and\ \bibinfo {author}
  {\bibfnamefont {S.}~\bibnamefont {Mukamel}},\ }\bibfield  {title} {\bibinfo
  {title} {Attosecond monitoring of nonadiabatic molecular dynamics by
  transient x-ray transmission spectroscopy},\ }\href
  {https://doi.org/10.1021/acs.jctc.3c00062} {\bibfield  {journal} {\bibinfo
  {journal} {Journal of Chemical Theory and Computation}\ }\textbf {\bibinfo
  {volume} {19}},\ \bibinfo {pages} {2327} (\bibinfo {year}
  {2023})}\BibitemShut {NoStop}%
\bibitem [{\citenamefont {Galv{\'{a}}n}\ \emph {et~al.}(2019)\citenamefont
  {Galv{\'{a}}n}, \citenamefont {Vacher}, \citenamefont {Alavi}, \citenamefont
  {Angeli}, \citenamefont {Aquilante}, \citenamefont {Autschbach},
  \citenamefont {Bao}, \citenamefont {Bokarev}, \citenamefont {Bogdanov},
  \citenamefont {Carlson}, \citenamefont {Chibotaru}, \citenamefont
  {Creutzberg}, \citenamefont {Dattani}, \citenamefont {Delcey}, \citenamefont
  {Dong}, \citenamefont {Dreuw}, \citenamefont {Freitag}, \citenamefont
  {Frutos}, \citenamefont {Gagliardi}, \citenamefont {Gendron}, \citenamefont
  {Giussani}, \citenamefont {Gonz{\'{a}}lez}, \citenamefont {Grell},
  \citenamefont {Guo}, \citenamefont {Hoyer}, \citenamefont {Johansson},
  \citenamefont {Keller}, \citenamefont {Knecht}, \citenamefont
  {Kova{\v{c}}evi{\'{c}}}, \citenamefont {K\"{a}llman}, \citenamefont {Manni},
  \citenamefont {Lundberg}, \citenamefont {Ma}, \citenamefont {Mai},
  \citenamefont {Malhado}, \citenamefont {Malmqvist}, \citenamefont
  {Marquetand}, \citenamefont {Mewes}, \citenamefont {Norell}, \citenamefont
  {Olivucci}, \citenamefont {Oppel}, \citenamefont {Phung}, \citenamefont
  {Pierloot}, \citenamefont {Plasser}, \citenamefont {Reiher}, \citenamefont
  {Sand}, \citenamefont {Schapiro}, \citenamefont {Sharma}, \citenamefont
  {Stein}, \citenamefont {S{\o}rensen}, \citenamefont {Truhlar}, \citenamefont
  {Ugandi}, \citenamefont {Ungur}, \citenamefont {Valentini}, \citenamefont
  {Vancoillie}, \citenamefont {Veryazov}, \citenamefont {Weser}, \citenamefont
  {Weso{\l}owski}, \citenamefont {Widmark}, \citenamefont {Wouters},
  \citenamefont {Zech}, \citenamefont {Zobel},\ and\ \citenamefont
  {Lindh}}]{FdezGalvn2019}%
  \BibitemOpen
  \bibfield  {author} {\bibinfo {author} {\bibfnamefont {I.~F.}\ \bibnamefont
  {Galv{\'{a}}n}}, \bibinfo {author} {\bibfnamefont {M.}~\bibnamefont
  {Vacher}}, \bibinfo {author} {\bibfnamefont {A.}~\bibnamefont {Alavi}},
  \bibinfo {author} {\bibfnamefont {C.}~\bibnamefont {Angeli}}, \bibinfo
  {author} {\bibfnamefont {F.}~\bibnamefont {Aquilante}}, \bibinfo {author}
  {\bibfnamefont {J.}~\bibnamefont {Autschbach}}, \bibinfo {author}
  {\bibfnamefont {J.~J.}\ \bibnamefont {Bao}}, \bibinfo {author} {\bibfnamefont
  {S.~I.}\ \bibnamefont {Bokarev}}, \bibinfo {author} {\bibfnamefont {N.~A.}\
  \bibnamefont {Bogdanov}}, \bibinfo {author} {\bibfnamefont {R.~K.}\
  \bibnamefont {Carlson}}, \bibinfo {author} {\bibfnamefont {L.~F.}\
  \bibnamefont {Chibotaru}}, \bibinfo {author} {\bibfnamefont {J.}~\bibnamefont
  {Creutzberg}}, \bibinfo {author} {\bibfnamefont {N.}~\bibnamefont {Dattani}},
  \bibinfo {author} {\bibfnamefont {M.~G.}\ \bibnamefont {Delcey}}, \bibinfo
  {author} {\bibfnamefont {S.~S.}\ \bibnamefont {Dong}}, \bibinfo {author}
  {\bibfnamefont {A.}~\bibnamefont {Dreuw}}, \bibinfo {author} {\bibfnamefont
  {L.}~\bibnamefont {Freitag}}, \bibinfo {author} {\bibfnamefont {L.~M.}\
  \bibnamefont {Frutos}}, \bibinfo {author} {\bibfnamefont {L.}~\bibnamefont
  {Gagliardi}}, \bibinfo {author} {\bibfnamefont {F.}~\bibnamefont {Gendron}},
  \bibinfo {author} {\bibfnamefont {A.}~\bibnamefont {Giussani}}, \bibinfo
  {author} {\bibfnamefont {L.}~\bibnamefont {Gonz{\'{a}}lez}}, \bibinfo
  {author} {\bibfnamefont {G.}~\bibnamefont {Grell}}, \bibinfo {author}
  {\bibfnamefont {M.}~\bibnamefont {Guo}}, \bibinfo {author} {\bibfnamefont
  {C.~E.}\ \bibnamefont {Hoyer}}, \bibinfo {author} {\bibfnamefont
  {M.}~\bibnamefont {Johansson}}, \bibinfo {author} {\bibfnamefont
  {S.}~\bibnamefont {Keller}}, \bibinfo {author} {\bibfnamefont
  {S.}~\bibnamefont {Knecht}}, \bibinfo {author} {\bibfnamefont
  {G.}~\bibnamefont {Kova{\v{c}}evi{\'{c}}}}, \bibinfo {author} {\bibfnamefont
  {E.}~\bibnamefont {K\"{a}llman}}, \bibinfo {author} {\bibfnamefont {G.~L.}\
  \bibnamefont {Manni}}, \bibinfo {author} {\bibfnamefont {M.}~\bibnamefont
  {Lundberg}}, \bibinfo {author} {\bibfnamefont {Y.}~\bibnamefont {Ma}},
  \bibinfo {author} {\bibfnamefont {S.}~\bibnamefont {Mai}}, \bibinfo {author}
  {\bibfnamefont {J.~P.}\ \bibnamefont {Malhado}}, \bibinfo {author}
  {\bibfnamefont {P.~{\AA}.}\ \bibnamefont {Malmqvist}}, \bibinfo {author}
  {\bibfnamefont {P.}~\bibnamefont {Marquetand}}, \bibinfo {author}
  {\bibfnamefont {S.~A.}\ \bibnamefont {Mewes}}, \bibinfo {author}
  {\bibfnamefont {J.}~\bibnamefont {Norell}}, \bibinfo {author} {\bibfnamefont
  {M.}~\bibnamefont {Olivucci}}, \bibinfo {author} {\bibfnamefont
  {M.}~\bibnamefont {Oppel}}, \bibinfo {author} {\bibfnamefont {Q.~M.}\
  \bibnamefont {Phung}}, \bibinfo {author} {\bibfnamefont {K.}~\bibnamefont
  {Pierloot}}, \bibinfo {author} {\bibfnamefont {F.}~\bibnamefont {Plasser}},
  \bibinfo {author} {\bibfnamefont {M.}~\bibnamefont {Reiher}}, \bibinfo
  {author} {\bibfnamefont {A.~M.}\ \bibnamefont {Sand}}, \bibinfo {author}
  {\bibfnamefont {I.}~\bibnamefont {Schapiro}}, \bibinfo {author}
  {\bibfnamefont {P.}~\bibnamefont {Sharma}}, \bibinfo {author} {\bibfnamefont
  {C.~J.}\ \bibnamefont {Stein}}, \bibinfo {author} {\bibfnamefont {L.~K.}\
  \bibnamefont {S{\o}rensen}}, \bibinfo {author} {\bibfnamefont {D.~G.}\
  \bibnamefont {Truhlar}}, \bibinfo {author} {\bibfnamefont {M.}~\bibnamefont
  {Ugandi}}, \bibinfo {author} {\bibfnamefont {L.}~\bibnamefont {Ungur}},
  \bibinfo {author} {\bibfnamefont {A.}~\bibnamefont {Valentini}}, \bibinfo
  {author} {\bibfnamefont {S.}~\bibnamefont {Vancoillie}}, \bibinfo {author}
  {\bibfnamefont {V.}~\bibnamefont {Veryazov}}, \bibinfo {author}
  {\bibfnamefont {O.}~\bibnamefont {Weser}}, \bibinfo {author} {\bibfnamefont
  {T.~A.}\ \bibnamefont {Weso{\l}owski}}, \bibinfo {author} {\bibfnamefont
  {P.-O.}\ \bibnamefont {Widmark}}, \bibinfo {author} {\bibfnamefont
  {S.}~\bibnamefont {Wouters}}, \bibinfo {author} {\bibfnamefont
  {A.}~\bibnamefont {Zech}}, \bibinfo {author} {\bibfnamefont {J.~P.}\
  \bibnamefont {Zobel}},\ and\ \bibinfo {author} {\bibfnamefont
  {R.}~\bibnamefont {Lindh}},\ }\bibfield  {title} {\bibinfo {title}
  {{OpenMolcas}: From source code to insight},\ }\href
  {https://doi.org/10.1021/acs.jctc.9b00532} {\bibfield  {journal} {\bibinfo
  {journal} {Journal of Chemical Theory and Computation}\ }\textbf {\bibinfo
  {volume} {15}},\ \bibinfo {pages} {5925} (\bibinfo {year}
  {2019})}\BibitemShut {NoStop}%
\bibitem [{\citenamefont {Aquilante}\ \emph {et~al.}(2020)\citenamefont
  {Aquilante}, \citenamefont {Autschbach}, \citenamefont {Baiardi},
  \citenamefont {Battaglia}, \citenamefont {Borin}, \citenamefont {Chibotaru},
  \citenamefont {Conti}, \citenamefont {Vico}, \citenamefont {Delcey},
  \citenamefont {Galv{\'{a}}n}, \citenamefont {Ferr{\'{e}}}, \citenamefont
  {Freitag}, \citenamefont {Garavelli}, \citenamefont {Gong}, \citenamefont
  {Knecht}, \citenamefont {Larsson}, \citenamefont {Lindh}, \citenamefont
  {Lundberg}, \citenamefont {Malmqvist}, \citenamefont {Nenov}, \citenamefont
  {Norell}, \citenamefont {Odelius}, \citenamefont {Olivucci}, \citenamefont
  {Pedersen}, \citenamefont {Pedraza-Gonz{\'{a}}lez}, \citenamefont {Phung},
  \citenamefont {Pierloot}, \citenamefont {Reiher}, \citenamefont {Schapiro},
  \citenamefont {Segarra-Mart{\'{\i}}}, \citenamefont {Segatta}, \citenamefont
  {Seijo}, \citenamefont {Sen}, \citenamefont {Sergentu}, \citenamefont
  {Stein}, \citenamefont {Ungur}, \citenamefont {Vacher}, \citenamefont
  {Valentini},\ and\ \citenamefont {Veryazov}}]{Aquilante2020}%
  \BibitemOpen
  \bibfield  {author} {\bibinfo {author} {\bibfnamefont {F.}~\bibnamefont
  {Aquilante}}, \bibinfo {author} {\bibfnamefont {J.}~\bibnamefont
  {Autschbach}}, \bibinfo {author} {\bibfnamefont {A.}~\bibnamefont {Baiardi}},
  \bibinfo {author} {\bibfnamefont {S.}~\bibnamefont {Battaglia}}, \bibinfo
  {author} {\bibfnamefont {V.~A.}\ \bibnamefont {Borin}}, \bibinfo {author}
  {\bibfnamefont {L.~F.}\ \bibnamefont {Chibotaru}}, \bibinfo {author}
  {\bibfnamefont {I.}~\bibnamefont {Conti}}, \bibinfo {author} {\bibfnamefont
  {L.~D.}\ \bibnamefont {Vico}}, \bibinfo {author} {\bibfnamefont
  {M.}~\bibnamefont {Delcey}}, \bibinfo {author} {\bibfnamefont {I.~F.}\
  \bibnamefont {Galv{\'{a}}n}}, \bibinfo {author} {\bibfnamefont
  {N.}~\bibnamefont {Ferr{\'{e}}}}, \bibinfo {author} {\bibfnamefont
  {L.}~\bibnamefont {Freitag}}, \bibinfo {author} {\bibfnamefont
  {M.}~\bibnamefont {Garavelli}}, \bibinfo {author} {\bibfnamefont
  {X.}~\bibnamefont {Gong}}, \bibinfo {author} {\bibfnamefont {S.}~\bibnamefont
  {Knecht}}, \bibinfo {author} {\bibfnamefont {E.~D.}\ \bibnamefont {Larsson}},
  \bibinfo {author} {\bibfnamefont {R.}~\bibnamefont {Lindh}}, \bibinfo
  {author} {\bibfnamefont {M.}~\bibnamefont {Lundberg}}, \bibinfo {author}
  {\bibfnamefont {P.~{\AA}.}\ \bibnamefont {Malmqvist}}, \bibinfo {author}
  {\bibfnamefont {A.}~\bibnamefont {Nenov}}, \bibinfo {author} {\bibfnamefont
  {J.}~\bibnamefont {Norell}}, \bibinfo {author} {\bibfnamefont
  {M.}~\bibnamefont {Odelius}}, \bibinfo {author} {\bibfnamefont
  {M.}~\bibnamefont {Olivucci}}, \bibinfo {author} {\bibfnamefont {T.~B.}\
  \bibnamefont {Pedersen}}, \bibinfo {author} {\bibfnamefont {L.}~\bibnamefont
  {Pedraza-Gonz{\'{a}}lez}}, \bibinfo {author} {\bibfnamefont {Q.~M.}\
  \bibnamefont {Phung}}, \bibinfo {author} {\bibfnamefont {K.}~\bibnamefont
  {Pierloot}}, \bibinfo {author} {\bibfnamefont {M.}~\bibnamefont {Reiher}},
  \bibinfo {author} {\bibfnamefont {I.}~\bibnamefont {Schapiro}}, \bibinfo
  {author} {\bibfnamefont {J.}~\bibnamefont {Segarra-Mart{\'{\i}}}}, \bibinfo
  {author} {\bibfnamefont {F.}~\bibnamefont {Segatta}}, \bibinfo {author}
  {\bibfnamefont {L.}~\bibnamefont {Seijo}}, \bibinfo {author} {\bibfnamefont
  {S.}~\bibnamefont {Sen}}, \bibinfo {author} {\bibfnamefont {D.-C.}\
  \bibnamefont {Sergentu}}, \bibinfo {author} {\bibfnamefont {C.~J.}\
  \bibnamefont {Stein}}, \bibinfo {author} {\bibfnamefont {L.}~\bibnamefont
  {Ungur}}, \bibinfo {author} {\bibfnamefont {M.}~\bibnamefont {Vacher}},
  \bibinfo {author} {\bibfnamefont {A.}~\bibnamefont {Valentini}},\ and\
  \bibinfo {author} {\bibfnamefont {V.}~\bibnamefont {Veryazov}},\ }\bibfield
  {title} {\bibinfo {title} {Modern quantum chemistry with [open]molcas},\
  }\href {https://doi.org/10.1063/5.0004835} {\bibfield  {journal} {\bibinfo
  {journal} {The Journal of Chemical Physics}\ }\textbf {\bibinfo {volume}
  {152}},\ \bibinfo {pages} {214117} (\bibinfo {year} {2020})}\BibitemShut
  {NoStop}%
\bibitem [{\citenamefont {Li~Manni}\ \emph {et~al.}(2023)\citenamefont
  {Li~Manni}, \citenamefont {Fdez.~Galván}, \citenamefont {Alavi},
  \citenamefont {Aleotti}, \citenamefont {Aquilante}, \citenamefont
  {Autschbach}, \citenamefont {Avagliano}, \citenamefont {Baiardi},
  \citenamefont {Bao}, \citenamefont {Battaglia}, \citenamefont {Birnoschi},
  \citenamefont {Blanco-González}, \citenamefont {Bokarev}, \citenamefont
  {Broer}, \citenamefont {Cacciari}, \citenamefont {Calio}, \citenamefont
  {Carlson}, \citenamefont {Carvalho~Couto}, \citenamefont {Cerdán},
  \citenamefont {Chibotaru}, \citenamefont {Chilton}, \citenamefont {Church},
  \citenamefont {Conti}, \citenamefont {Coriani}, \citenamefont
  {Cuéllar-Zuquin}, \citenamefont {Daoud}, \citenamefont {Dattani},
  \citenamefont {Decleva}, \citenamefont {de~Graaf}, \citenamefont {Delcey},
  \citenamefont {De~Vico}, \citenamefont {Dobrautz}, \citenamefont {Dong},
  \citenamefont {Feng}, \citenamefont {Ferré}, \citenamefont {Filatov(Gulak)},
  \citenamefont {Gagliardi}, \citenamefont {Garavelli}, \citenamefont
  {González}, \citenamefont {Guan}, \citenamefont {Guo}, \citenamefont
  {Hennefarth}, \citenamefont {Hermes}, \citenamefont {Hoyer}, \citenamefont
  {Huix-Rotllant}, \citenamefont {Jaiswal}, \citenamefont {Kaiser},
  \citenamefont {Kaliakin}, \citenamefont {Khamesian}, \citenamefont {King},
  \citenamefont {Kochetov}, \citenamefont {Krośnicki}, \citenamefont {Kumaar},
  \citenamefont {Larsson}, \citenamefont {Lehtola}, \citenamefont {Lepetit},
  \citenamefont {Lischka}, \citenamefont {López~Ríos}, \citenamefont
  {Lundberg}, \citenamefont {Ma}, \citenamefont {Mai}, \citenamefont
  {Marquetand}, \citenamefont {Merritt}, \citenamefont {Montorsi},
  \citenamefont {M\"{o}rchen}, \citenamefont {Nenov}, \citenamefont {Nguyen},
  \citenamefont {Nishimoto}, \citenamefont {Oakley}, \citenamefont {Olivucci},
  \citenamefont {Oppel}, \citenamefont {Padula}, \citenamefont {Pandharkar},
  \citenamefont {Phung}, \citenamefont {Plasser}, \citenamefont {Raggi},
  \citenamefont {Rebolini}, \citenamefont {Reiher}, \citenamefont {Rivalta},
  \citenamefont {Roca-Sanjuán}, \citenamefont {Romig}, \citenamefont {Safari},
  \citenamefont {Sánchez-Mansilla}, \citenamefont {Sand}, \citenamefont
  {Schapiro}, \citenamefont {Scott}, \citenamefont {Segarra-Martí},
  \citenamefont {Segatta}, \citenamefont {Sergentu}, \citenamefont {Sharma},
  \citenamefont {Shepard}, \citenamefont {Shu}, \citenamefont {Staab},
  \citenamefont {Straatsma}, \citenamefont {Sørensen}, \citenamefont
  {Tenorio}, \citenamefont {Truhlar}, \citenamefont {Ungur}, \citenamefont
  {Vacher}, \citenamefont {Veryazov}, \citenamefont {Voß}, \citenamefont
  {Weser}, \citenamefont {Wu}, \citenamefont {Yang}, \citenamefont {Yarkony},
  \citenamefont {Zhou}, \citenamefont {Zobel},\ and\ \citenamefont
  {Lindh}}]{LiManni2023}%
  \BibitemOpen
  \bibfield  {author} {\bibinfo {author} {\bibfnamefont {G.}~\bibnamefont
  {Li~Manni}}, \bibinfo {author} {\bibfnamefont {I.}~\bibnamefont
  {Fdez.~Galván}}, \bibinfo {author} {\bibfnamefont {A.}~\bibnamefont
  {Alavi}}, \bibinfo {author} {\bibfnamefont {F.}~\bibnamefont {Aleotti}},
  \bibinfo {author} {\bibfnamefont {F.}~\bibnamefont {Aquilante}}, \bibinfo
  {author} {\bibfnamefont {J.}~\bibnamefont {Autschbach}}, \bibinfo {author}
  {\bibfnamefont {D.}~\bibnamefont {Avagliano}}, \bibinfo {author}
  {\bibfnamefont {A.}~\bibnamefont {Baiardi}}, \bibinfo {author} {\bibfnamefont
  {J.~J.}\ \bibnamefont {Bao}}, \bibinfo {author} {\bibfnamefont
  {S.}~\bibnamefont {Battaglia}}, \bibinfo {author} {\bibfnamefont
  {L.}~\bibnamefont {Birnoschi}}, \bibinfo {author} {\bibfnamefont
  {A.}~\bibnamefont {Blanco-González}}, \bibinfo {author} {\bibfnamefont
  {S.~I.}\ \bibnamefont {Bokarev}}, \bibinfo {author} {\bibfnamefont
  {R.}~\bibnamefont {Broer}}, \bibinfo {author} {\bibfnamefont
  {R.}~\bibnamefont {Cacciari}}, \bibinfo {author} {\bibfnamefont {P.~B.}\
  \bibnamefont {Calio}}, \bibinfo {author} {\bibfnamefont {R.~K.}\ \bibnamefont
  {Carlson}}, \bibinfo {author} {\bibfnamefont {R.}~\bibnamefont
  {Carvalho~Couto}}, \bibinfo {author} {\bibfnamefont {L.}~\bibnamefont
  {Cerdán}}, \bibinfo {author} {\bibfnamefont {L.~F.}\ \bibnamefont
  {Chibotaru}}, \bibinfo {author} {\bibfnamefont {N.~F.}\ \bibnamefont
  {Chilton}}, \bibinfo {author} {\bibfnamefont {J.~R.}\ \bibnamefont {Church}},
  \bibinfo {author} {\bibfnamefont {I.}~\bibnamefont {Conti}}, \bibinfo
  {author} {\bibfnamefont {S.}~\bibnamefont {Coriani}}, \bibinfo {author}
  {\bibfnamefont {J.}~\bibnamefont {Cuéllar-Zuquin}}, \bibinfo {author}
  {\bibfnamefont {R.~E.}\ \bibnamefont {Daoud}}, \bibinfo {author}
  {\bibfnamefont {N.}~\bibnamefont {Dattani}}, \bibinfo {author} {\bibfnamefont
  {P.}~\bibnamefont {Decleva}}, \bibinfo {author} {\bibfnamefont
  {C.}~\bibnamefont {de~Graaf}}, \bibinfo {author} {\bibfnamefont {M.~G.}\
  \bibnamefont {Delcey}}, \bibinfo {author} {\bibfnamefont {L.}~\bibnamefont
  {De~Vico}}, \bibinfo {author} {\bibfnamefont {W.}~\bibnamefont {Dobrautz}},
  \bibinfo {author} {\bibfnamefont {S.~S.}\ \bibnamefont {Dong}}, \bibinfo
  {author} {\bibfnamefont {R.}~\bibnamefont {Feng}}, \bibinfo {author}
  {\bibfnamefont {N.}~\bibnamefont {Ferré}}, \bibinfo {author} {\bibfnamefont
  {M.}~\bibnamefont {Filatov(Gulak)}}, \bibinfo {author} {\bibfnamefont
  {L.}~\bibnamefont {Gagliardi}}, \bibinfo {author} {\bibfnamefont
  {M.}~\bibnamefont {Garavelli}}, \bibinfo {author} {\bibfnamefont
  {L.}~\bibnamefont {González}}, \bibinfo {author} {\bibfnamefont
  {Y.}~\bibnamefont {Guan}}, \bibinfo {author} {\bibfnamefont {M.}~\bibnamefont
  {Guo}}, \bibinfo {author} {\bibfnamefont {M.~R.}\ \bibnamefont {Hennefarth}},
  \bibinfo {author} {\bibfnamefont {M.~R.}\ \bibnamefont {Hermes}}, \bibinfo
  {author} {\bibfnamefont {C.~E.}\ \bibnamefont {Hoyer}}, \bibinfo {author}
  {\bibfnamefont {M.}~\bibnamefont {Huix-Rotllant}}, \bibinfo {author}
  {\bibfnamefont {V.~K.}\ \bibnamefont {Jaiswal}}, \bibinfo {author}
  {\bibfnamefont {A.}~\bibnamefont {Kaiser}}, \bibinfo {author} {\bibfnamefont
  {D.~S.}\ \bibnamefont {Kaliakin}}, \bibinfo {author} {\bibfnamefont
  {M.}~\bibnamefont {Khamesian}}, \bibinfo {author} {\bibfnamefont {D.~S.}\
  \bibnamefont {King}}, \bibinfo {author} {\bibfnamefont {V.}~\bibnamefont
  {Kochetov}}, \bibinfo {author} {\bibfnamefont {M.}~\bibnamefont
  {Krośnicki}}, \bibinfo {author} {\bibfnamefont {A.~A.}\ \bibnamefont
  {Kumaar}}, \bibinfo {author} {\bibfnamefont {E.~D.}\ \bibnamefont {Larsson}},
  \bibinfo {author} {\bibfnamefont {S.}~\bibnamefont {Lehtola}}, \bibinfo
  {author} {\bibfnamefont {M.-B.}\ \bibnamefont {Lepetit}}, \bibinfo {author}
  {\bibfnamefont {H.}~\bibnamefont {Lischka}}, \bibinfo {author} {\bibfnamefont
  {P.}~\bibnamefont {López~Ríos}}, \bibinfo {author} {\bibfnamefont
  {M.}~\bibnamefont {Lundberg}}, \bibinfo {author} {\bibfnamefont
  {D.}~\bibnamefont {Ma}}, \bibinfo {author} {\bibfnamefont {S.}~\bibnamefont
  {Mai}}, \bibinfo {author} {\bibfnamefont {P.}~\bibnamefont {Marquetand}},
  \bibinfo {author} {\bibfnamefont {I.~C.~D.}\ \bibnamefont {Merritt}},
  \bibinfo {author} {\bibfnamefont {F.}~\bibnamefont {Montorsi}}, \bibinfo
  {author} {\bibfnamefont {M.}~\bibnamefont {M\"{o}rchen}}, \bibinfo {author}
  {\bibfnamefont {A.}~\bibnamefont {Nenov}}, \bibinfo {author} {\bibfnamefont
  {V.~H.~A.}\ \bibnamefont {Nguyen}}, \bibinfo {author} {\bibfnamefont
  {Y.}~\bibnamefont {Nishimoto}}, \bibinfo {author} {\bibfnamefont {M.~S.}\
  \bibnamefont {Oakley}}, \bibinfo {author} {\bibfnamefont {M.}~\bibnamefont
  {Olivucci}}, \bibinfo {author} {\bibfnamefont {M.}~\bibnamefont {Oppel}},
  \bibinfo {author} {\bibfnamefont {D.}~\bibnamefont {Padula}}, \bibinfo
  {author} {\bibfnamefont {R.}~\bibnamefont {Pandharkar}}, \bibinfo {author}
  {\bibfnamefont {Q.~M.}\ \bibnamefont {Phung}}, \bibinfo {author}
  {\bibfnamefont {F.}~\bibnamefont {Plasser}}, \bibinfo {author} {\bibfnamefont
  {G.}~\bibnamefont {Raggi}}, \bibinfo {author} {\bibfnamefont
  {E.}~\bibnamefont {Rebolini}}, \bibinfo {author} {\bibfnamefont
  {M.}~\bibnamefont {Reiher}}, \bibinfo {author} {\bibfnamefont
  {I.}~\bibnamefont {Rivalta}}, \bibinfo {author} {\bibfnamefont
  {D.}~\bibnamefont {Roca-Sanjuán}}, \bibinfo {author} {\bibfnamefont
  {T.}~\bibnamefont {Romig}}, \bibinfo {author} {\bibfnamefont {A.~A.}\
  \bibnamefont {Safari}}, \bibinfo {author} {\bibfnamefont {A.}~\bibnamefont
  {Sánchez-Mansilla}}, \bibinfo {author} {\bibfnamefont {A.~M.}\ \bibnamefont
  {Sand}}, \bibinfo {author} {\bibfnamefont {I.}~\bibnamefont {Schapiro}},
  \bibinfo {author} {\bibfnamefont {T.~R.}\ \bibnamefont {Scott}}, \bibinfo
  {author} {\bibfnamefont {J.}~\bibnamefont {Segarra-Martí}}, \bibinfo
  {author} {\bibfnamefont {F.}~\bibnamefont {Segatta}}, \bibinfo {author}
  {\bibfnamefont {D.-C.}\ \bibnamefont {Sergentu}}, \bibinfo {author}
  {\bibfnamefont {P.}~\bibnamefont {Sharma}}, \bibinfo {author} {\bibfnamefont
  {R.}~\bibnamefont {Shepard}}, \bibinfo {author} {\bibfnamefont
  {Y.}~\bibnamefont {Shu}}, \bibinfo {author} {\bibfnamefont {J.~K.}\
  \bibnamefont {Staab}}, \bibinfo {author} {\bibfnamefont {T.~P.}\ \bibnamefont
  {Straatsma}}, \bibinfo {author} {\bibfnamefont {L.~K.}\ \bibnamefont
  {Sørensen}}, \bibinfo {author} {\bibfnamefont {B.~N.~C.}\ \bibnamefont
  {Tenorio}}, \bibinfo {author} {\bibfnamefont {D.~G.}\ \bibnamefont
  {Truhlar}}, \bibinfo {author} {\bibfnamefont {L.}~\bibnamefont {Ungur}},
  \bibinfo {author} {\bibfnamefont {M.}~\bibnamefont {Vacher}}, \bibinfo
  {author} {\bibfnamefont {V.}~\bibnamefont {Veryazov}}, \bibinfo {author}
  {\bibfnamefont {T.~A.}\ \bibnamefont {Voß}}, \bibinfo {author}
  {\bibfnamefont {O.}~\bibnamefont {Weser}}, \bibinfo {author} {\bibfnamefont
  {D.}~\bibnamefont {Wu}}, \bibinfo {author} {\bibfnamefont {X.}~\bibnamefont
  {Yang}}, \bibinfo {author} {\bibfnamefont {D.}~\bibnamefont {Yarkony}},
  \bibinfo {author} {\bibfnamefont {C.}~\bibnamefont {Zhou}}, \bibinfo {author}
  {\bibfnamefont {J.~P.}\ \bibnamefont {Zobel}},\ and\ \bibinfo {author}
  {\bibfnamefont {R.}~\bibnamefont {Lindh}},\ }\bibfield  {title} {\bibinfo
  {title} {The openmolcas web: A community-driven approach to advancing
  computational chemistry},\ }\href {https://doi.org/10.1021/acs.jctc.3c00182}
  {\bibfield  {journal} {\bibinfo  {journal} {Journal of Chemical Theory and
  Computation}\ }\textbf {\bibinfo {volume} {19}},\ \bibinfo {pages}
  {6933–6991} (\bibinfo {year} {2023})}\BibitemShut {NoStop}%
\bibitem [{\citenamefont {Carlini}\ \emph {et~al.}(2023)\citenamefont
  {Carlini}, \citenamefont {Montorsi}, \citenamefont {Wu}, \citenamefont
  {Bolognesi}, \citenamefont {Borrego-Varillas}, \citenamefont {Casavola},
  \citenamefont {Castrovilli}, \citenamefont {Chiarinelli}, \citenamefont
  {Mocci}, \citenamefont {Vismarra}, \citenamefont {Lucchini}, \citenamefont
  {Nisoli}, \citenamefont {Mukamel}, \citenamefont {Garavelli}, \citenamefont
  {Richter}, \citenamefont {Nenov},\ and\ \citenamefont
  {Avaldi}}]{Carlini2023}%
  \BibitemOpen
  \bibfield  {author} {\bibinfo {author} {\bibfnamefont {L.}~\bibnamefont
  {Carlini}}, \bibinfo {author} {\bibfnamefont {F.}~\bibnamefont {Montorsi}},
  \bibinfo {author} {\bibfnamefont {Y.}~\bibnamefont {Wu}}, \bibinfo {author}
  {\bibfnamefont {P.}~\bibnamefont {Bolognesi}}, \bibinfo {author}
  {\bibfnamefont {R.}~\bibnamefont {Borrego-Varillas}}, \bibinfo {author}
  {\bibfnamefont {A.~R.}\ \bibnamefont {Casavola}}, \bibinfo {author}
  {\bibfnamefont {M.~C.}\ \bibnamefont {Castrovilli}}, \bibinfo {author}
  {\bibfnamefont {J.}~\bibnamefont {Chiarinelli}}, \bibinfo {author}
  {\bibfnamefont {D.}~\bibnamefont {Mocci}}, \bibinfo {author} {\bibfnamefont
  {F.}~\bibnamefont {Vismarra}}, \bibinfo {author} {\bibfnamefont
  {M.}~\bibnamefont {Lucchini}}, \bibinfo {author} {\bibfnamefont
  {M.}~\bibnamefont {Nisoli}}, \bibinfo {author} {\bibfnamefont
  {S.}~\bibnamefont {Mukamel}}, \bibinfo {author} {\bibfnamefont
  {M.}~\bibnamefont {Garavelli}}, \bibinfo {author} {\bibfnamefont
  {R.}~\bibnamefont {Richter}}, \bibinfo {author} {\bibfnamefont
  {A.}~\bibnamefont {Nenov}},\ and\ \bibinfo {author} {\bibfnamefont
  {L.}~\bibnamefont {Avaldi}},\ }\bibfield  {title} {\bibinfo {title} {Electron
  and ion spectroscopy of azobenzene in the valence and core shells},\ }\href
  {https://doi.org/10.1063/5.0133824} {\bibfield  {journal} {\bibinfo
  {journal} {The Journal of Chemical Physics}\ }\textbf {\bibinfo {volume}
  {158}},\ \bibinfo {pages} {054201} (\bibinfo {year} {2023})}\BibitemShut
  {NoStop}%
\bibitem [{\citenamefont {Delcey}\ \emph {et~al.}(2019)\citenamefont {Delcey},
  \citenamefont {S{\o}rensen}, \citenamefont {Vacher}, \citenamefont {Couto},\
  and\ \citenamefont {Lundberg}}]{Delcey2019}%
  \BibitemOpen
  \bibfield  {author} {\bibinfo {author} {\bibfnamefont {M.~G.}\ \bibnamefont
  {Delcey}}, \bibinfo {author} {\bibfnamefont {L.~K.}\ \bibnamefont
  {S{\o}rensen}}, \bibinfo {author} {\bibfnamefont {M.}~\bibnamefont {Vacher}},
  \bibinfo {author} {\bibfnamefont {R.~C.}\ \bibnamefont {Couto}},\ and\
  \bibinfo {author} {\bibfnamefont {M.}~\bibnamefont {Lundberg}},\ }\bibfield
  {title} {\bibinfo {title} {Efficient calculations of a large number of highly
  excited states for multiconfigurational wavefunctions},\ }\href
  {https://doi.org/10.1002/jcc.25832} {\bibfield  {journal} {\bibinfo
  {journal} {Journal of Computational Chemistry}\ }\textbf {\bibinfo {volume}
  {40}},\ \bibinfo {pages} {1789} (\bibinfo {year} {2019})}\BibitemShut
  {NoStop}%
\bibitem [{\citenamefont {Montorsi}\ \emph {et~al.}(2022)\citenamefont
  {Montorsi}, \citenamefont {Segatta}, \citenamefont {Nenov}, \citenamefont
  {Mukamel},\ and\ \citenamefont {Garavelli}}]{Montorsi2022}%
  \BibitemOpen
  \bibfield  {author} {\bibinfo {author} {\bibfnamefont {F.}~\bibnamefont
  {Montorsi}}, \bibinfo {author} {\bibfnamefont {F.}~\bibnamefont {Segatta}},
  \bibinfo {author} {\bibfnamefont {A.}~\bibnamefont {Nenov}}, \bibinfo
  {author} {\bibfnamefont {S.}~\bibnamefont {Mukamel}},\ and\ \bibinfo {author}
  {\bibfnamefont {M.}~\bibnamefont {Garavelli}},\ }\bibfield  {title} {\bibinfo
  {title} {Soft x-ray spectroscopy simulations with multiconfigurational wave
  function theory: Spectrum completeness, sub-{eV} accuracy, and quantitative
  reproduction of line shapes},\ }\href
  {https://doi.org/10.1021/acs.jctc.1c00566} {\bibfield  {journal} {\bibinfo
  {journal} {Journal of Chemical Theory and Computation}\ }\textbf {\bibinfo
  {volume} {18}},\ \bibinfo {pages} {1003} (\bibinfo {year}
  {2022})}\BibitemShut {NoStop}%
\bibitem [{\citenamefont {Pedersen}\ \emph {et~al.}(2009)\citenamefont
  {Pedersen}, \citenamefont {Aquilante},\ and\ \citenamefont
  {Lindh}}]{Pedersen2009}%
  \BibitemOpen
  \bibfield  {author} {\bibinfo {author} {\bibfnamefont {T.~B.}\ \bibnamefont
  {Pedersen}}, \bibinfo {author} {\bibfnamefont {F.}~\bibnamefont
  {Aquilante}},\ and\ \bibinfo {author} {\bibfnamefont {R.}~\bibnamefont
  {Lindh}},\ }\bibfield  {title} {\bibinfo {title} {Density fitting with
  auxiliary basis sets from cholesky decompositions},\ }\href
  {https://doi.org/10.1007/s00214-009-0608-y} {\bibfield  {journal} {\bibinfo
  {journal} {Theoretical Chemistry Accounts}\ }\textbf {\bibinfo {volume}
  {124}},\ \bibinfo {pages} {1} (\bibinfo {year} {2009})}\BibitemShut {NoStop}%
\bibitem [{\citenamefont {Zobel}\ \emph {et~al.}(2019)\citenamefont {Zobel},
  \citenamefont {Widmark},\ and\ \citenamefont {Veryazov}}]{Zobel2019}%
  \BibitemOpen
  \bibfield  {author} {\bibinfo {author} {\bibfnamefont {J.~P.}\ \bibnamefont
  {Zobel}}, \bibinfo {author} {\bibfnamefont {P.-O.}\ \bibnamefont {Widmark}},\
  and\ \bibinfo {author} {\bibfnamefont {V.}~\bibnamefont {Veryazov}},\
  }\bibfield  {title} {\bibinfo {title} {The {ANO}-r basis set},\ }\href
  {https://doi.org/10.1021/acs.jctc.9b00873} {\bibfield  {journal} {\bibinfo
  {journal} {Journal of Chemical Theory and Computation}\ }\textbf {\bibinfo
  {volume} {16}},\ \bibinfo {pages} {278} (\bibinfo {year} {2019})}\BibitemShut
  {NoStop}%
\bibitem [{\citenamefont {Peng}\ and\ \citenamefont {Reiher}(2012)}]{Peng2012}%
  \BibitemOpen
  \bibfield  {author} {\bibinfo {author} {\bibfnamefont {D.}~\bibnamefont
  {Peng}}\ and\ \bibinfo {author} {\bibfnamefont {M.}~\bibnamefont {Reiher}},\
  }\bibfield  {title} {\bibinfo {title} {Exact decoupling of the relativistic
  fock operator},\ }\bibfield  {journal} {\bibinfo  {journal} {Theoretical
  Chemistry Accounts}\ }\textbf {\bibinfo {volume} {131}},\ \href
  {https://doi.org/10.1007/s00214-011-1081-y} {10.1007/s00214-011-1081-y}
  (\bibinfo {year} {2012})\BibitemShut {NoStop}%
\end{thebibliography}%

\end{document}


\title{Supporting Information:\\ Stimulated X-ray Raman scattering for selective preparation of \emph{dark} states bypassing optical selection rules}
\author{Francesco Montorsi}
\affiliation{Dipartimento di Chimica Industriale ``Toso Montanari'', Universit\`a di Bologna - Alma Mater Studiorum, Via Piero Gobetti 85, 40129 - Bologna, Italy}
\author{Shaul Mukamel}
\affiliation{Department of Chemistry and Department of Physics and Astronomy, University of California, Irvine, 92697, USA}
\author{Filippo Tamassia}
\author{Marco Garavelli}
\affiliation{Dipartimento di Chimica Industriale ``Toso Montanari'', Universit\`a di Bologna - Alma Mater Studiorum, Via Piero Gobetti 85, 40129 - Bologna, Italy}
\author{Francesco Segatta}
\email{francesco.segatta@unibo.it}
\affiliation{Dipartimento di Chimica Industriale ``Toso Montanari'', Universit\`a di Bologna - Alma Mater Studiorum, Via Piero Gobetti 85, 40129 - Bologna, Italy}
\author{Artur Nenov}
\email{artur.nenov@unibo.it}
\affiliation{Dipartimento di Chimica Industriale ``Toso Montanari'', Universit\`a di Bologna - Alma Mater Studiorum, Via Piero Gobetti 85, 40129 - Bologna, Italy}
\maketitle

\section{Model Hamiltonian}
The systems are described by the following Hamiltonian:

\begin{equation}
    \begin{aligned}
        \hat{H} & = \underbrace{\sum_{a\in \{\mathcal{G},\mathcal{E},\mathcal{C} \}} \epsilon_{a} \ket{a}\bra{a} }_{\hat{H}_0}+ \hat{H}_{\text{int}}
    \end{aligned}
\end{equation}
where $\hat{H}_0$ is the field-free Hamiltonian (or \emph{molecular} Hamiltonian) written in the electronic states basis, $\hat{H}_{\text{int}}$ is the (field-matter) interaction Hamiltonian, $\epsilon_{a}$ is the energy of the electronic state $a$, and $\mathcal{G},\mathcal{E},\mathcal{C}$ refer to the different electronic states manifolds considered in the main text (ground, valence excited and core-excited states, respectively). In what follows, we will label generic states (spanning the three manifolds) with indexes $a,b$; \emph{valence} states (i.e., the ground state $g$ or states in $\mathcal{E}$) with $v$, $v^{\prime}$; valence \emph{excited} states (belonging to $\mathcal{E}$), with $e$, $e^{\prime}$; finally, we label with $c$ (\emph{core}) the states belonging to the $\mathcal{C}$ manifold.

The interaction Hamiltonian can be described under the dipole approximation\cite{MukamelBook} as:
\begin{equation}\label{eq:int_ham}
    \begin{aligned}
        \hat{H}_{\text{int}} & = -\boldsymbol{E}(t)\cdot\boldsymbol{\hat{\mu}}
    \end{aligned}
\end{equation}
where $\boldsymbol{E}(t)$ is the electric field which can be generically described by the following expression: 
\begin{equation}\label{eq:E_fourier}
    \boldsymbol{E}(t) =  \boldsymbol{\hat{\epsilon}} \, \mathcal{E}(t) e^{-i\omega_0t} +  \boldsymbol{\hat{\epsilon}}^* \, \mathcal{E}^*(t) e^{+i\omega_0t}
\end{equation}
where $\boldsymbol{\hat{\epsilon}}$ is the polarization vector, $\omega_0$ is the carrier frequency and $\mathcal{E}(t)$ is the pulse envelope in the time domain. For the purpose of this paper, it is convenient to recast this expression inserting the Fourier transform of the pulse envelope (i.e, the envelope expressed in the frequency domain $\mathcal{E}(\omega)$). 
\begin{equation}
    \boldsymbol{E}(t) = \int_{-\infty}^{+\infty}\frac{d\omega}{2\pi}\, \boldsymbol{\hat{\epsilon}} \, \mathcal{E}(\omega) e^{-i(\omega+\omega_0)t} + \int_{-\infty}^{+\infty}\frac{d\omega}{2\pi}\, \boldsymbol{\hat{\epsilon}}^* \, \mathcal{E}^*(\omega) e^{+i(\omega+\omega_0)t}
    \label{eq:field}
\end{equation}
The transition dipole operator $\boldsymbol{\hat{\mu}}$ is defined as:
\begin{equation}\label{eq:dipole}
    \begin{aligned}
        \boldsymbol{\hat{\mu}} & = \sum_{v\in\{\mathcal{G},\mathcal{E}\}, c\in\mathcal{C}} \boldsymbol{\mu}_{vc}\ket{v}\bra{c} \ + 
        \boldsymbol{\mu}^\dagger_{vc}\ket{c}\bra{v}
    \end{aligned}
\end{equation}
where $\boldsymbol{\mu}_{vc}$ is the transition dipole moment vector associated to the $v\rightarrow c$ transition and $\boldsymbol{\mu}_{vc}^\dagger$ is its transpose conjugate. Note that, given this definition, we are considering dipole coupling only \textit{between} manifolds, and not \textit{within} manifolds. Using such a general notation allows to consider also complex transition dipole moments (which is, e.g., the case for transitions between spin-orbit coupled states).

\subsection{Solution of the Schr\"odinger equation: technical details}

\noindent The model Hamiltonian described in the last section has been also employed for the quantum dynamical simulation performed with the QuTiP python package\cite{Johansson2012,Johansson2013}. In both examples discussed in the main text, a Gaussian pulse characterized by a full width at half maximum (FWHM) of 500 as and by an intensity of $\sim 4\times10^{17}\,$W/cm$^{2}$ has been employed, which is consistent with the one reported in the experiment of Cryan et. al \cite{Cryan2020}, and that also takes into account the intensity averaging over the focal volume (as reported in the same experiment). The carrier frequency of the field has been instead tuned near resonance with the nitrogen K-edge ($\omega_0=396$ eV) and with the sulfur L-edge ($\omega_0=160$ eV) in the two examples, respectively. The following polarization vector has been considered for the field in both the two scenarios: $\boldsymbol{\hat{\epsilon}} = 1/\sqrt{3}(1,1,1)$. Since a realistic description of an isotropic sample requires to perform a rotational averaging over all possible samples configurations, for simplicity we considered here a sample of oriented molecules, where only the dot product between $\boldsymbol{\hat{\epsilon}}$ and $\boldsymbol{\mu}_{ab}$ has to be computed.

Population decay and coherence dephasing terms ($\gamma_{aa}$ and $\gamma_{ab}$, respectively) are assumed to be negligibly small, which is a sensible approximation in the few-fs time-scale explored.

A time step of 0.1 as has been used to propagate the wave-function.

\section{$e-g$ Coherence preparation mechanism}

\noindent The second order perturbative expansion of the density matrix can be performed employing the interaction Hamiltonian of eq. \ref{eq:int_ham}, so that 

\begin{equation}
    \begin{aligned}
        \hat{\rho}^{(2)}(t) = & -\frac{1}{\hbar^2} \int_{-\infty}^{t}dt_2\int_{-\infty}^{t_2} dt_1 \boldsymbol{E}(t_2) \, \boldsymbol{E}(t_1) \, \mathbb{G}(t)\left[\boldsymbol{\hat{\mu}}(t_2),\left[\boldsymbol{\hat{\mu}}(t_1),\hat{\rho}(-\infty)\right]\right]
    \end{aligned}
    \label{eq:rho2}
\end{equation}
where $\boldsymbol{\hat{\mu}}(t)$ is the time evolution of the transition dipole operator in the interaction picture ($\boldsymbol{\hat{\mu}}(t) = e^{+i/\hbar \, \hat{H}_0t} \, \boldsymbol{\hat{\mu}} \, e^{-i/\hbar \, \hat{H}_0t}$) and $\mathbb{G}(t)$ is the time-evolution super-operator ($\mathbb{G}(t)\, \bullet = e^{-i/\hbar \,\hat{H}_0t} \bullet\, e^{+i/\hbar \, \hat{H}_0t}$) which describes the field-free evolution of the density matrix prepared by the transition dipole operator.\cite{MukamelBook} Let us focus on the $\rho^{(2)}_{ab}(t)$ element of the second order density matrix, i.e.
\begin{equation}
    \begin{aligned}
        \rho^{(2)}_{ab}(t)=\braket{a|\hat{\rho}^{(2)}(t)|b} = & -\frac{1}{\hbar^2} \int_{-\infty}^{t}dt_2\int_{-\infty}^{t_2} dt_1 \boldsymbol{E}(t_2) \, \boldsymbol{E}(t_1) \, \braket{a|\mathbb{G}(t)\left[\boldsymbol{\hat{\mu}}(t_2),\left[\boldsymbol{\hat{\mu}}(t_1),\hat{\rho}(-\infty)\right]\right]|b}
    \end{aligned}
\end{equation}
By developing the commutators of the previous equation we get the following four terms:
\begin{equation}
    \begin{aligned}
        \braket{a|\mathbb{G}(t)\left[\boldsymbol{\hat{\mu}}(t_2),\left[\boldsymbol{\hat{\mu}}(t_1),\hat{\rho}(-\infty)\right]\right]|b} = & + \braket{a|\mathbb{G}(t) \boldsymbol{\hat{\mu}}(t_2)\boldsymbol{\hat{\mu}}(t_1)\hat{\rho}(-\infty)|b} \\
        & - \braket{a|\mathbb{G}(t) \boldsymbol{\hat{\mu}}(t_2)\hat{\rho}(-\infty)\boldsymbol{\hat{\mu}}(t_1)|b}\\
        & - \braket{a|\mathbb{G}(t) \boldsymbol{\hat{\mu}}(t_1)\hat{\rho}(-\infty)\boldsymbol{\hat{\mu}}(t_2)|b} \\
        & + \braket{a|\mathbb{G}(t) \hat{\rho}(-\infty)\boldsymbol{\hat{\mu}}(t_1)\boldsymbol{\hat{\mu}}(t_2)|b}
    \end{aligned}
    \label{eq:paths}
\end{equation}
The four resulting Liouville pathways can be further simplified putting $\hat{\rho}(-\infty) = \ket{g}\bra{g}$:
\begin{align}
    \braket{a|\mathbb{G}(t) \boldsymbol{\hat{\mu}}(t_2)\boldsymbol{\hat{\mu}}(t_1)\hat{\rho}(-\infty)|b} & = e^{-i\omega_{ga}t}\braket{a|\boldsymbol{\hat{\mu}}(t_2)\boldsymbol{\hat{\mu}}(t_1)|g}  \\
    \braket{a|\mathbb{G}(t) \boldsymbol{\hat{\mu}}(t_2)\hat{\rho}(-\infty)\boldsymbol{\hat{\mu}}(t_1)|b} & = e^{-i\omega_{ba}t} \braket{a| \boldsymbol{\hat{\mu}}(t_2)|g}\braket{g|\boldsymbol{\hat{\mu}}(t_1)|b} \label{eq:t2t1} \\
    \braket{a|\mathbb{G}(t) \boldsymbol{\hat{\mu}}(t_1)\hat{\rho}(-\infty)\boldsymbol{\hat{\mu}}(t_2)|b} & = e^{-i\omega_{ba}t} \braket{a| \boldsymbol{\hat{\mu}}(t_1)|g}\braket{g|\boldsymbol{\hat{\mu}}(t_2)|b}\label{eq:t1t2} \\ 
    \braket{a|\mathbb{G}(t) \hat{\rho}(-\infty)\boldsymbol{\hat{\mu}}(t_1)\boldsymbol{\hat{\mu}}(t_2)|b} & = e^{-i\omega_{bg}t}\braket{g|\boldsymbol{\hat{\mu}}(t_1)\boldsymbol{\hat{\mu}}(t_2)|b} 
\end{align}
where $\omega_{ij}=(\epsilon_j-\epsilon_i)/\hbar$, and we  made use of the fact that the first and the last term are different from zero only if $b=g$ or $a=g$, respectively. These two pathways are in fact contributing only to the  $\rho_{ag}(t)$ and $\rho_{gb}(t)$ elements of the total density matrix. The relation between this four pathways and the Feynman diagrams reported in Figure 2 of the main text is made apparent in what follows. Here, we develop and discuss only the first term. All the other pathways can be similarly treated. Let us focus on the $\braket{a|\boldsymbol{\hat{\mu}}(t_2)\boldsymbol{\hat{\mu}}(t_1)|g}$ term, and rewrite each of the dipole operator in the product as prescribed by equation \ref{eq:dipole}. It reads:
\begin{equation}
    \begin{aligned}
        \braket{a|\boldsymbol{\hat{\mu}}(t_2)\boldsymbol{\hat{\mu}}(t_1)|g} & = \sum_{v,v^\prime,c,c^\prime} \bra{a} e^{+i/\hbar \, \hat{H}_0t_2} \boldsymbol{\mu}_{vc}\ket{v}\bra{c} e^{-i/\hbar \, \hat{H}_0(t_2-t_1)}  \boldsymbol{\mu}^\dagger_{v^\prime c^\prime} \ket{c^\prime}\bra{v^\prime} e^{-i/\hbar \, \hat{H}_0t_1}\ket{g} \\
        & = \sum_{v,v^\prime,c,c^\prime} \boldsymbol{\mu}_{vc} \boldsymbol{\mu}^\dagger_{v^\prime c^\prime} e^{+i\epsilon_a/\hbar \,  t_2} \braket{a|v}\bra{c} e^{-i/\hbar \, \hat{H}_0(t_2-t_1)} \ket{c^\prime}\braket{v^\prime |g} e^{+i\epsilon_g/\hbar \, t_1}
        \end{aligned}
\end{equation}\noindent where, of the four terms that would appear from the dipole operator product $\boldsymbol{\hat{\mu}}(t_2)\boldsymbol{\hat{\mu}}(t_1)$, only one survives. Moreover, this term is different from zero if and only if $v^\prime=g$ and $a=v$. As a consequence of this, the generic state $a$ has to belong to the $\mathcal{G}$ or to the $\mathcal{E}$ manifold (a consequence of the way we defined the dipole operator, that couples the $\mathcal{G}/\mathcal{E}$ manifolds with the $\mathcal{C}$ manifold, but does not couple couple states within the $\mathcal{E}$ manifold). Therefore one is left with the following two pathways:
\begin{align}
    \braket{g|\boldsymbol{\hat{\mu}}(t_2)\boldsymbol{\hat{\mu}}(t_1)|g} &= \sum_{c} \boldsymbol{\mu}_{gc} \boldsymbol{\mu}^\dagger_{gc} e^{+i\epsilon_g/\hbar \,  t_2}  e^{-i/\hbar \, \epsilon_c(t_2-t_1)}  e^{+i\epsilon_g/\hbar \, t_1} = \sum_{c} \boldsymbol{\mu}_{gc} \boldsymbol{\mu}^\dagger_{gc} e^{-i\omega_{gc}(t_2-t_1)}\label{eq:gg} \\ 
    \braket{e|\boldsymbol{\hat{\mu}}(t_2)\boldsymbol{\hat{\mu}}(t_1)|g} & = \sum_{c} \boldsymbol{\mu}_{ec} \boldsymbol{\mu}^\dagger_{gc} e^{+i\epsilon_e/\hbar \,  t_2}  e^{-i/\hbar \, \epsilon_c(t_2-t_1)}  e^{+i\epsilon_g/\hbar \, t_1} = \sum_{c} \boldsymbol{\mu}_{ec} \boldsymbol{\mu}^\dagger_{gc} e^{-i\omega_{ec}t_2}e^{-i\omega_{gc}t_1} \label{eq:eg}
\end{align}
The first term (eq. \ref{eq:gg}) highlights the second-order X-ray Raman process that brings population back to the ground state; the second term (eq. \ref{eq:eg}) highlights  the possibility to create an $eg$ coherence through the X-ray Raman process. Note that $-\omega_{gc}$ and $+\omega_{ec}$ are the frequencies of the absorbed and emitted photons during the interaction with the X-ray field. By inserting $1 = e^{i\epsilon_g/\hbar \, t_2 -i\epsilon_g/\hbar \, t_2}$ and replacing $\omega_{ab}$ with the complex frequency $\tilde{\omega}_{ab} = \omega_{ab} - i\gamma_{ab}/2$ (in order to account for the dephasing of the created coherence), one obtains: 
\begin{equation}\label{eq.eg_element}
    \begin{aligned}
        \braket{e|\boldsymbol{\hat{\mu}}(t_2)\boldsymbol{\hat{\mu}}(t_1)|g} & = \sum_{c} \boldsymbol{\mu}_{ec}\boldsymbol{\mu}^\dagger_{gc}  e^{+i \tilde{\omega}_{ge} t_2}e^{-i \tilde{\omega}_{gc} (t_2 - t_1)} 
    \end{aligned}
\end{equation}
Following similar steps, one realizes that that the terms \ref{eq:t2t1} and \ref{eq:t1t2} can only contribute to $cc^\prime$ elements of the density matrix.

\begin{figure}[t]
\centering
\includegraphics[width=0.7\textwidth,angle=0]{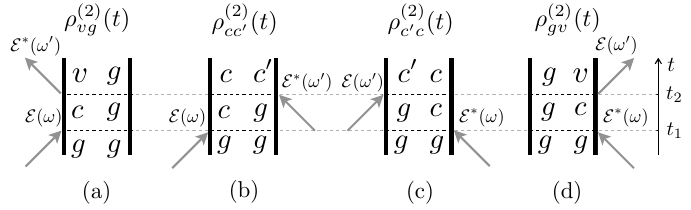}
\caption{Feynman diagrams associated with the four Liouville paths developed in equation \ref{eq:paths}. Diagram (a) is the one used to develop the Raman amplitude.}
\label{fig:second-order}
\end{figure}

Employing eq. \ref{eq.eg_element}, we can thus write  $\rho^{(2)}_{eg}(t)$ as
\begin{equation}
    \rho^{(2)}_{eg}(t) =  -\frac{1}{\hbar^2} \sum_{c} \boldsymbol{\mu}_{ec}\boldsymbol{\mu}^\dagger_{gc} \int_{-\infty}^{t}dt_2\int_{-\infty}^{t_2} dt_1 \boldsymbol{E}(t_2) \, \boldsymbol{E}(t_1) \,  e^{-i\tilde{\omega}_{ge}(t-t_2)}e^{-i\tilde{\omega}_{gc}(t_2-t_1)}
    \label{eq:rho_eg}
\end{equation}
Eq. \ref{eq:rho_eg} can be expressed in terms of the Feynman diagram a) of figure \ref{fig:second-order}.
Similar expressions can be written by applying the same procedure on the three remaining Liouville paths of eq. \ref{eq:paths}, whose Feynman diagrams are again reported in figure \ref{fig:second-order}.

We now introduce a physically motivated approximation, that greatly simplifies the integrals. We assume that the pulse duration to be extremely short in time (i.e., we assume any other ``time'' to be long compared to the pulse duration). Let us refer to the pulse time duration as $\sigma_t$ (which is reminiscent of the standard deviation of a Gaussian like pulse). At this point, one can safely assume $t\gg \sigma_t$, which corresponds to look at the prepared coherence only after the completion of such preparation process (i.e., after the pulse has completely passed through).\footnote{Note that a similar physical motivation is used to introduce the so-called Doorway-Window approximation in nonlinear spectroscopy.\cite{MukamelBook}} This allows to push the upper integration limit of the $t_2$ integral to $+\infty$ without changing the integral value. Moreover, leveraging on such a very short duration of the pulse, we can also assume the $eg$ coherence dephasing time to be much longer than the preparation time, so that it does not change the integrand (i.e., $\gamma_{ge}\sim 0$, so that $\displaystyle e^{-\gamma_{ge}/2t_2}\sim1$ and $e^{+i(\omega_{ge}-i\gamma_{ge}/2)t_2}\sim e^{+i\omega_{ge}t_2}$). Taking this into account, eq. \ref{eq:rho_eg} becomes
\begin{equation}
  \rho^{(2)}_{eg}(t) =  -\frac{1}{\hbar^2} e^{-i\tilde{\omega}_{ge}t} \sum_{c} \boldsymbol{\mu}_{ec}\boldsymbol{\mu}^\dagger_{gc} \int_{-\infty}^{+\infty}dt_2\int_{-\infty}^{t_2} dt_1 \boldsymbol{E}(t_2) \, \boldsymbol{E}(t_1) \,  e^{+i\omega_{ge}t_2}e^{-i\tilde{\omega}_{gc}(t_2-t_1)} \label{eq:rho_eg_appox}
\end{equation}

The integral reported in the last equation can be more easily solved carrying out the following change of variables: $\tau = t_2-t_1$ (\emph{interaction time}) and $T = (t_2 + t_1)/2$ (\emph{mean time}). In this new coordinate system the integral part of eq. \ref{eq:rho_eg_appox} reads:
\begin{equation}
    \int_{-\infty}^{+\infty}dT e^{+i\omega_{ge}T} \int_{0}^{+\infty} d\tau \boldsymbol{E}(T+\tau/2) \, \boldsymbol{E}(T-\tau/2) \,  e^{-i(2\tilde{\omega}_{gc}-\omega_{ge})\tau/2}
    \label{eq:ref_int_rho_2}
\end{equation}
This integral is our starting point since it can be developed differently according to the pulse scheme used to induce the Raman process.

\subsection{The single pulse limit}
\noindent In the single pulse limit the two interaction with the X-ray field occurs under the time envelope of the same pulse. By inserting the Fourier component of the electric field (eq. \ref{eq:E_fourier}) and keeping only the terms which survives the rotating wave approximation, one gets
\begin{equation}
    \int_{-\infty}^{+\infty}\frac{d\omega}{2\pi} \mathcal{E}(\omega) \int_{-\infty}^{+\infty} \frac{d\omega^\prime}{2\pi}\mathcal{E}^*(\omega^\prime)
    \int_{-\infty}^{+\infty}dT e^{i(\omega^\prime-\omega+\omega_{ge})T} \int_{0}^{+\infty} d\tau e^{+i(\omega + \omega^\prime + 2\omega_0 - 2\tilde{\omega}_{gc} +\omega_{ge})\tau/2}
\end{equation}
where both the integrals over $T$ and $\tau$ can be solved analytically leading to:
\begin{equation}
    \int_{-\infty}^{+\infty}\frac{d\omega}{2\pi} \mathcal{E}(\omega) \int_{-\infty}^{+\infty} \frac{d\omega^\prime}{2\pi}
    \frac{-4\pi \mathcal{E}^*(\omega^\prime) \delta(\omega^\prime-\omega+\omega_{ge})}{i(\omega + \omega^\prime + 2\omega_0 - 2\tilde{\omega}_{cg} + \omega_{ge})} 
\end{equation}
This finally leads to:
\begin{equation}
    \rho^{(2)}_{eg}(t) =  -\frac{i}{\hbar^2} e^{-i\tilde{\omega}_{ge}t} \sum_{c} (\boldsymbol{\hat{\epsilon}}^*\cdot\boldsymbol{\mu}_{ec})(\boldsymbol{\hat{\epsilon}}\cdot\boldsymbol{\mu}^{\dagger}_{gc}) \int_{-\infty}^{+\infty}\frac{d\omega}{2\pi} 
    \frac{\mathcal{E}(\omega) \mathcal{E}^*(\omega-\omega_{ge})}{\omega +  \omega_0 - \tilde{\omega}_{gc}} 
\end{equation}
which is the equation reported in the main text. Moreover, this last equation can be recast in an alternative form which express the entire Raman process as single event. 
 To that end it is possible to introduce the polarizability operator $\hat{\alpha}$ in the dipole approximation\cite{Mukamel2013}:    
\begin{equation*}
    \rho^{(2)}_{eg}(t) =  -\frac{i}{\hbar^2} e^{-i\tilde{\omega}_{ge}t}\alpha^{(2)}_{eg}
\end{equation*}
where $\alpha^{(2)}_{eg}$ corresponds to the $eg$ element of the polarizability defined by the following sum over state expression:
\begin{equation}
    \alpha^{(2)}_{eg} = \sum_{c} (\boldsymbol{\hat{\epsilon}}^*\cdot\boldsymbol{\mu}_{ec})(\boldsymbol{\hat{\epsilon}}\cdot\boldsymbol{\mu}^{\dagger}_{gc}) \int_{-\infty}^{+\infty}\frac{d\omega}{2\pi} 
    \frac{\mathcal{E}(\omega) \mathcal{E}^*(\omega-\omega_{ge})}{\omega +  \omega_0 - \tilde{\omega}_{gc}}
\end{equation}
In concluding we note that a more general expression for the molecular polarizabilitty can be obtained employing the so-called minimal coupling light-matter Hamiltonian\cite{Cavaletto2023}
which goes beyond the dipole approximation. 

\section{Population preparation mechanism}
\noindent As discussed in the main text, perturbation theory up to the fourth order has to be invoked in order to describe the (X-ray) Raman induced population of a valence excited states:

\begin{equation}
    \begin{aligned}
        \hat{\rho}^{(4)}(t) = & \frac{1}{\hbar^4} \int_{-\infty}^{t}dt_4\int_{-\infty}^{t_4} dt_3 \int_{-\infty}^{t_3} dt_2 \int_{-\infty}^{t_2} dt_1\boldsymbol{E}(t_4) \, \boldsymbol{E}(t_3) \, \boldsymbol{E}(t_2) \, \boldsymbol{E}(t_1) \, \times \\
        & \times \mathbb{G}(t)\left[\boldsymbol{\hat{\mu}}(t_4),\left[\boldsymbol{\hat{\mu}}(t_3),\left[\boldsymbol{\hat{\mu}}(t_2),\left[\boldsymbol{\hat{\mu}}(t_1),\hat{\rho}(-\infty)\right]\right]\right]\right]
    \end{aligned}
    \label{eq:rho4}
\end{equation}

\begin{figure}[t]
\centering
\includegraphics[width=\textwidth,angle=0]{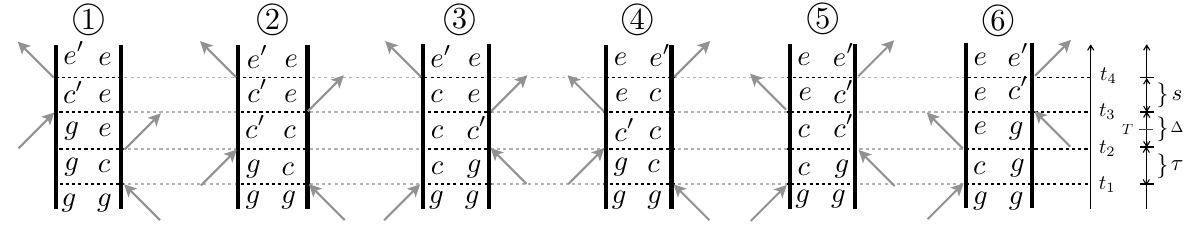}
\caption{Feynman diagrams associated with the six Liouville paths that can give rise to the population of the $e$ state (i.e, when $e=e^\prime$ in the diagrams). The definition of the time variable employed to solve the integrals of eq. \ref{eq:rho4} (i.e, $T$, $\Delta$, $s$ and $\tau$) is reported on the right of the figure.}
\label{fig:fourth-order}
\end{figure}

In this case we are interest in the matrix elements of the density matrix given by $\rho^{(4)}_{ee}$, which, for each given $e\in\mathcal{E}$, encodes the probability to populate such valence excited states. Applying a diagrammatic approach (reported in figure \ref{fig:fourth-order}) one can notice that only six Liouville pathways are responsible for the population of such $e$ states; these are:
\begin{align}
    \bra{e} \mathbb{G}(t) \boldsymbol{\hat{\mu}}(t_4)\boldsymbol{\hat{\mu}}(t_3) \hat{\rho}(-\infty) \boldsymbol{\hat{\mu}}(t_1)\boldsymbol{\hat{\mu}}(t_2)\ket{e} & = e^{-\gamma_{ee}t}\bra{e} \boldsymbol{\hat{\mu}}(t_4)\boldsymbol{\hat{\mu}}(t_3) \ket{g}\bra{g} \boldsymbol{\hat{\mu}}(t_1)\boldsymbol{\hat{\mu}}(t_2)\ket{e} \qquad\,\,\,\text{\raisebox{.5pt}{\textcircled{\raisebox{-.9pt} {1}}}} \nonumber  \\
    \bra{e} \mathbb{G}(t) \boldsymbol{\hat{\mu}}(t_4)\boldsymbol{\hat{\mu}}(t_2) \hat{\rho}(-\infty) \boldsymbol{\hat{\mu}}(t_1)\boldsymbol{\hat{\mu}}(t_3)\ket{e} & = e^{-\gamma_{ee}t} \bra{e} \boldsymbol{\hat{\mu}}(t_4)\boldsymbol{\hat{\mu}}(t_2) \ket{g}\bra{g} \boldsymbol{\hat{\mu}}(t_1)\boldsymbol{\hat{\mu}}(t_3)\ket{e} \qquad\,\,\,\text{\raisebox{.5pt}{\textcircled{\raisebox{-.9pt} {2}}}} \nonumber \\
    \bra{e} \mathbb{G}(t) \boldsymbol{\hat{\mu}}(t_4)\boldsymbol{\hat{\mu}}(t_1) \hat{\rho}(-\infty) \boldsymbol{\hat{\mu}}(t_2)\boldsymbol{\hat{\mu}}(t_3)\ket{e} & = e^{-\gamma_{ee}t} \bra{e} \boldsymbol{\hat{\mu}}(t_4)\boldsymbol{\hat{\mu}}(t_1) \ket{g} \bra{g} \boldsymbol{\hat{\mu}}(t_2)\boldsymbol{\hat{\mu}}(t_3)\ket{e} \qquad\,\,\,\text{\raisebox{.5pt}{\textcircled{\raisebox{-.9pt} {3}}}} \nonumber \\
    \bra{e} \mathbb{G}(t) \boldsymbol{\hat{\mu}}(t_3)\boldsymbol{\hat{\mu}}(t_2) \hat{\rho}(-\infty) \boldsymbol{\hat{\mu}}(t_1)\boldsymbol{\hat{\mu}}(t_4)\ket{e} & = e^{-\gamma_{ee}t} \bra{e} \boldsymbol{\hat{\mu}}(t_3)\boldsymbol{\hat{\mu}}(t_2) \ket{g} \bra{g} \boldsymbol{\hat{\mu}}(t_1)\boldsymbol{\hat{\mu}}(t_4)\ket{e} \qquad\,\,\,\text{\raisebox{.5pt}{\textcircled{\raisebox{-.9pt} {4}}}} \nonumber \\
    \bra{e} \mathbb{G}(t) \boldsymbol{\hat{\mu}}(t_3)\boldsymbol{\hat{\mu}}(t_1) \hat{\rho}(-\infty) \boldsymbol{\hat{\mu}}(t_2)\boldsymbol{\hat{\mu}}(t_4)\ket{e} & = e^{-\gamma_{ee}t} \bra{e} \boldsymbol{\hat{\mu}}(t_3)\boldsymbol{\hat{\mu}}(t_1) \ket{g}\bra{g} \boldsymbol{\hat{\mu}}(t_2)\boldsymbol{\hat{\mu}}(t_4)\ket{e} \qquad\,\,\,\text{\raisebox{.5pt}{\textcircled{\raisebox{-.9pt} {5}}}} \nonumber \\
    \bra{e} \mathbb{G}(t) \boldsymbol{\hat{\mu}}(t_2)\boldsymbol{\hat{\mu}}(t_1) \hat{\rho}(-\infty) \boldsymbol{\hat{\mu}}(t_3)\boldsymbol{\hat{\mu}}(t_4)\ket{e} & = e^{-\gamma_{ee}t} \bra{e} \boldsymbol{\hat{\mu}}(t_2)\boldsymbol{\hat{\mu}}(t_1) \ket{g} \bra{g} \boldsymbol{\hat{\mu}}(t_3)\boldsymbol{\hat{\mu}}(t_4)\ket{e} \qquad\,\,\,\text{\raisebox{.5pt}{\textcircled{\raisebox{-.9pt} {6}}}} \nonumber
\end{align}

\noindent where $\hat{\rho}(-\infty) = \ket{g}\bra{g}$. By looking at these expressions, one notices that \text{\raisebox{.5pt}{\textcircled{\raisebox{-.9pt} {6}}}}, \text{\raisebox{.5pt}{\textcircled{\raisebox{-.9pt} {5}}}} and \text{\raisebox{.5pt}{\textcircled{\raisebox{-.9pt} {4}}}}
are the complex conjugate of \text{\raisebox{.5pt}{\textcircled{\raisebox{-.9pt} {1}}}}, \text{\raisebox{.5pt}{\textcircled{\raisebox{-.9pt} {2}}}} and \text{\raisebox{.5pt}{\textcircled{\raisebox{-.9pt} {3}}}}, respectively (as it is apparent by looking the diagrams of figure \ref{fig:fourth-order}). We thus label the three inequivalent paths as  \text{\raisebox{.5pt}{\textcircled{\raisebox{-.9pt} {\tiny{I}}}}},\text{\raisebox{.5pt}{\textcircled{\raisebox{-.9pt} {\tiny{II}}}}},\text{\raisebox{.5pt}{\textcircled{\raisebox{-.9pt} {\tiny{III}}}}} and derive the corresponding expression for the four point dipole interaction (by inserting the definitions of the various terms ($\boldsymbol{E}(t)$, $\mathbb{G}(t)$ and $\hat{\boldsymbol{\mu}}(t)$) in the previous expressions). We thus get
\begin{equation}
    \begin{aligned}
        \text{\raisebox{.5pt}{\textcircled{\raisebox{-.9pt} {\tiny{I}}}}} & \Rightarrow e^{-\gamma_{ee}t} \, 2\Re\left[\sum_{c,c^\prime} \boldsymbol{\mu}_{gc}\boldsymbol{\mu}^\dagger_{ec}\boldsymbol{\mu}_{ec^\prime}\boldsymbol{\mu}^\dagger_{gc^\prime} e^{+\gamma_{ee}t_4} e^{-i\tilde{\omega}_{e c^\prime }(t_4-t_3)} e^{+i\tilde{\omega}_{ge}(t_3-t_2)}e^{+i\tilde{\omega}_{gc}(t_2-t_1)}\right] = 2\Re\left[\mathcal{R}_{\text{\raisebox{.5pt}{\textcircled{\raisebox{-.9pt} {\tiny{I}}}}}}\right]\\
        \text{\raisebox{.5pt}{\textcircled{\raisebox{-.9pt} {\tiny{II}}}}} & \Rightarrow e^{-\gamma_{ee}t} \, 2\Re\left[ \sum_{c,c^\prime} \boldsymbol{\mu}_{gc}\boldsymbol{\mu}^\dagger_{ec}\boldsymbol{\mu}_{ec^\prime}\boldsymbol{\mu}^\dagger_{gc^\prime} e^{+\gamma_{ee}t_4} e^{-i\tilde{\omega}_{ec^\prime}(t_4-t_3)}  e^{-i\tilde{\omega}_{cc^\prime}(t_3-t_2)} e^{+i\tilde{\omega}_{gc}(t_2-t_1)} \right]= 2\Re\left[\mathcal{R}_{\text{\raisebox{.5pt}{\textcircled{\raisebox{-.9pt} {\tiny{II}}}}}}\right]\\
        \text{\raisebox{.5pt}{\textcircled{\raisebox{-.9pt} {\tiny{III}}}}} & \Rightarrow e^{-\gamma_{ee}t} \, 2\Re\left[ \sum_{c,c^\prime} \boldsymbol{\mu}_{gc^\prime}\boldsymbol{\mu}^\dagger_{ec^\prime}\boldsymbol{\mu}_{ec}\boldsymbol{\mu}^\dagger_{gc} e^{+\gamma_{ee}t_4} e^{-i\tilde{\omega}_{ec}(t_4-t_3)}  e^{+i\tilde{\omega}_{cc^\prime}(t_3-t_2)} e^{-i\tilde{\omega}_{gc}(t_2-t_1)} \right] = 2\Re\left[\mathcal{R}_{\text{\raisebox{.5pt}{\textcircled{\raisebox{-.9pt} {\tiny{III}}}}}}\right]
    \end{aligned}
\end{equation} 

\noindent which, as expected by population terms, are real numbers (a consequence of the summation of each term with its complex conjugate).

\noindent Going back to eq. \ref{eq:rho4}, we apply the same approximations discussed for the coherence preparation mechanism ($t\gg\sigma_t$) so that, the upper integration limit of the $dt_4$ integral can be  set to $+\infty$. We also assume the lifetime of the $e$ state to be much longer than the preparation time so that $\gamma_{ee}\sim 0$ and  $e^{-\gamma_{ee}t_4}\sim 1$.  The following time variables substitution is then employed, defining: $T=(t_3+t_2)/2$, $\Delta=t_3-t_2$, $s=t_4-t_3$ and $\tau=t_2-t_1$ (and $|J(T,\Delta,s,\tau)|=1$). 

Thus, eq. \ref{eq:rho4} reads:
\begin{equation}
    \begin{aligned}
        \rho^{(4)}_{ee}(t) = & \frac{1}{\hbar^4} \int_{-\infty}^{+\infty}dT\int_{0}^{+\infty} d\Delta \int_{0}^{+\infty} d\tau \int_{0}^{+\infty} ds \, \boldsymbol{E}(T+\Delta/2+s) \, \boldsymbol{E}(T+\Delta/2) \, \times \\
        & \times \boldsymbol{E}(T-\Delta/2) \, \boldsymbol{E}(T-\Delta/2-\tau) \, 2\Re\left[\mathcal{R}_{\text{\raisebox{.5pt}{\textcircled{\raisebox{-.9pt} {\tiny{I}}}}}}+\mathcal{R}_{\text{\raisebox{.5pt}{\textcircled{\raisebox{-.9pt} {\tiny{II}}}}}}+\mathcal{R}_{\text{\raisebox{.5pt}{\textcircled{\raisebox{-.9pt} {\tiny{III}}}}}}\right]
    \end{aligned}
    \label{eq:rho4_time_sub}
\end{equation}

\subsection{Single pulse limit}
We again focus on the single pulse limit, that is described in the main text. In this case, all the required field-matter interactions occur under the time envelope of the same electromagnetic field. By inserting the Fourier components of the field defined in eq. \ref{eq:field}, the forth-order population terms of eq. \ref{eq:rho4_time_sub} become:
\begin{equation}
    \begin{aligned}
        \rho^{(4)-\text{\raisebox{.5pt}{\textcircled{\raisebox{-.9pt} {\tiny{I}}}}}}_{ee}(t) = & \frac{e^{-\gamma_{ee}t}}{\hbar^4}2\Re\sum_{c,c^\prime} (\boldsymbol{\hat{\epsilon}}^*\cdot\boldsymbol{\mu}_{gc})(\boldsymbol{\hat{\epsilon}}\cdot\boldsymbol{\mu}^\dagger_{ec})(\boldsymbol{\hat{\epsilon}}\cdot\boldsymbol{\mu}^\dagger_{gc^\prime})(\boldsymbol{\hat{\epsilon}}^*\cdot\boldsymbol{\mu}_{ec^\prime}) \int_{-\infty}^{+\infty} \frac{d\omega}{2\pi} \mathcal{E}^*(\omega) \\
        & \int_{-\infty}^{+\infty} \frac{d\omega^\prime}{2\pi} \mathcal{E}(\omega^\prime)\int_{-\infty}^{+\infty} \frac{d\omega}{2\pi} \mathcal{E}(\omega^{\prime\prime}) 
         \int_{-\infty}^{+\infty} \frac{d\omega^{\prime\prime\prime}}{2\pi} \mathcal{E}^*(\omega^{\prime\prime\prime})\int_{-\infty}^{+\infty}dT e^{i(\omega^{\prime\prime\prime}-\omega^{\prime\prime}-\omega^{\prime}+\omega)T} \\
        & \int_{0}^{+\infty} d\Delta e^{+i(\omega^{\prime\prime\prime}-\omega^{\prime\prime}+\omega^{\prime}-\omega +2\tilde{\omega}_{ge})\Delta/2} \int_{0}^{+\infty} ds e^{i(\omega^{\prime\prime\prime}+\omega_0-\tilde{\omega}_{ec^\prime})s} \int_{0}^{+\infty} d\tau e^{-i(\omega+\omega_0-\tilde{\omega}_{gc})\tau}
    \end{aligned}
\end{equation}

\begin{equation}
    \begin{aligned}
        \rho^{(4)-\text{\raisebox{.5pt}{\textcircled{\raisebox{-.9pt} {\tiny{II}}}}}}_{ee}(t) = & \frac{e^{-\gamma_{ee}t}}{\hbar^4}2\Re\sum_{c,c^\prime} (\boldsymbol{\hat{\epsilon}}^*\cdot\boldsymbol{\mu}_{gc})(\boldsymbol{\hat{\epsilon}}\cdot\boldsymbol{\mu}^\dagger_{gc^\prime})(\boldsymbol{\hat{\epsilon}}\cdot\boldsymbol{\mu}^\dagger_{ec})(\boldsymbol{\hat{\epsilon}}^*\cdot\boldsymbol{\mu}_{ec^\prime})\int_{-\infty}^{+\infty} \frac{d\omega}{2\pi} \mathcal{E}^*(\omega) \\
        & \int_{-\infty}^{+\infty} \frac{d\omega^\prime}{2\pi} \mathcal{E}(\omega^\prime)\int_{-\infty}^{+\infty} \frac{d\omega}{2\pi} \mathcal{E}(\omega^{\prime\prime}) \int_{-\infty}^{+\infty} \frac{d\omega^{\prime\prime\prime}}{2\pi} \mathcal{E}^*(\omega^{\prime\prime\prime})\int_{-\infty}^{+\infty}dT e^{i(\omega^{\prime\prime\prime}-\omega^{\prime\prime}-\omega^{\prime}+\omega)T}\\
        & \int_{0}^{+\infty} d\Delta e^{+i(\omega^{\prime\prime\prime}-\omega^{\prime\prime}+\omega^{\prime}-\omega -2\tilde{\omega}_{cc^\prime})\Delta/2}\int_{0}^{+\infty} ds e^{i(\omega^{\prime\prime\prime}+\omega_0-\tilde{\omega}_{ec^\prime})s} \int_{0}^{+\infty} d\tau e^{-i(\omega+\omega_0-\tilde{\omega}_{gc})\tau}
    \end{aligned}
\end{equation}

\begin{equation}
    \begin{aligned}
        \rho^{(4)-\text{\raisebox{.5pt}{\textcircled{\raisebox{-.9pt} {\tiny{III}}}}}}_{ee}(t) = & \frac{e^{-\gamma_{ee}t}}{\hbar^4}2\Re\sum_{c,c^\prime}
        (\boldsymbol{\hat{\epsilon}}\cdot\boldsymbol{\mu}^\dagger_{gc})(\boldsymbol{\hat{\epsilon}}^*\cdot\boldsymbol{\mu}_{gc^\prime})(\boldsymbol{\hat{\epsilon}}\cdot\boldsymbol{\mu}^\dagger_{ec^\prime})(\boldsymbol{\hat{\epsilon}}^*\cdot\boldsymbol{\mu}_{ec})\int_{-\infty}^{+\infty} \frac{d\omega}{2\pi} \mathcal{E}(\omega) \\
        & \int_{-\infty}^{+\infty} \frac{d\omega^\prime}{2\pi} \mathcal{E}^*(\omega^\prime)\int_{-\infty}^{+\infty} \frac{d\omega}{2\pi} \mathcal{E}(\omega^{\prime\prime})\int_{-\infty}^{+\infty} \frac{d\omega^{\prime\prime\prime}}{2\pi} \mathcal{E}^*(\omega^{\prime\prime\prime})\int_{-\infty}^{+\infty}dT e^{i(\omega^{\prime\prime\prime}-\omega^{\prime\prime}+\omega^{\prime}-\omega)T}\\
        & \int_{0}^{+\infty} d\Delta e^{+i(\omega^{\prime\prime\prime}-\omega^{\prime\prime}-\omega^{\prime}+\omega +2\tilde{\omega}_{cc^\prime})\Delta/2}\int_{0}^{+\infty} ds e^{i(\omega^{\prime\prime\prime}+\omega_0-\tilde{\omega}_{ec})s} \int_{0}^{+\infty} d\tau e^{+i(\omega+\omega_0-\tilde{\omega}_{gc})\tau}
    \end{aligned}
\end{equation}

\noindent These equations can be further simplified via analytical evaluation of some of the integrals, obtaining the expressions reported in the main text of the paper. These are:  

\begin{equation}
    \begin{aligned}
        \rho^{(4)-\text{\raisebox{.5pt}{\textcircled{\raisebox{-.9pt} {\tiny{I}}}}}}_{ee}(t) = & \frac{e^{-\gamma_{ee}t}}{\hbar^4}2\Re\sum_{c,c^\prime}(\boldsymbol{\hat{\epsilon}}^*\cdot\boldsymbol{\mu}_{gc})(\boldsymbol{\hat{\epsilon}}\cdot\boldsymbol{\mu}^\dagger_{ec})(\boldsymbol{\hat{\epsilon}}\cdot\boldsymbol{\mu}^\dagger_{gc^\prime})(\boldsymbol{\hat{\epsilon}}^*\cdot\boldsymbol{\mu}_{ec^\prime})
        \int_{-\infty}^{+\infty} \frac{d\omega}{2\pi} \int_{-\infty}^{+\infty} \frac{d\omega^\prime}{2\pi}\\
        & \int_{-\infty}^{+\infty} \frac{d\omega^{\prime\prime}}{2\pi}\frac{\mathcal{E}^*(\omega)\mathcal{E}(\omega^\prime)\mathcal{E}(\omega^{\prime\prime})\mathcal{E}^*(\omega - \omega^{\prime}-\omega^{\prime\prime})}{i(\omega^{\prime\prime}+\omega^{\prime}-\omega+\omega_0-\tilde{\omega}_{ec^\prime})(\omega^\prime-\omega+\tilde{\omega}_{ge})(\omega+\omega_0-\tilde{\omega}_{gc})}
    \end{aligned}
\end{equation}

\begin{equation}
    \begin{aligned}
        \rho^{(4)-\text{\raisebox{.5pt}{\textcircled{\raisebox{-.9pt} {\tiny{II}}}}}}_{ee}(t) = & \frac{e^{-\gamma_{ee}t}}{\hbar^4}2\Re\sum_{c,c^\prime} (\boldsymbol{\hat{\epsilon}}^*\cdot\boldsymbol{\mu}_{gc})(\boldsymbol{\hat{\epsilon}}\cdot\boldsymbol{\mu}^\dagger_{gc^\prime})(\boldsymbol{\hat{\epsilon}}\cdot\boldsymbol{\mu}^\dagger_{ec})(\boldsymbol{\hat{\epsilon}}^*\cdot\boldsymbol{\mu}_{ec^\prime})
        \int_{-\infty}^{+\infty} \frac{d\omega}{2\pi} \int_{-\infty}^{+\infty} \frac{d\omega^\prime}{2\pi} \\
        &\int_{-\infty}^{+\infty} \frac{d\omega^{\prime\prime}}{2\pi} \frac{\mathcal{E}^*(\omega)\mathcal{E}(\omega^\prime)\mathcal{E}(\omega^{\prime\prime})\mathcal{E}^*(\omega - \omega^{\prime}-\omega^{\prime\prime})}{i(\omega^{\prime\prime}+\omega^{\prime}-\omega+\omega_0-\tilde{\omega}_{ec^\prime})(\omega^\prime-\omega-\tilde{\omega}_{cc^\prime})(\omega+\omega_0-\tilde{\omega}_{gc})}
    \end{aligned}
\end{equation}

\begin{equation}
    \begin{aligned}
        \rho^{(4)-\text{\raisebox{.5pt}{\textcircled{\raisebox{-.9pt} {\tiny{III}}}}}}_{ee}(t) = & \frac{e^{-\gamma_{ee}t}}{\hbar^4}2\Re\sum_{c,c^\prime}
        (\boldsymbol{\hat{\epsilon}}\cdot\boldsymbol{\mu}^\dagger_{gc})(\boldsymbol{\hat{\epsilon}}^*\cdot\boldsymbol{\mu}_{gc^\prime})(\boldsymbol{\hat{\epsilon}}\cdot\boldsymbol{\mu}^\dagger_{ec^\prime})(\boldsymbol{\hat{\epsilon}}^*\cdot\boldsymbol{\mu}_{ec})
        \int_{-\infty}^{+\infty} \frac{d\omega}{2\pi} \int_{-\infty}^{+\infty} \frac{d\omega^\prime}{2\pi} \\
        & \int_{-\infty}^{+\infty} \frac{d\omega^{\prime\prime}}{2\pi}\frac{\mathcal{E}(\omega)\mathcal{E}^*(\omega^\prime)\mathcal{E}(\omega^{\prime\prime})\mathcal{E}^*(\omega - \omega^{\prime}+\omega^{\prime\prime})}{i(\omega^{\prime\prime}-\omega^{\prime}+\omega+\omega_0-\tilde{\omega}_{ec})(\omega-\omega^\prime+\tilde{\omega}_{cc^\prime})(\omega+\omega_0-\tilde{\omega}_{gc})}
    \end{aligned}
\end{equation}
In analogy with the coherence preparation mechanism, by replacing the sum over states part of the last equations with the corresponding element of the polarizability operator, we obtain the compact expressions:
\begin{align}
    \rho^{(4)-\text{\raisebox{.5pt}{\textcircled{\raisebox{-.9pt} {\tiny{I}}}}}}_{ee}(t) & = \frac{1}{\hbar^4} e^{-\gamma_{ee}t} 2\Re\left[\alpha^{(4)}_{ee,I}\right] \\
    \rho^{(4)-\text{\raisebox{.5pt}{\textcircled{\raisebox{-.9pt} {\tiny{II}}}}}}_{ee}(t) & = \frac{1}{\hbar^4} e^{-\gamma_{ee}t} 2\Re\left[\alpha^{(4)}_{ee,II}\right] \\
    \rho^{(4)-\text{\raisebox{.5pt}{\textcircled{\raisebox{-.9pt} {\tiny{III}}}}}}_{ee}(t) & = \frac{1}{\hbar^4} e^{-\gamma_{ee}t} 2\Re\left[\alpha^{(4)}_{ee,III}\right] 
\end{align}

We conclude this section by noticing that $e-e^{\prime}$ coherences will  also be obtained by the same 4$_{\text{th}}$ order Raman process (as it clearly appears in the development of the Feynman diagrams of Figure \ref{fig:fourth-order}, with expressions that strictly resemble those obtained above.

\section{Electronic structure: additional details}

Ab-initio multireference electronic structure calculations were performed to obtain energies, transition dipole moments and spin-orbit couplings for valence and core-excited singlet (and triplet) manifolds of the two molecular systems explored in this letter. All the calculation  were carried out with the quantum chemistry package OpenMolcas\cite{FdezGalvn2019,Aquilante2020,LiManni2023}

\begin{figure}[h]
\centering
\includegraphics[width=\textwidth,angle=0]{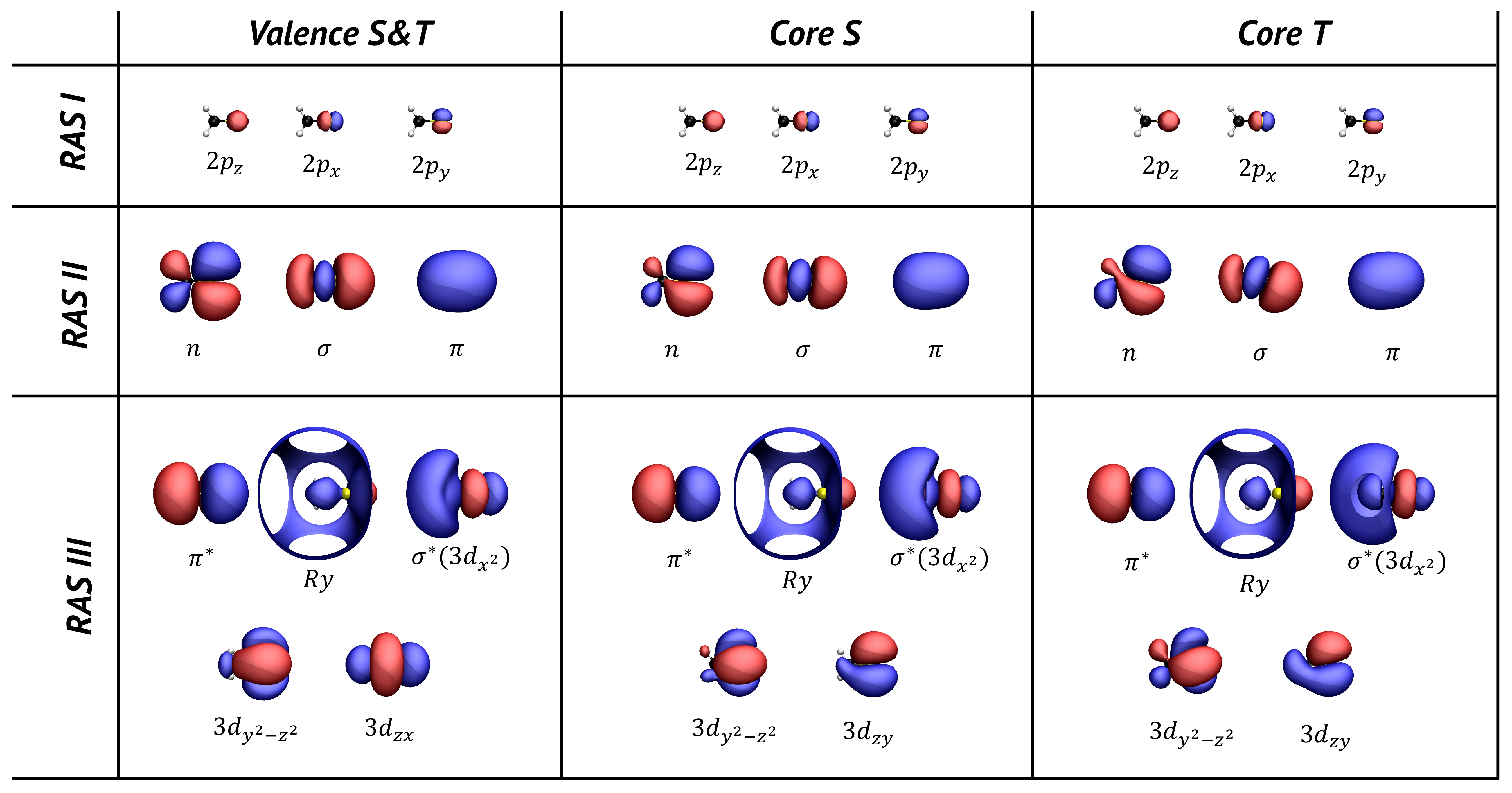}
\caption{Thio-formaldehyde active spaces employed for the computation of: valence excited state singlet and triplet manifolds (\emph{left} column); core-excited state singlet (\emph{central} column) and triplet (\emph{right} column) manifolds. Note that singlet and triplet core-excited states present a slightly different molecular orbital relaxation induced by the core hole formation.}
\label{fig:active_space}
\end{figure}

\noindent The active space (AS) employed to compute core-excited states (for both K- and L-edges) was designed as follows. It includes: (i) the core orbital of interest (i.e., the $1s$ orbital for the K-edge and the $2p$ orbitals for the L-edge), placed in RAS1; (ii) valence doubly occupied and virtual orbitals requested to target valence excited singlet and triplet states (placed in RAS2 and RAS3, respectively). The AS employed to compute K-edge states of \emph{trans}-azobenzene is the same reported in ref. \cite{Carlini2023} while the calculations on the thio-formaldehyde system are performed using the active space reported in figure \ref{fig:active_space}. Core-excited singlets and triplets states are computed in the core-valence separation (CVS) framework through a projection scheme (HEXS\cite{Delcey2019}), which sets to zero the CI-coefficients of those configuration state functions with a maximum occupation in the RAS1 subspace, thus projecting them out of the wave-function (and therefore allows to target core-hole states). Valence excited singlet and triplet manifold are obtained in a separate calculation in which the core(s) is/are doubly occupied. A previous benchmark of this protocol has demonstrated a sub-eV accuracy of core-excited  transition energies as well as a good reproduction of electronic excitations relative intensities.\cite{Montorsi2022} The same protocol has already been  successfully applied to study the nitrogen K-edge of \emph{trans}-azobenzene with accurate reproduction of the experimental XAS spectra,\cite{Carlini2023} while for thio-formaldehyde sulfur L-edge no XAS experimental spectra are available, to the best of our knowledge.

Transition dipole moments and spin-orbit couplings were obtained via the RAS state interaction (RASSI) OpenMolcas module. All calculations employed a density fitting approximation known as Cholesky decomposition\cite{Pedersen2009} and were performed utilizing a relativistically corrected atomic natural orbitals basis set (ANO-R2\cite{Zobel2019}). Scalar relativistic effects were taken into account through a decoupling of the relativistic Dirac Hamiltonian via the exact-two-component (X2C) approach\cite{Peng2012}. 

\newpage

\subsection{Thio-formaldhyde: active space}

\newpage

\subsection{Thio-formaldhyde: electronic state characterization}

\begin{table}[h]
\caption{Spin free valence states of thio-formaldehyde included in the model Hamiltonian. Dominant configuration state functions (and weights) for the various valence excited states are reported. The labels of the different molecular orbitals involved in the dominant configurations are consistent with those reported  in Figure \ref{fig:active_space}.}\label{tab1}%
\begin{tabular}{@{}ccc@{}}
\toprule
 SF state  &$\,\,$ Energy (eV) $\,\,$ & Configuration (weight)  \\
\midrule\midrule
$S_0$ & 0.00 & closed shell (0.92\%)\\
\midrule
$T_1$ & 1.80 & $n^{[1]}\pi^{*[1]}$ (0.94\%)\\
\midrule
$S_1$ & 2.16 & $n^{[1]}\pi^{*[1]}$ (0.87\%)\\
      &      & $n^{[1]}\pi^{[1]}\pi^{*[2]}$ (0.07\%)\\
\midrule
T$_2$ & 3.39 & $\pi^{[1]}\pi^{*[1]}$ (0.97\%)\\
\botrule
\end{tabular}
\end{table}

\newpage

\begin{table}[h!]
\caption{Spin free core-excited singlet states ($S^{(c)}_n$) of thio-formaldehyde included in the model Hamiltonian. The energy gap with respect to $S_0$ (in eV) and the oscillator strengths ($\mathcal{F}$) for both the $S_0\rightarrow S^{(c)}_n$ and the $S_1\rightarrow S^{(c)}_n$ transitions (in eV) are also reported here. Only $\mathcal{F}$ values grater than $10^{-5}$ are reported. The labels of the different molecular orbitals involved in the dominant configuration state functions are consistent with those reported  in Figure \ref{fig:active_space}.}\label{VStoCs}%
\begin{tabular}{@{}ccccc@{}}
\toprule
 SF state & $\,\,$ Energy (eV) $\,\,$ & $\mathcal{F}$ ($S_0\rightarrow S^{(c)}_n$) & $\mathcal{F}$ ($S_1\rightarrow S^{(c)}_n$) & Configuration (weight)  \\
 \midrule\midrule
$S^{(c)}_1$ & 161.55 & / & 6.41$\times10^{-5}$ & $2p_y^{[1]}\pi^{*[1]}$ (0.76\%)\\
            &        &   & & $2p_y^{[1]}\pi^{[1]}\pi^{*[2]}$ (0.16\%)\\
\midrule
$S^{(c)}_2$ & 161.75 & 3.41$\times10^{-4}$ & / & $2p_z^{[1]}\pi^{*[1]}$ (0.77\%)\\
            &        & & & $2p_z^{[1]}\pi^{[1]}\pi^{*[2]}$ (0.21\%)\\
\midrule
$S^{(c)}_3$ & 161.78& 7.27$\times10^{-4}$ & / & $2p_x^{[1]}\pi^{*[1]}$ (0.76\%) \\
            &        & & & $2p_x^{[1]}\pi^{[1]}\pi^{*[2]}$ (0.16\%)\\
\midrule
$S^{(c)}_4$ & 166.54& 5.13$\times10^{-4}$ & / & $2p_y^{[1]}Ry^{[1]}$ (0.76\%) \\
            &        & & & $2p_y^{[1]}\pi^{[1]}\pi^{*[1]}Ry^{[1]}$ (0.12\%)\\
\midrule
$S^{(c)}_5$ & 166.66& 3.61$\times10^{-4}$ & / & $2p_z^{[1]}Ry^{[1]}$ (0.76\%)\\
\midrule
$S^{(c)}_6$ & 166.92& 1.89$\times10^{-4}$ & / & $2p_x^{[1]}Ry^{[1]}$ (0.76\%)\\
            &        & & & $2p_x^{[1]}\pi^{[1]}\pi^{*[1]}Ry^{[1]}$ (0.11\%)\\
\botrule
\end{tabular}
\end{table}

\newpage

\begin{table}[h!]
\caption{Spin free core-excited triplet states ($T^{(c)}_n$) of thio-formaldehyde included in the model Hamiltonian. The energy gap with respect to $S_0$ (in eV) and the absolute value of the oscillator strengths ($\mathcal{F}$) for both the $T_1\rightarrow T^{(c)}_n$ and the $T_2\rightarrow T^{(c)}_n$ transitions (in eV) are also reported here. Only $\mathcal{F}$ values grater than $10^{-5}$ are reported. The labels of the different molecular orbitals involved in the dominant configuration state functions are consistent with those reported  in Figure \ref{fig:active_space}.}\label{VTtoCT}%
\begin{tabular}{@{}ccccc@{}}
\toprule
 SF state & $\,\,$ Energy (eV) $\,\,$ & $\mathcal{F}$ ($T_1\rightarrow T^{(c)}_n$) & $\mathcal{F}$ ($T_2\rightarrow T^{(c)}_n$) & Configuration (weight)  \\
 \midrule\midrule
$T^{(c)}_1$ & 161.48 & / & 3.66$\times10^{-4}$ & $2p_z^{[1]}\pi^{*[1]}$ (0.77\%) \\
            &        & & & $2p_z^{[1]}\pi^{[1]}\pi^{*[2]}$ (0.13\%)\\
\midrule
$T^{(c)}_2$ & 161.50 & 1.16$\times10^{-4}$ & / & $2p_y^{[1]}\pi^{*[1]}$ (0.76\%)\\
            &        & & & $2p_y^{[1]}\pi^{[1]}\pi^{*[2]}$ (0.14\%)\\
\midrule
$T^{(c)}_3$ & 161.71 & / & 1.19$\times10^{-4}$ & $2p_x^{[1]}\pi^{*[1]}$ (0.77\%) \\
            &        &  & & $2p_x^{[1]}\pi^{[1]}\pi^{*[2]}$ (0.14\%)\\
\midrule
$T^{(c)}_4$ & 166.46 & / & / & $2p_z^{[1]}Ry^{[1]}$ (0.76\%)\\
            &        &  &  & $2p_z^{[1]}\pi^{[1]}\pi^{*[1]}Ry^{[1]}$ (0.10\%)\\
\midrule
$T^{(c)}_5$ & 166.59 & / & / & $2p_x^{[1]}Ry^{[1]}$ (0.76\%)\\
\midrule
$T^{(c)}_6$ & 166.69 & / & / & $2p_y^{[1]}Ry^{[1]}$ (0.76\%)\\
            &        &  &  & $2p_y^{[1]}\pi^{[1]}\pi^{*[1]}Ry^{[1]}$ (0.10\%)\\
\botrule
\end{tabular}
\end{table}

\newpage

\begin{longtable}{@{}ccc@{}}
\toprule
 SO state  & $\,\,$ Energy (eV) $\,\,$ & Composition (weight)  \\
\midrule\midrule
$R_0$ & 0.00 & $S_0$ (99\%)\\
\midrule
$R_1$ & 1.80 & $T_1$ (99\%)\\
\midrule
$R_2$ & 1.80 & $T_1$ (99\%)\\
\midrule
$R_3$ & 1.80 & $T_1$ (99\%)\\
\midrule
$R_4$ & 2.16 & $S_1$ (99\%)\\
\midrule
$R_5$ & 3.39 & $T_2$ (99\%)\\
\midrule
$R_6$ & 3.39 & $T_2$ (99\%)\\
\midrule
$R_7$ & 3.39 & $T_2$ (99\%)\\
\midrule
$R_8$ & 161.12 & $T^{(c)}_1$ (54\%) + $T^{(c)}_2$ (46\%)\\
\midrule
$R_9$ & 161.12 & $T^{(c)}_1$ (53\%) + $T^{(c)}_2$ (47\%)\\
\midrule
\red{$R_{10}$} & 161.14 & $T^{(c)}_1$ (61\%) + $S^{(c)}_1$ (37\%)\\
\midrule
$R_{11}$ & 161.19 & $T^{(c)}_2$ (74\%) + $T^{(c)}_3$ (17\%)\\
\midrule
$R_{12}$ & 161.25 & $T^{(c)}_2$ (25\%) + $T^{(c)}_3$ (57\%) + $T^{(c)}_1$ (18\%)\\
\midrule
\red{$R_{13}$} & 161.27 & $T^{(c)}_3$ (57\%) + $S^{(c)}_1$ (32\%) + $T^{(c)}_1$ (11\%)\\
\midrule
\red{$R_{14}$} & 161.29 & $S^{(c)}_3$ (54\%) + $T^{(c)}_2$ (26\%) + $T^{(c)}_1$ (13\%)\\
\midrule
\red{$R_{15}$} & 161.36 & $S^{(c)}_2$ (52\%) + $T^{(c)}_3$ (47\%)\\
\midrule
$R_{16}$ & 162.33 & $T^{(c)}_3$ (43\%) + $T^{(c)}_2$ (29\%) + $T^{(c)}_1$ (28\%)\\
\midrule
\red{$R_{17}$} & 162.34 & $T^{(c)}_3$ (42\%) + $S^{(c)}_1$ (31\%) + $T^{(c)}_1$ (28\%) \\
\midrule
\red{$R_{18}$} & 162.36 & $S^{(c)}_3$ (46\%) + $T^{(c)}_2$ (28\%) + $T^{(c)}_1$ (27\%)\\
\midrule
\red{$R_{19}$} & 162.42 & $S^{(c)}_3$ (38\%) + $T^{(c)}_3$ (36\%) + $T^{(c)}_2$ (25\%)\\
\midrule
$R_{20}$ & 166.13 & $T^{(c)}_4$ (70\%) +  $T^{(c)}_5$ (24\%)\\
\midrule
$R_{21}$ & 166.14 & $T^{(c)}_4$ (66\%) +  $T^{(c)}_5$ (33\%)\\
\midrule
\red{$R_{22}$} & 166.15 & $T^{(c)}_4$ (74\%) + $S^{(c)}_5$ (15\%) + $T^{(c)}_6$ (11\%)\\
\midrule
\red{$R_{23}$} & 166.19 & $S^{(c)}_4$ (63\%) + $T^{(c)}_5$ (35\%) \\
\midrule
$R_{24}$ & 166.27 & $T^{(c)}_5$ (43\%) + $T^{(c)}_6$ (54\%) \\
\midrule
$R_{25}$ & 166.27 & $T^{(c)}_5$ (33\%) + $T^{(c)}_6$ (59\%) \\
\midrule
\red{$R_{26}$} & 166.30 &  $T^{(c)}_6$ (51\%) + $S^{(c)}_5$ (49\%)\\
\midrule
\red{$R_{27}$} & 166.39 & $S^{(c)}_6$ (48\%) + $T^{(c)}_5$ (40\%) + $T^{(c)}_4$ (12\%)\\
\midrule
$R_{28}$ & 167.34 & $T^{(c)}_6$ (40\%) + $T^{(c)}_5$ (33\%) + $T^{(c)}_4$ (27\%)\\
\midrule
\red{$R_{29}$} & 167.36 & $T^{(c)}_6$ (38\%) + $T^{(c)}_5$ (32\%) + $S^{(c)}_4$ (29\%)\\
\midrule
\red{$R_{30}$} & 167.36 & $T^{(c)}_6$ (38\%) + $S^{(c)}_5$ (36\%) + $T^{(c)}_4$ (26\%)\\
\midrule
\red{$R_{31}$} & 167.44  & $S^{(c)}_6$ (51\%) + $T^{(c)}_5$ (27\%) + $T^{(c)}_4$ (22\%)\\
\botrule
\caption{Spin orbit states ($R_n$ where $R$ means \emph{root}) of thio-formaldehyde included in the model Hamiltonian and their composition as linear combination of SF states. Linear combination of singlet and triplets core-excited states are labeled in red.}\label{SOstates}%
\end{longtable}

\begin{longtable}{@{}ccc@{}}
\toprule
 $\,\,$ Initial SO state ($i$) $\,\,$ & $\,\,$ Final SO state ($f$)$\,\,$  & $\,\,$ $\mathcal{F}$ ($i\rightarrow f$) $\,\,$  \\
\midrule\midrule
 $R_{0}$ ($S_0$) & $R_{11}$ &  3.21$\times10^{-5}$\\
\midrule
 $R_{0}$ ($S_0$) & $R_{14}$ & 3.97$\times10^{-4}$ \\
 \midrule
 $R_{0}$ ($S_0$) & $R_{15}$ & 1.82$\times10^{-4}$ \\
 \midrule
 $R_{0}$ ($S_0$) & $R_{18}$ & 3.29$\times10^{-4}$ \\
 \midrule
 $R_{0}$ ($S_0$) & $R_{19}$ & 1.26$\times10^{-4}$ \\
 \midrule
 $R_{0}$ ($S_0$) & $R_{21}$ & 2.05$\times10^{-5}$ \\
 \midrule
 $R_{0}$ ($S_0$) & $R_{22}$ & 5.10$\times10^{-5}$ \\
 \midrule
 $R_{0}$ ($S_0$) & $R_{23}$ & 3.29$\times10^{-4}$ \\
 \midrule
 $R_{0}$ ($S_0$) & $R_{25}$ & 4.18$\times10^{-5}$ \\
 \midrule
 $R_{0}$ ($S_0$) & $R_{26}$ & 1.74$\times10^{-4}$ \\
 \midrule
 $R_{0}$ ($S_0$) & $R_{27}$ & 8.34$\times10^{-4}$ \\
 \midrule
 $R_{0}$ ($S_0$) & $R_{29}$ & 1.42$\times10^{-4}$ \\
 \midrule
 $R_{0}$ ($S_0$) & $R_{30}$ & 1.37$\times10^{-4}$ \\
 \midrule
 $R_{0}$ ($S_0$) & $R_{31}$ & 1.04$\times10^{-3}$ \\
 \midrule\midrule
 $R_{1}$ ($T_1$) & $R_{8}$ & 5.28$\times10^{-5}$ \\
 \midrule
 $R_{1}$ ($T_1$) & $R_{12}$ & 2.65$\times10^{-5}$ \\
 \midrule 
 $R_{1}$ ($T_1$) & $R_{16}$ & 2.65$\times10^{-5}$ \\
 \midrule
 $R_{1}$ ($T_1$) & $R_{12}$ & 3.65$\times10^{-5}$ \\
 \midrule
 $R_{2}$ ($T_1$) & $R_{9}$ & 5.35$\times10^{-5}$ \\
 \midrule
 $R_{2}$ ($T_1$) & $R_{14}$ & 2.72$\times10^{-5}$ \\
 \midrule
 $R_{2}$ ($T_1$) & $R_{18}$ & 3.51$\times10^{-5}$ \\
 \midrule
 $R_{3}$ ($T_1$) & $R_{11}$ & 8.21$\times10^{-5}$ \\
 \midrule
 $R_{3}$ ($T_1$) & $R_{19}$ & 3.31$\times10^{-5}$ \\
 \midrule\midrule
  $R_{4}$ ($S_1$) & $R_{10}$ & 2.62$\times10^{-5}$ \\
 \midrule
  $R_{4}$ ($S_1$) & $R_{13}$ & 1.93$\times10^{-5}$ \\
 \midrule
  $R_{4}$ ($S_1$) & $R_{17}$ & 1.84$\times10^{-5}$ \\
\midrule\midrule
  $R_{5}$ ($T_2$) & $R_{9}$ & 1.94$\times10^{-4}$ \\
 \midrule
   $R_{5}$ ($T_2$) & $R_{11}$ & 1.96$\times10^{-5}$ \\
 \midrule
   $R_{5}$ ($T_2$) & $R_{13}$ & 7.43$\times10^{-5}$ \\
 \midrule
   $R_{5}$ ($T_2$) & $R_{15}$ & 5.62$\times10^{-5}$ \\
 \midrule
   $R_{5}$ ($T_2$) & $R_{16}$ & 9.77$\times10^{-5}$ \\
 \midrule
    $R_{5}$ ($T_2$) & $R_{18}$ & 4.33$\times10^{-5}$ \\
 \midrule
    $R_{6}$ ($T_2$) & $R_{8}$ & 1.96$\times10^{-4}$ \\
 \midrule
    $R_{6}$ ($T_2$) & $R_{12}$ & 6.76$\times10^{-5}$ \\
 \midrule
    $R_{6}$ ($T_2$) & $R_{13}$ & 6.74$\times10^{-5}$ \\
 \midrule
    $R_{6}$ ($T_2$) & $R_{16}$ & 1.02$\times10^{-4}$ \\
 \midrule
    $R_{6}$ ($T_2$) & $R_{17}$ & 5.00$\times10^{-5}$ \\
 \midrule
    $R_{7}$ ($T_2$) & $R_{10}$ & 2.23$\times10^{-4}$ \\
 \midrule
    $R_{7}$ ($T_2$) & $R_{12}$ & 6.77$\times10^{-5}$ \\
 \midrule
    $R_{7}$ ($T_2$) & $R_{13}$ & 4.18$\times10^{-5}$ \\
 \midrule
    $R_{7}$ ($T_2$) & $R_{16}$ & 5.13$\times10^{-5}$ \\
 \midrule
    $R_{7}$ ($T_2$) & $R_{17}$ & 1.01$\times10^{-4}$ \\
\botrule
\caption{Oscillator strengths $\mathcal{F}$ in atomic units of the $i\rightarrow f$ transition between spin-orbit states (SO). Only $\mathcal{F}$ values grater than $10^{-5}$ are reported. }\label{tab1}%
\end{longtable}

\newpage
\section{ISXRS selection rules at the L-edge (thio-formaldehyde)}

\noindent The ISXRS selection rules for processes involving L-edge intermediate states can be rationalized by combining the following concepts:
\begin{itemize}
    \item Due to to the pronounced localization of core orbitals, core excitations can be approximated as atomic transitions;
    \item $p \rightarrow p$ transitions are forbidden by symmetry and hence the brightness of a general $\chi_c\rightarrow\chi_{v^*}$ transition at the L-edge is granted by the n$s$ and/or n$d$ (n being the principal quantum number) components of $\chi_{v^*}$;
    \item The intermediate L-edge states are given by linear combinations of singlets and triplets core-excited states ($\Psi_{\mathcal{L}} = c_T\psi_{\mathcal{L}}^T+c_S\psi_{\mathcal{L}}^S $), while both the initial and the final valence states are here \emph{pure} spin-free states (i.e., they are either singlets or triplets).
\end{itemize}
When ISXRS starts from the ground (singlet) state $\Psi_\mathcal{G}$, the brightness of the excitation process (that creates a core-excited state) is dictated by the $\mathcal{G}\rightarrow\mathcal{L}$ transition dipole moment ($\boldsymbol{\hat{\mu}}_{\mathcal{GL}}$) which reads:
\begin{equation}
    \boldsymbol{\hat{\mu}}_{\mathcal{GL}} = \bra{\Psi_\mathcal{G}} \boldsymbol{\hat{\mu}} \ket{\Psi_\mathcal{L}} = c_S\bra{\Psi_\mathcal{G}} \boldsymbol{\hat{\mu}} \ket{\psi^S_\mathcal{L}}
\end{equation} \noindent where only the singlet part of the core-excited state wave function contributes to the transition brightness, since this is the only spin-allowed transition. From such core-excited state, the transition towards a given valence  state depends upon the spin of the valence state reached through the stokes process. When an L-edge state is de-excited into a valence triplet state $\psi_\mathcal{E}^T$, the transition dipole moment of the $\mathcal{L}\rightarrow\mathcal{E}$ transition ($\boldsymbol{\hat{\mu}}_{\mathcal{LE}}$) reads:  
\begin{equation}
    \boldsymbol{\hat{\mu}}_{\mathcal{GL}} = \bra{\Psi_\mathcal{L}} \boldsymbol{\hat{\mu}} \ket{\Psi_\mathcal{E}^T} = c_T\bra{\psi^T_\mathcal{L}} \boldsymbol{\hat{\mu}} \ket{\Psi^T_\mathcal{E}}
\end{equation}
\noindent and only the triplet part of the core-excited state wave function gives rise to a non-zero transition dipole interaction. Conversely, when the system is de-excited into a valence singlet sate $\Psi_\mathcal{E}^S$, the transition dipole coupling $\boldsymbol{\hat{\mu}}_{\mathcal{LE}}$ becomes:
\begin{equation}
    \boldsymbol{\hat{\mu}}_{\mathcal{GL}} = \bra{\Psi_\mathcal{L}} \boldsymbol{\hat{\mu}} \ket{\Psi_\mathcal{E}^S} = c_S\bra{\psi^S_\mathcal{L}} \boldsymbol{\hat{\mu}} \ket{\Psi^S_\mathcal{E}}
\end{equation}
This give rise to the following combined ISXRS transition elements $\boldsymbol{\hat{\mu}}_{\mathcal{GL}}\boldsymbol{\hat{\mu}}_{\mathcal{LE}}$ for the two different cases:
\begin{align}
    \Psi_\mathcal{G}\rightarrow \left(\Psi_{\mathcal{L}}\rightarrow \right)\Psi_\mathcal{E}^T: & \ c_Sc_T\bra{\Psi_\mathcal{G}} \boldsymbol{\hat{\mu}} \ket{\psi^S_\mathcal{L}}\bra{\psi^T_\mathcal{L}} \boldsymbol{\hat{\mu}} \ket{\Psi^T_\mathcal{E}}\\
    \Psi_\mathcal{G}\rightarrow \left(\Psi_{\mathcal{L}}\rightarrow \right) \Psi_\mathcal{E}^S: & \ |c_S|^2\bra{\Psi_\mathcal{G}} \boldsymbol{\hat{\mu}} \ket{\psi^S_\mathcal{L}} \bra{\psi^S_\mathcal{L}} \boldsymbol{\hat{\mu}} \ket{\Psi^S_\mathcal{E}}
\end{align}\noindent which explains how a singlet to triplet transition is eventually achieved via the Raman process.

Keeping this in mind, we now try to give a deeper look into the thio-formaldehyde example proposed in the main text. In particular, the ISXRS process on that molecular system was shown to be selective towards the population of  two valence triplet states, with negligibly small population of the $S_1$ singlet. Let us analyze such result.

\subsection*{Darkness of the singlet channel}
Table \ref{VStoCs} reports the oscillator strengths in atomic units for both the $S_0 \rightarrow S^{(c)}_n$ and $S_1 \rightarrow S^{(c)}_n$ transitions, which are  respectively related to the $\bra{\Psi_\mathcal{G}} \boldsymbol{\hat{\mu}} \ket{\psi^S_\mathcal{L}}$ and $\bra{\psi^S_\mathcal{L}} \boldsymbol{\hat{\mu}} \ket{\Psi^S_\mathcal{E}}$ transition dipole moments. The three lowest energy core-excited states $S_n^{(c)}$ accessible from the ground state $S_0$ are all associated to $2p\rightarrow\pi^*$ transitions. In particular, only $2p_z\rightarrow\pi^*$ and $2p_x\rightarrow\pi^*$ transitions appears to be bright while $2p_y\rightarrow\pi^*$ is dark, as expected by symmetry selection rules.

\noindent As previously discussed,  the brightness of the $2p\rightarrow\pi^*$ transition is originated by the participation of sulfur $3s$/$3d$ atomic contributions to the $\pi^*$ molecular orbital. However, due to symmetry constrains, only $d$-type orbitals can contribute to the $\pi^*$. Moreover, since the molecule lies on the $xy$ plane, with the CS bond oriented along the $x$ axis, only the $3d_{xz}$ and $3d_{z^2}$ atomic orbitals will have the proper symmetry to be involved in the $\pi^*$. This explains the observed transition selectivity, since only $2p_x\rightarrow3d_{xz}$ and $2p_z\rightarrow3d_{z^2}$ transitions are symmetry allowed (both on the $z$ component of the transition dipole operator).

\noindent When the system is de-excited into $S_1$ ($n\pi^*$ state) from a $2p^{[1]}\pi^{*[1]}$ core-excited state, an $n\rightarrow2p$ electronic transition should occur. Again, due to symmetry restrictions, only the $3d_{xy}$ atomic orbital can participate to the $n$ orbital and hence only $3d_{xy}\rightarrow 2p_y$  transitions will be allowed (along the $x$ component of dipole coupling).

\noindent Taking all of this into account, only $2p_{x/z}^{[1]}\pi^{*[1]}$ states can be created by the pump interaction, but no $n\rightarrow 2p_{x/z}$ transition will be available for the stokes process, so that eventually the full pump-stokes Raman process will not be effective in producing the desired $n\pi^*$ singlet state, as shown in the main text.

\subsection*{Brightness of the triplet channel}

\noindent The same selection rules do apply for electronic transition between valence excited ($T_n$) and core-excited ($T^{(C)}_n$) triplets as shown in the Table \ref{VTtoCT}. However, in this case the combined ISXRS probability depends on the $\bra{\Psi_\mathcal{G}} \boldsymbol{\hat{\mu}} \ket{\psi^S_\mathcal{L}}\bra{\psi^T_\mathcal{L}} \boldsymbol{\hat{\mu}} \ket{\Psi^T_\mathcal{E}}$ product which involve both the singlet and the triplet part of the core-excited state wave function. 

Notably, the spin-orbit coupling operator cannot create linear combinations of singlet and triplet states with arbitrary angular momenta. Indeed, when combining such states, the total angular momentum (i.e., the sum of the spin and orbital angular momenta) must be conserved. In particular, the linearly combined singlet and triplet states must have different orbital angular momenta, as this should compensate their different spin angular momenta. This limits the electronic configurations of spin-free states that can combine and give rise to intermediate L-edge core-excited states.

\noindent Let us consider, for instance, a singlet state described by the following one electron excitation: $\ket{2p_x^{\alpha}\pi^{*\beta}}$. Following the conservation of total angular momentum, this state can only couple with a triplet state showing a different orbital angular momentum. If the coupled core-excited state involves the same valence orbitals this implies that the $m$ quantum number of the \emph{atomic} $2p$ core orbital must change, eventually leading to the following spin-coupled wave functions: $c_S\ket{2p_x^{\alpha}\pi^{*\beta}}+c_T\ket{2p_y^{\alpha}\pi^{*\alpha}}$ or $c_S\ket{2p_x^{\alpha}\pi^{*\beta}}+c_T\ket{2p_z^{\alpha}\pi^{*\alpha}}$.

\noindent Focusing on $c_S\ket{2p_x^{\alpha}\pi^{*\beta}}+c_T\ket{2p_y^{\alpha}\pi^{*\alpha}}$, this state can be transiently excited by the pump interaction relying on the bright $2p_x\rightarrow\pi^*$ transition on the singlet \emph{side} of the wave function; still, the stokes interaction will be able to de-excite an electron from the (doubly-occupied) $n$ orbital to the the $2p_y$ hole present in the triplet \emph{side} of the wave function, a transition that is now bright.  This explains the non-zero population that is obtained on the triplet channel of the ISXRS process for, e.g., the $n\pi^*$ valence triplet state. Similar reasoning apply also for the $\pi\pi^*$ triplet state.

\noindent We note that the proposed analysis might appear restricted to the specific case of thio-formaldehyde. However, since it focused on standard types of molecular orbitals involved in the low lying electronic transitions of the vast majority of organic molecular systems (namely, $2p$, $n$ and $\pi$ type orbitals), a similar rationale can be drawn for a large ensemble of molecules.

\newpage
\section{Field intensity: perturbative and non-perturbative regimes}

\begin{figure}[h]
\centering
\includegraphics[width=\textwidth,angle=0]{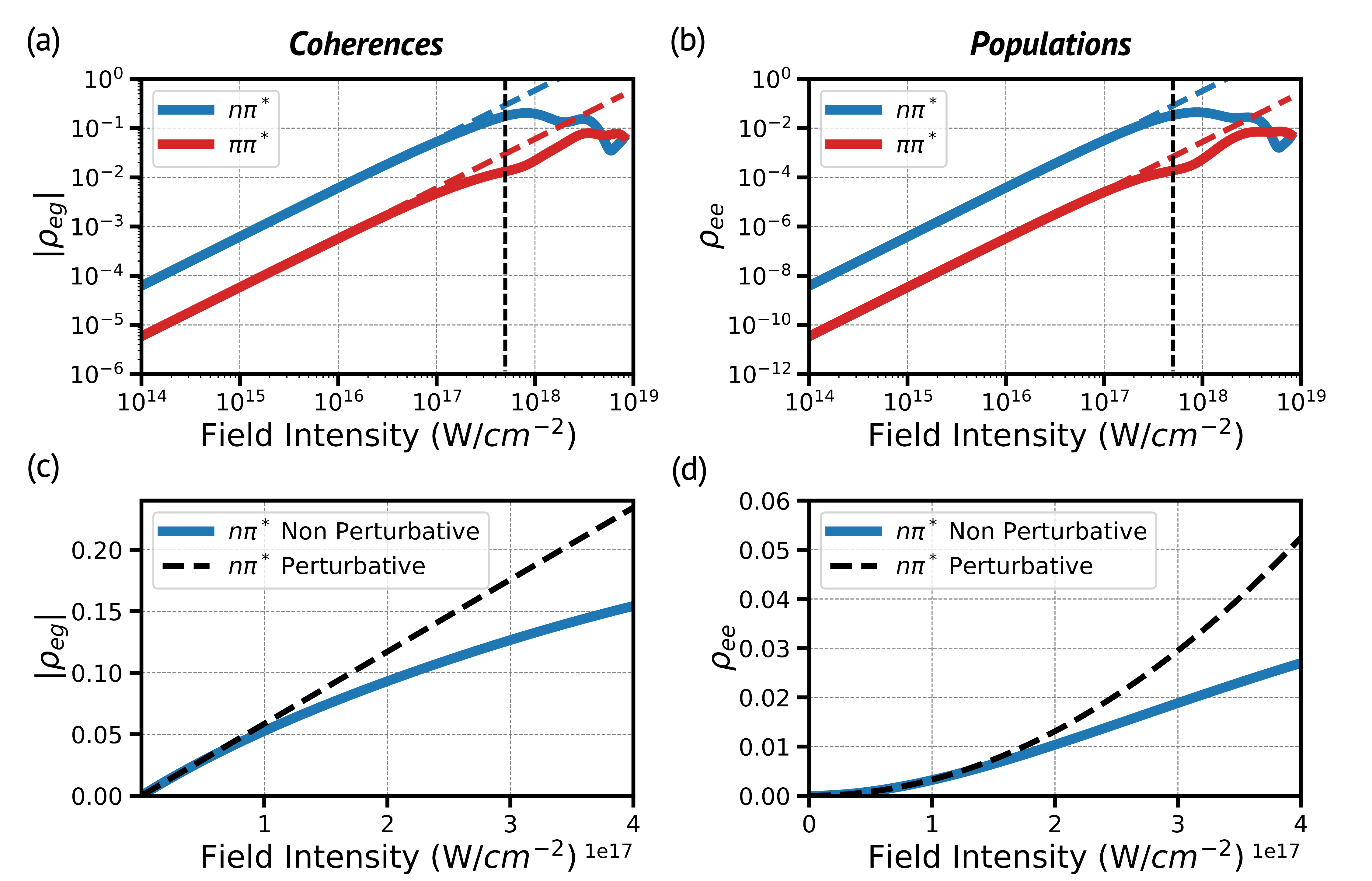}
\caption{ISXRS induced variation of (a) coherence module ($|\rho_{eg}|$) and (b) population magnitudes ($\rho_{ee}$) for various  intensities of a 500 as X-ray pulse resonant with the azobenzene nitrogen K-edge ($\omega_0 = 396$ eV). The plots are performed in logarithmic scales. The vertical lines indicates the intensity employed in the ISXRS simulations reported in the main text. Dashed lines (full lines) indicate the perturbative (non-perturbative) evaluation of coherences and populations as prescribed by eq. (2)-(6) in the main text. Panels (c) and (d) show the linear and quadratic dependence of, respectively, coherences and populations, against the field intensity (linear scales). They also highlight the regions in which the non-perturbative approach starts to deviate from the perturbative one.}
\label{fig:int_scan_azo}
\end{figure}

\begin{figure}[t]
\centering
\includegraphics[width=\textwidth,angle=0]{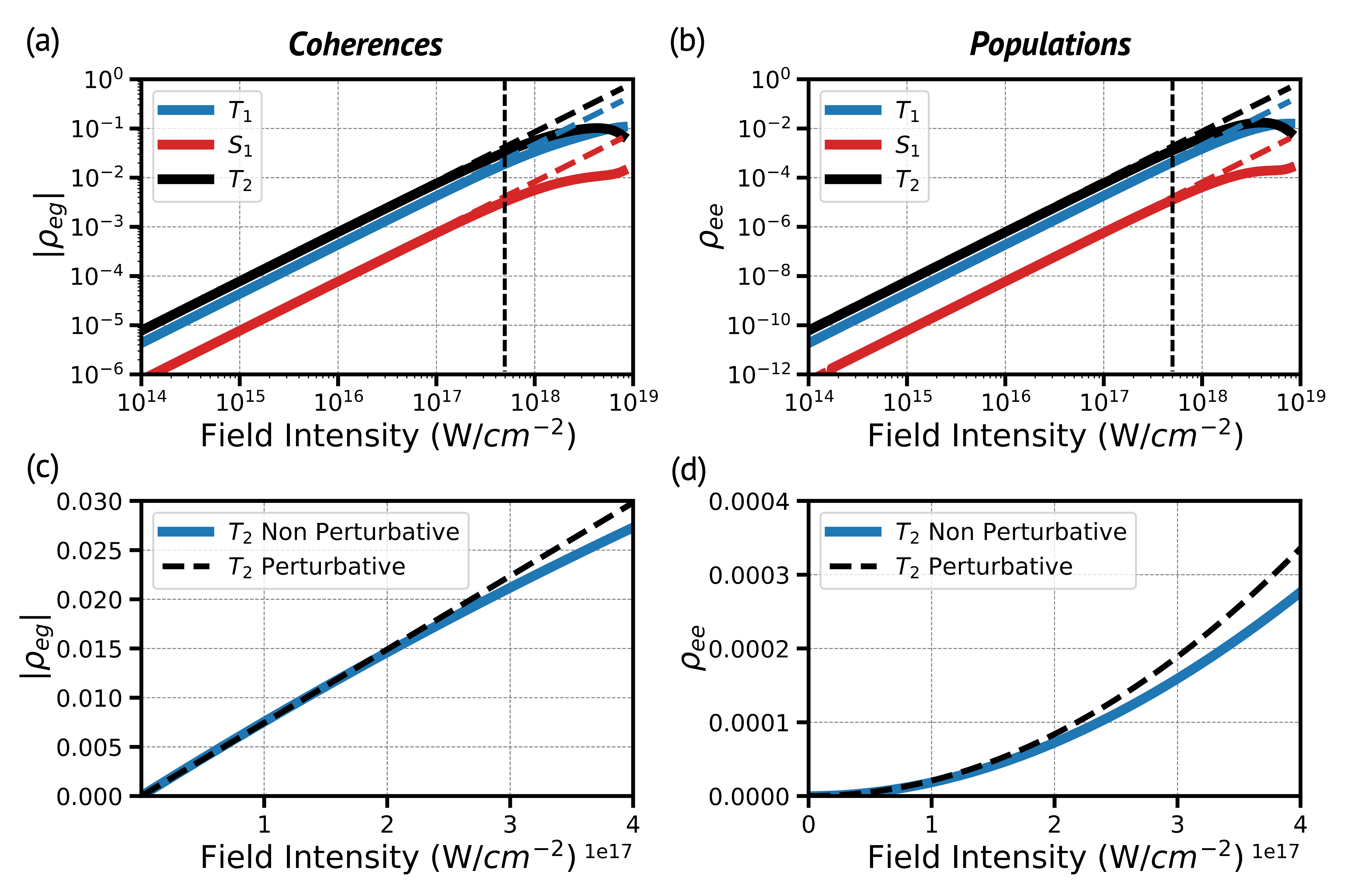}
\caption{ISXRS induced variation of (a) coherence module ($|\rho_{eg}|$) and (b) population magnitudes ($\rho_{ee}$)  for various intensities of a 500 as X-ray pulse resonant with the thio-formaldehyde sulfur L-edge ($\omega_0 = 160$ eV). The plots are performed in logarithmic scales. The vertical lines indicates the intensity employed in the ISXRS simulations reported in the main text. Dashed lines (full lines) indicate the perturbative (non-perturbative) evaluation of coherences and populations as prescribed by eq. (2)-(6) in the main text. Panels (c) and (d) show the linear and quadratic dependence of, respectively, coherences and populations, against the field intensity (linear scales). They also highlight the regions in which the non-perturbative approach starts to deviate from the perturbative one.}
\label{fig:int_scan_thio}
\end{figure}

In Figures (\ref{fig:int_scan_azo}) and (\ref{fig:int_scan_thio}) we reported the coherence and population magnitudes variations in the described ISXRS process against the field intensity for a field with the same shape of the ones employed for the ISXRS simulations reported in the main text (whose intensity of about $4\times 10^{17}$ W/cm$^{-2}$ is indicated by the vertical lines). These figures compare perturbative (dashed lines) and non-perturbative (full lines) results, and highlight the linear and quadratic dependence of, respectively, coherence and population contributions until non-perturbative effects become relevant.

Notably, deviations from the perturbative regime occur at different field intensities for the two systems. In fact, these occur earlier in azobenzene than in thio-formaldehyde, a fact that can be rationalized in terms of the larger core-to-valence transition dipole moments in the former system. For azobenzene, such deviations become relevant above the intensity employed in the ISXRS simulations reported in the main text, and produce an inversion of the $\pi\pi^*$ and $n\pi^*$ states at a field intensity of about $10^{19}$ W/cm$^{-2}$. At variance, weaker deviations are observed for thio-formaldehyde, for which the perturbative regime is still valid for a larger intensity window (and also at the field intensity value employed in the ISXRS results reported in this letter). Still, also in this case, an inversion of the triplet populations is observed at large intensity values. 

Finally, one notes that such deviations do not occur at the same intensity values for coherences and populations.

\bibliography{bib}